\documentclass[10pt]{article}
\pdfoutput=1

\usepackage{amsmath,amsfonts,amssymb,bm}

\usepackage[utf8]{inputenc}
\usepackage[T1]{fontenc}
\usepackage{hyperref}
\usepackage[pdftex]{graphicx}
\usepackage{bm,amsmath,color}
\usepackage{bbold}
\usepackage{wasysym}
\usepackage{verbatim}
\usepackage{appendix}
\usepackage{epstopdf}
\usepackage{animate}
\usepackage{xmpmulti}
\usepackage{centernot}
\usepackage{multirow}
\usepackage{tikz}
\usepackage{graphicx}
\usepackage{float}
\usepackage{mathtools}
\usepackage{comment}
\usepackage{braket}
\usepackage{tensor}

\makeatletter

\newcommand{\be}{\begin{equation}}
\newcommand{\ee}{\end{equation}}
\newcommand{\beq}{\begin{eqnarray}}
\newcommand{\eeq}{\end{eqnarray}}
\newcommand{\ba}{\begin{align}}
\newcommand{\ea}{\end{align}}

\begin{document}

\

\begin{center}
\baselineskip 24 pt 
{\LARGE \bf Thermodynamics of the   
\texorpdfstring{$q$}{q}-deformed Kittel--Shore model for spin-1 particles}
\end{center}

\bigskip
\bigskip
\begin{center}

{\sc Víctor Mariscal $^{1}$ and J. Javier Relancio$^{1,2}$}

\medskip
{$^1$Departamento de Matem\'aticas y Computaci\'on, Universidad de Burgos, 
09001 Burgos, Spain}

{$^2$Centro de Astropart\'{\i}culas y F\'{\i}sica de Altas Energ\'{\i}as (CAPA),
Universidad de Zaragoza, Zaragoza 50009, Spain}
\medskip
 
e-mail: {\href{mailto:angelb@ubu.es}{angelb@ubu.es}, \href{mailto:igsagredo@ubu.es}{igsagredo@ubu.es}, \href{mailto:vmariscal@ubu.es}{vmariscal@ubu.es}, \href{mailto:jjrelancio@ubu.es}{jjrelancio@ubu.es}}

\end{center}
 
\medskip

\begin{abstract}
Quantum algebras ($q$-algebras) have been used in the literature to deform spin-chain models. In particular, in previous works by the present authors, the Kittel--Shore Hamiltonian was deformed using $\mathfrak{su}_q(2)$ algebra. The deformed Hamiltonian was presented for a general spin, and the thermodynamic properties were studied for spin-1/2 particles. In this paper, we focus on the $q$-deformation of the Kittel--Shore model for spin-1 systems. We analyse the impact of this deformation on the thermodynamic properties and phase transitions of the system as a function of the deformation parameter $q$. Specifically, we study the specific heat, magnetic susceptibility, and magnetisation, including a detailed analysis of the Curie temperature characterising the ferromagnetic phase. Furthermore, a finite-size scaling analysis is conducted for both the ferromagnetic and antiferromagnetic cases. Results are systematically compared with those of the undeformed model to reveal the physical effects of the $q$-deformation.
\end{abstract}

\medskip
\medskip

\noindent
PACS:   
\bigskip



\tableofcontents

\section{Introduction}
The Kittel--Shore (KS) model~\cite{kittel1965development} represents an infinite-range spin model in which $N$ spins are identically coupled to each other. One of its fundamental advantages lies in the possibility of solving exactly this model in terms of statistical physics. While several studies~\cite{al1998exact,van1993heisenberg,czachor2002verification} have addressed various properties and applications of this model, most existing literature focuses almost exclusively on fermionic spin systems ($j=1/2$).

The Kittel--Shore (KS) model can be exactly solved because it is a maximally superintegrable system, where the Hamiltonian is expressed in terms of the total angular momentum quadratic Casimir, resulting from the decomposition of the tensor product of $N$ irreducible $U(\mathfrak{su}(2))$ representations into invariant subspaces~\cite{kittel1965development}. With a maximal set of commuting conserved quantities, it serves as a canonical and analytically tractable paradigm for long-range interacting spin systems. It exhibits a ferromagnetic phase transition that develops gradually as the system size increases~\cite{kittel1965development}, supporting finite-size scaling analyses in infinitely coordinated networks~\cite{botet1982size,botet1983large} in accordance with Fisher’s and Barber’s scaling hypotheses~\cite{cardy1988current}. Arbitrary spin extensions were developed in~\cite{al1998exact,van1993heisenberg,czachor2002verification}, and antiferromagnetic couplings were examined in~\cite{al1998exact,czachor2008energy}.

It is worth emphasising that the KS Hamiltonian corresponds to the Heisenberg XXX model on the complete graph~\cite{bjornberg2020quantum}; that is, a long-range generalisation of the XXX model with a fully symmetric, constant all-to-all coupling that reduces to a standard mean-field Hamiltonian once $I \to I/N$~\cite{fannes1980equilibrium}. Models on the complete graph exhibit exceptional analytical accessibility. In both the spin-$1/2$ and spin-$1$ cases, integral formulas for the thermodynamic-limit partition function in the ferromagnetic regime, explicit magnetisation, free energy, and critical exponents are available~\cite{van1993heisenberg,bjornberg2020quantum,ryan2023class,toth1990phase,penrose1991bose,bjornberg2016free,alon2021mean,bjornberg2023heisenberg}. In the spin-$1$ setting, KS-type Hamiltonians have also been studied as bilinear--biquadratic models~\cite{papanicolaou1986ground,jakab2018bilinear}. Rotational $SU(2)$ and permutational $S_N$ symmetries are fundamental in deriving most of these results.

Quantum groups, such as $U_{q}(\mathfrak{su}(2))$, provide a fundamental algebraic framework underlying the symmetries of integrable and deformed quantum spin systems~\cite{vladimir1986drinfeld,jimbo1985q,sklyanin1988boundary,pasquier1990common,gomez1996quantum,kulish1991general}. Various $q$-deformed and long-range systems with quantum group symmetry have been developed, including the braid-translated $q$-deformed periodic XXZ chain~\cite{martin1993algebraic,martin1994blob}, the $q$-deformed Haldane--Shastry model~\cite{lamers2022spin}, systems with quantum affine algebra symmetry~\cite{hakobyan1996spin}, as well as elliptic extensions and integrable deformations~\cite{matushko2022elliptic,klabbers2024deformed}. It is important to note that the algebraic $q$-deformation of the isotropic Heisenberg XXX model is fundamentally different from the standard anisotropic XXZ chain. While the XXZ model introduces a local spatial anisotropy between spin components, the $q$-deformation built from quantum group coproducts leads to nonlocal, site-dependent effective couplings.

The purpose of this paper is to describe the $q$-deformation~\cite{biedenharn1995quantum,curtright1991quantum} of the KS model for spin-1 particles. This deformation is described by the mathematical scheme of Hopf algebras~\cite{majid2000foundations}. The $q$-deformation of the KS model was constructed in \cite{ballesteros2025quantum}, where its coalgebra symmetry and integrability properties were studied.
The thermodynamic properties of the KS model were initially studied for the fermionic ($j=1/2$) case \cite{al1998exact,kac1968statistical}, as its $q$-deformation was carried out in~\cite{mariscal2025thermodynamics}. In this paper, we will focus on its extension to $j=1$, for both undeformed and deformed scenarios. The main features of this model, including the behaviour of the thermodynamic quantities, have not been studied previously for bosons, even in the undeformed case.   

Building upon the methodology established in our previous work~\cite{mariscal2025thermodynamics}, this study explores the thermodynamic properties of the system under $q$-deformation in both the ferromagnetic and antiferromagnetic regimes. Specifically, we systematically examined key thermodynamic observables, including the specific heat, magnetic susceptibility, magnetisation, Curie temperature, and phase transitions. Notably, the low-temperature approximation used in \cite{mariscal2025thermodynamics} and based on the most probable energy levels must be fundamentally reassessed, as the spin-1 ($j=1$) nature of the constituents significantly alters the underlying level statistics and state degeneracies compared with the spin-1/2 case.

In the antiferromagnetic regime, we derive explicit analytical expressions for the primary thermodynamic observables in both undeformed and $q$-deformed scenarios. Given the large physical energy scales of the system, these low-temperature approximations remain exceptionally robust across the relevant thermal ranges, providing highly precise tools for realistic laboratory measurements. Conversely, the ferromagnetic regime resists low-level analytical approximations because of the large energy gaps and the dominant weight of the excited multiplets at intermediate temperatures, requiring a combined numerical and qualitative investigation.

Furthermore, we systematically investigate the thermodynamic limit ($N \to \infty$) and perform a comprehensive finite-size scaling (FSS) analysis across both coupling regimes. In the ferromagnetic case, $q$-deformation breaks the particle-number independence observed in the undeformed limit, prompting a detailed statistical mechanics analysis to derive analytical expressions for the Curie temperature using complementary approaches. In the antiferromagnetic case, we evaluate how $q$-deformation reshapes the high-temperature convergence of thermodynamic properties.

The remainder of this paper is organised as follows. After introducing the algebraic framework of \(U_q(\mathfrak{su}(2))\) and the spin-1 Hamiltonian in Section \ref{sec:ksk}, the thermodynamic behaviour and finite-size scaling of the undeformed model are presented in Sections \ref{fer} and \ref{anti} for ferromagnetic and antiferromagnetic couplings, respectively. The physical impact of \(q\)-deformation is addressed sequentially in both the regimes. The ferromagnetic phase is first explored through its small-system thermodynamics, the derivation of the Curie temperature in the thermodynamic limit, and its finite-size scaling features, all within Section \ref{fer}. Next, the antiferromagnetic counterpart is detailed, covering the deformation effects in short chains, infinite-system asymptotics, and finite-size response in Section \ref{anti}. Finally, Section \ref{sec:conclusions} summarises the main conclusions, discusses the potential physical applications, and outlines future perspectives.

\section{The Kittel-Shore model and its \texorpdfstring{$q$}{q}-deformation}
\label{sec:ksk}
In this section, we present the Kittel-Shore model and examine its generic deformation. This framework is subsequently applied to the specific case of \(j=1\) to analyse both the energy spectrum and energy densities.

\subsection{The Kittel-Shore model}
We introduce the KS model~\cite{kittel1965development} for a system of $N$ spins subjected to an external magnetic field.
\begin{equation}
    H_{KS}=- I \sum_{i<j}^{N}\Vec{J}^{(i)}\cdot\Vec{J}^{(j)}-\gamma h\sum_{{k}=1}
    ^{N}J_{z}^{({k})}, \label{eq:hamiltonian1}
  \end{equation}
where $I$ denotes the spin-spin interaction constant, $\gamma = g\mu_{B}$ represents the product in standard notation, and $h$ is the external magnetic field. The sign of the interaction constant dictates the magnetic nature of the system, distinguishing between ferromagnetic ($I>0$) and antiferromagnetic ($I<0$) regimes.

Using the ladder operators $J_\pm = J_x \pm i J_y$, following~\cite{ballesteros2025quantum}, the Hamiltonian in Eq.~\eqref{eq:hamiltonian1} can be expressed as
  \begin{align}
    H_{KS} & = -\frac{I}{2}\left( \sum_{i=1}^{N}J_{-}^{(i)}J_{+}^{(i)}+ \sum_{{w}=1}^{N-1}\sum_{r={w}+1}^{N}\left( J_{-}^{({w})}J_{+}^{(r)}+ J_{+}^{({w})}J_{-}^{(r)}\right) + \sum_{{l}=1}^{N}J_{z}^{({l})}\sum_{s=1}^{N}\left(J_{z}^{(s)}+\mathbb{I}\right)- \sum_{{t}=1}^{N}C^{({t})}\right) -\gamma h \sum_{{k}=1}^{N}J_{z}^{({k})}, \label{finalKSpm}
  \end{align}
where $C^{(i)}$ is the Casimir of the $\mathfrak{su}(2)$ algebra, given by
  \begin{equation}
    C^{(i)}= J_{-}^{(i)}J_{+}^{(i)}+J_{z}^{(i)}(J_{z}^{(i)}+\mathbb{I})=J_{+}^{(i)}J_{-}^{(i)}
    + J_{z}^{(i)}(J_{z}^{(i)}-\mathbb{I}).
  \end{equation}
Accordingly, the eigenvalues of the Hamiltonian~\eqref{finalKSpm} for a homogeneous spin chain with constituent spin-$j$ particles take the form
  \begin{equation}
    E_{N,J,m}=-\frac{I}{2}\left(J(J+1)-N j(j+1)\right)-\gamma h m, \label{eq:energy}
  \end{equation}
where $J=\sum_{i=1}^{N}j$ and $m=\sum_{i=1}^{N}m_{i}$, being $m_{i}$ is the eigenvalue of the operator $J_{i}^{z}$. Consequently, the eigenstates of the Hamiltonian~\eqref{finalKSpm} correspond to the basis defined by the Clebsch–Gordan coefficients.

The partition function for the $N$-spins problem can be obtained from
  \begin{equation}
    Z_{N}=\sum_{J}\sum_{m=-J}^{m=J}d_{N,J}\, \text{exp}(-\beta E_{N,J,m}),
  \end{equation}
where $\beta=1/k_{B}T$ and the coefficients $d_{N,J}$ represent the multiplicity of the state of $N$ spins with total angular momentum $J$ and quantum number $m$. These coefficients are given by~\cite{morse1932theory}
  \begin{equation}
    d_{N,J}=\Omega(N,J)-\Omega(N,J+1), \quad \text{with}\quad \Omega(N,J)=\text{coefficient
    of }x^{J}\text{ in }\left(x^{j}+x^{j-1}+\cdots +x^{-j}\right)^{N}.
    \label{eq:degeneracy_Morse}
  \end{equation}

For the specific case of spin-1 particles ($j=1$), the Cartesian components $J_x, J_y, J_z$ in the standard basis $\{|1, 1\rangle, |1, 0\rangle, |1, -1\rangle\}$ are represented by the $3 \times 3$ matrices
\begin{equation}
J_x =\frac{1}{\sqrt{2}}
\begin{pmatrix}
0 & 1 & 0 \\
1 & 0 & 1 \\
0 & 1 & 0 \\
\end{pmatrix}, \qquad
J_y =\frac{1}{\sqrt{2}}
\begin{pmatrix}
0 & -i & 0 \\
i & 0 & -i \\
0 & i & 0 \\
\end{pmatrix}, \qquad
J_z = 
\begin{pmatrix}
1 & 0 & 0 \\
0 & 0 & 0 \\
0 & 0 & -1 \\
\end{pmatrix}.
\label{eq:pauli}
\end{equation}
Using the ladder operators $J_\pm = J_x \pm i J_y$, we can equivalently express the matrix representation of $\{J_+, J_-, J_z\}$ as
\begin{equation}
\label{eq:jmp}
J_+ = 
\begin{pmatrix}
0 & \sqrt{2} & 0 \\
0 & 0 & \sqrt{2} \\
0 & 0 & 0 \\
\end{pmatrix}, \qquad
J_- =
\begin{pmatrix}
0 & 0 & 0 \\
\sqrt{2} & 0 & 0 \\
0 & \sqrt{2} & 0 \\
\end{pmatrix}, \qquad
J_z = \frac{1}{2} 
\begin{pmatrix}
1 & 0 & 0 \\
0 & 0 & 0 \\
0 & 0 & -1 \\
\end{pmatrix}.
\end{equation}

\begin{figure}[H]
\centering
  \begin{minipage}{0.2\textwidth}
    \includegraphics[scale=0.7]{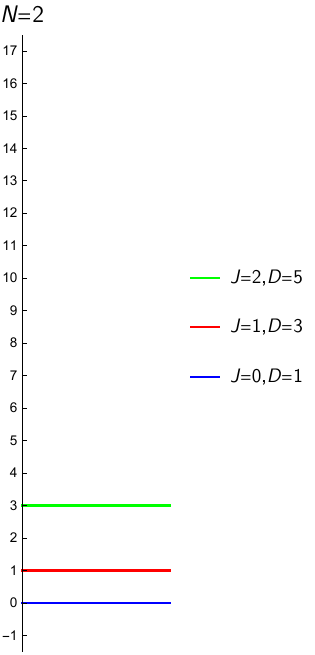}
  \end{minipage}
  \hspace{1mm}
  \begin{minipage}{0.2\textwidth}
    \includegraphics[scale=0.7]{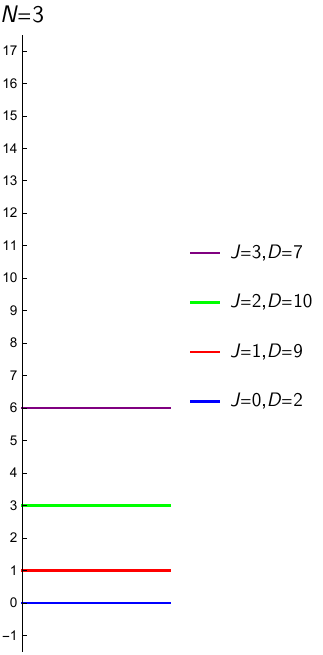}
  \end{minipage}
   \hspace{1mm}
   \begin{minipage}{0.2\textwidth}
    \includegraphics[scale=0.7]{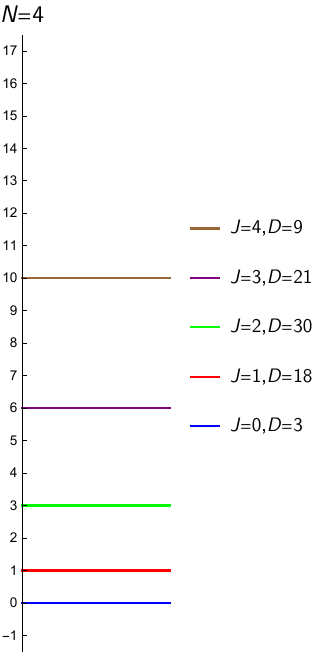}
  \end{minipage}
   \hspace{1mm}
  \begin{minipage}{0.2\textwidth}
    \includegraphics[scale=0.7]{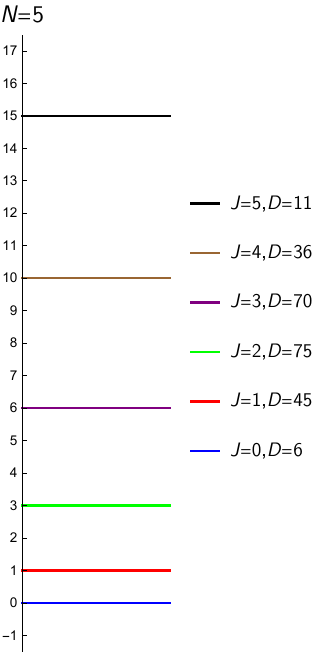}(a)
  \end{minipage}

\centering
  \begin{minipage}{0.2\textwidth}
    \includegraphics[scale=0.7]{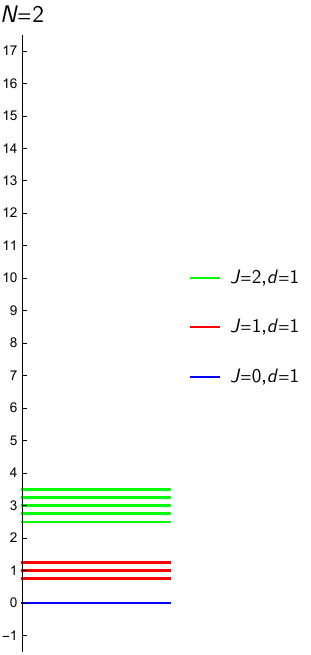}
  \end{minipage}
  \hspace{1mm}
  \begin{minipage}{0.2\textwidth}
    \includegraphics[scale=0.7]{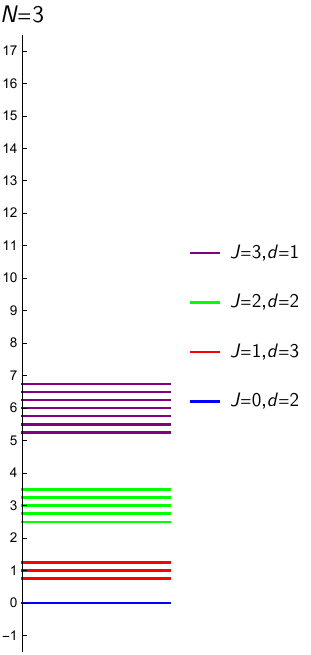}
  \end{minipage}
   \hspace{1mm}
   \begin{minipage}{0.2\textwidth}
    \includegraphics[scale=0.7]{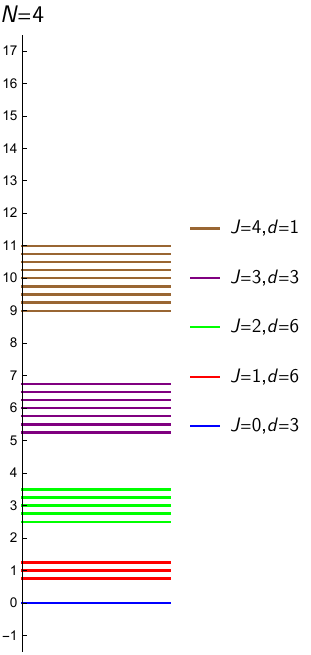}
  \end{minipage}
   \hspace{1mm}
  \begin{minipage}{0.2\textwidth}
    \includegraphics[scale=0.7]{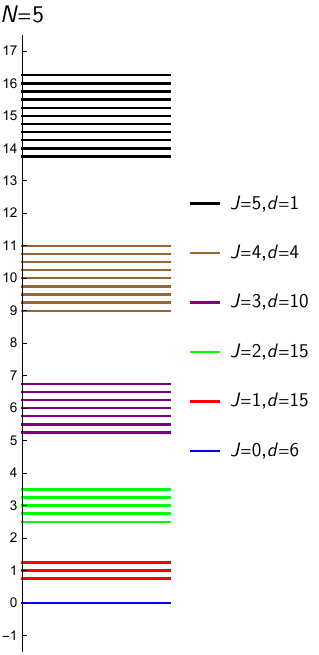}(b)
  \end{minipage}
  \caption{Energy levels (in the units $|I|$) and their degeneracies for $N=2,3,4,5$ spins with $j=1$ coupled antiferromagnetically (a) without an external magnetic field $(h=0)$ and (b) with an external magnetic field $(h=0.25)$. {Note that $1^{\otimes 2} = 2 \oplus 1 \oplus 0$, $1^{\otimes 3} = 3 \oplus 2^{\oplus2} \oplus 1^{\oplus3} \oplus 0^{\oplus2}$, $1^{\otimes 4} = 4 \oplus 3^{\oplus3} \oplus 2^{\oplus6} \oplus 1^{\oplus6} \oplus 0^{\oplus3}$ and $1^{\otimes 5} = 5 \oplus 4^{\oplus4} \oplus 3^{\oplus10} \oplus 2^{\oplus15} \oplus 1^{\oplus15} \oplus 0^{\oplus6}$.}}
  \label{qkskj1eigenB}
\end{figure}

\begin{figure}[H]
\centering
  \begin{minipage}{0.2\textwidth}
    \includegraphics[scale=0.7]{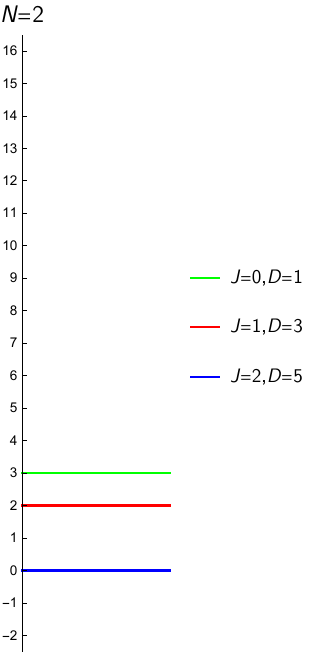}
  \end{minipage}
  \hspace{1mm}
  \begin{minipage}{0.2\textwidth}
    \includegraphics[scale=0.7]{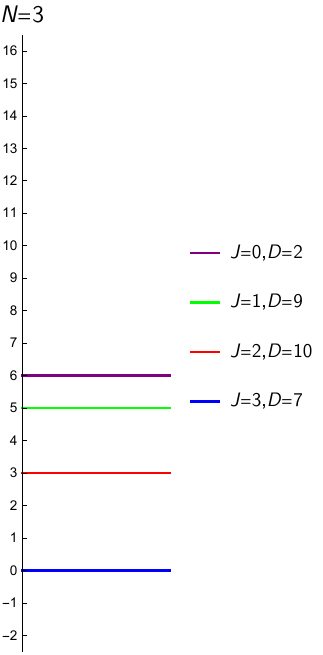}
  \end{minipage}
  \hspace{1mm}
  \begin{minipage}{0.2\textwidth}
    \includegraphics[scale=0.7]{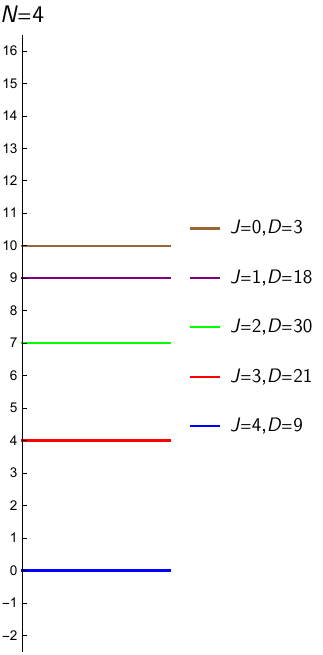}
  \end{minipage}
   \hspace{1mm}
  \begin{minipage}{0.2\textwidth}
    \includegraphics[scale=0.7]{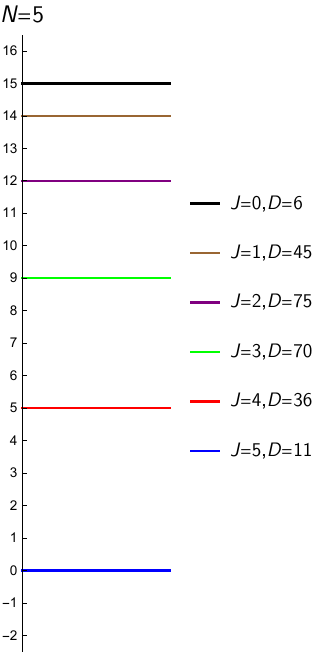}(a)
  \end{minipage}

\centering
  \begin{minipage}{0.2\textwidth}
    \includegraphics[scale=0.7]{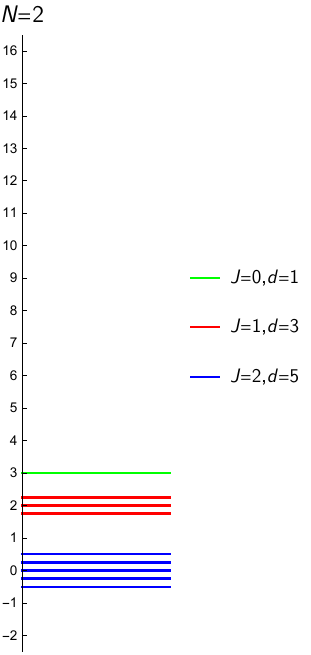}
  \end{minipage}
  \hspace{1mm}
  \begin{minipage}{0.2\textwidth}
    \includegraphics[scale=0.7]{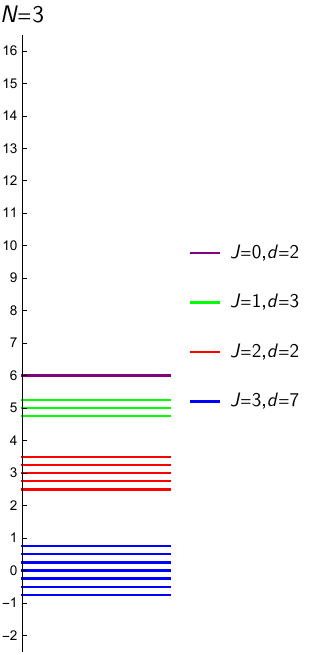}
  \end{minipage}
  \hspace{1mm}
  \begin{minipage}{0.2\textwidth}
    \includegraphics[scale=0.7]{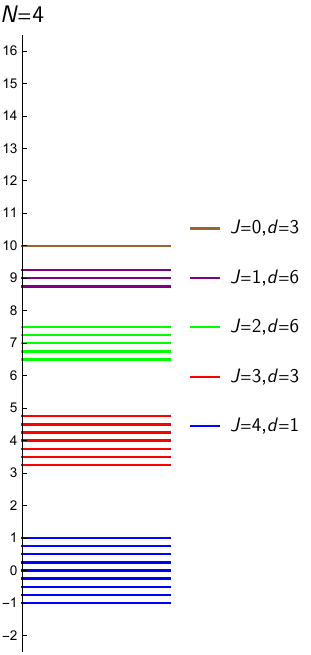}
  \end{minipage}
   \hspace{1mm}
  \begin{minipage}{0.2\textwidth}
    \includegraphics[scale=0.7]{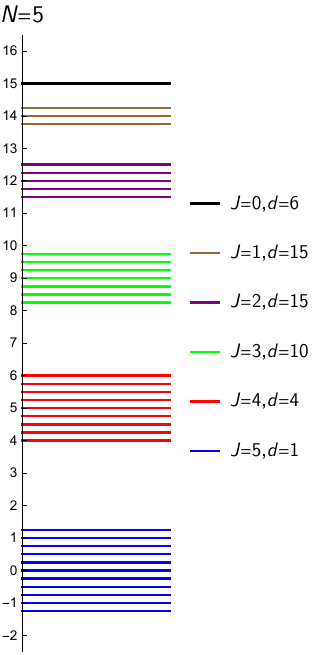}(b)
  \end{minipage}
  \caption{Energy levels (in the units $|I|$) and their degeneracies for $N=2,3,4,5$ spins with $j=1$ coupled ferromagnetically (a) without an external magnetic field $(h=0)$ and (b) with an external magnetic field $(h=0.25)$. {Note that $1^{\otimes 2} = 2 \oplus 1 \oplus 0$, $1^{\otimes 3} = 3 \oplus 2^{\oplus2} \oplus 1^{\oplus3} \oplus 0^{\oplus2}$, $1^{\otimes 4} = 4 \oplus 3^{\oplus3} \oplus 2^{\oplus6} \oplus 1^{\oplus6} \oplus 0^{\oplus3}$ and $1^{\otimes 5} = 5 \oplus 4^{\oplus4} \oplus 3^{\oplus10} \oplus 2^{\oplus15} \oplus 1^{\oplus15} \oplus 0^{\oplus6}$.}}
  \label{qkskj1eigenBFerro}
\end{figure}

The previous expression for the multiplicity of the state of $N$ spins with total angular momentum $J$ and quantum number $m$ spin-1 particles can be written as (see Appendix.~\ref{sec:appendix} for a derivation) 
\begin{equation}
    d_{N,J}= {}_2F_1\left(\frac{J-N}{2},\frac{1}{2} (J-N+1);J+1;4\right) \binom{N}{J} - {}_2F_1\left(\frac{1}{2} (J-N+1),\frac{1}{2} (J-N+2);J+2;4\right) \binom{N}{J+1},
    \label{eq:degeneracy}
\end{equation}
where ${}_{2}F_{1}$ represents Gauss's hypergeometric function,

In this $j=1$ model, the energy levels in Eq.~\eqref{eq:energy} for \(N=2,3,4,5\) are plotted for both the antiferromagnetic and ferromagnetic cases in Figs.~\ref{qkskj1eigenB} and \ref{qkskj1eigenBFerro}, respectively.

\subsection{\texorpdfstring{$q$}{q}-deformation of the Kittel-Shore model}
By virtue of its coalgebra symmetry~\cite{ballesteros1998systematic}, the Hamiltonian~\eqref{finalKSpm} can be deformed. It is worth noting that while the generators remain undeformed for $j = 1/2$, they become explicitly $q$-dependent for any higher spin representation $j > 1/2$~\cite{biedenharn1995quantum}. These generators satisfy the commutation relations (see \cite{biedenharn1995quantum}):
\begin{equation}
    [\bar{J}_z,\bar{J}_\pm]= \pm \bar{J}_\pm,\qquad [\bar{J}_+,\bar{J}_-]= \left[ 2 \bar{J}_z\right]_q = \frac{q^{\bar{J}_z}-q^{-\bar{J}_z}}{q^{1/2}-q^{-1/2}} ,
    \label{suq2}
\end{equation}
where  we have made use of the $q$-symbol
\begin{equation}
    [n]_{q}\coloneqq \frac{q^{n/2}-q^{-n/2}}{q^{1/2}-q^{-1/2}}. 
    \label{eq:qnumber}
\end{equation}
Specifically, an underlying $U_{q}(\mathfrak{su}(2))$ deformation yields a new superintegrable model (which becomes maximally superintegrable in the absence of an external magnetic field) expressed as~\cite{ballesteros2025quantum}:
\begin{align}
    \tilde H_{KS}^{q}= & -\frac{I}{2}\left(\sum_{i=1}^{N}\text{ exp}\left[-\eta \sum_{j=1}^{i-1}\bar{J}_{z}^{(j)}\right]\bar{J}_{-}^{(i)}\bar{J}_{+}^{(i)}\text{ exp}\left[\eta\sum_{h=i+1}^{N}\bar{J}_{z}^{(h)}\right]+\sum_{{w}=1}^{N-1}\sum_{r={w}+1}^{N}\left(e^{\eta/2}\bar{J}_{-}^{({w})}\bar{J}_{+}^{(r)}+e^{-\eta/2}\bar{J}_{+}^{({w})}\bar{J}_{-}^{(r)}\right)\cdot\right. \notag                                             \\
  & \cdot\text{ exp}\left[-\eta \frac{\bar{J}_{z}^{({w})}}{2}\right]\cdot\text{ exp}\left[\eta \frac{\bar{J}_{z}^{(r)}}{2}\right]\cdot\prod_{t=1}^{{w}-1}\text{ exp}\left[-\eta \bar{J}_{z}^{(t)}\right]\prod_{k=r+1}^{N}\text{ exp}\left[\eta \bar{J}_{z}^{(k)}\right]+\left[\sum_{{l}=1}^{N}\bar{J}_{z}^{({l})}\right]_{q}\left[\sum_{s=1}^{N}\bar{J}_{z}^{(s)}+\mathbb{I}\right]_{q}- \notag \\
  & -\sum_{{a}=1}^{N}C_{q}^{({a})}\Biggl)-\gamma h \sum_{{o}=1}^{N}\bar{J}_{z}^{({o})}, \label{finalqKSpm}
\end{align}
where $q=e^{\eta}$ and $C_{q}^{(i)}$ is the Casimir operator for the algebra $\mathfrak{su}_{q}(2)$
  \begin{equation}
    C_{q}^{(i)}= \bar{J}_{-}^{(i)}\bar{J}_{+}^{(i)}+ [\bar{J}_{z}^{(i)}]_{q}[\bar{J}_{z}^{(i)}+\mathbb{I}
    ]_{q}=\bar{J}_{+}^{(i)}\bar{J}_{-}^{(i)}+ [\bar{J}_{z}^{(i)}]_{q}[\bar{J}_{z}^{(i)}-\mathbb{I}]_{q},
    \label{qcas}
  \end{equation}
Then, for $q=1$ (equivalently, $\eta=0$), the deformed Hamiltonian~\eqref{finalqKSpm} leads to the undeformed KS Hamiltonian~\eqref{finalKSpm}.

By applying the action rules of the deformed angular momentum operators, the corresponding eigenvalues of this Hamiltonian are straightforwardly derived as
  \begin{equation}
    E^{q}_{N,J,m}=-\frac{I}{2}\left([J]_{q}[J+1]_{q}-N [j]_{q}[j+1]_{q}\right)-\gamma
    h m. \label{eq:qenergy}
  \end{equation}
The energy spectrum of the deformed model reveals that the deformation increases the spacing between adjacent energy levels~\cite{ballesteros2025quantum}. This widening of the level separation becomes more pronounced in the high-$J$ regime.

Accordingly, we employ the following expression to define the partition function of the deformed system:
  \begin{equation}
    Z_{N}^{q}=\sum_{J}\sum_{m=-J}^{m=+J}d_{N,J}\, \text{exp}(-\beta E^{q}_{N,J,m}),
    \label{partition}
  \end{equation}
Notably, for generic values (not roots of unity) of $q$ the degeneracies of this model remain identical to those of the undeformed case.

This partition function can be used to obtain the Helmholtz free energy, $F=-(k_{B}T/N)\log(Z_{N}^{q})$. All thermodynamic functions can be obtained by differentiating this energy. In particular, the specific heat, magnetic susceptibility, and magnetisation are given by
\begin{align}
    C_{V}&=-T\left.\frac{\partial^{2}F}{\partial T^{2}}\right|_{h=0},\\
    \chi&=-\left.\frac{\partial^{2}F}{\partial H^{2}}\right|_{h=0},\\
    M&=-\frac{\partial F}{\partial H}.
    \label{eq:thermody}
\end{align}

The invariance of the model under the transformation $q \rightarrow q^{-1}$ across both its underlying
algebraic structure and its thermodynamic behaviour was established in~\cite{mariscal2025thermodynamics}.

For the specific spin-1 case ($j=1$), the matrix representation in the $\{|1, 1\rangle, |1, 0\rangle, |1, -1\rangle\}$ basis is given by
\begin{equation}
\bar{J}_+ =
\begin{pmatrix}
0 & \sqrt{[2]_{q}} & 0 \\
0 & 0 & \sqrt{[2]_{q}} \\
0 & 0 & 0 \\
\end{pmatrix}, \qquad
\bar{J}_- =
\begin{pmatrix}
0 & 0 & 0 \\
\sqrt{[2]_{q}} & 0 & 0 \\
0 & \sqrt{[2]_{q}} & 0 \\
\end{pmatrix}, \qquad
\bar{J}_z = 
\begin{pmatrix}
1 & 0 & 0 \\
0 & 0 & 0 \\
0 & 0 & -1 \\
\end{pmatrix}.
\end{equation}

It was shown in~\cite{ballesteros2025quantum} that the arbitrary-spin generators of the $q$-deformed algebra $U_q(\mathfrak{su}(2))$ can be explicitly expressed in terms of the physical spin operators of $U(\mathfrak{su}(2))$. Indeed, for $j=1$, we find

\begin{equation}
    \bar{J}_z =J_z \, ,\qquad
 \bar{J}_+ =\sqrt{\frac{[2]_q}{2}} J_+  \, ,\qquad
\bar{J}_- =\sqrt{\frac{[2]_q}{2}} J_-\, ,
\end{equation}

\begin{figure}[H]
\centering
  \begin{minipage}{0.2\textwidth}
    \includegraphics[scale=0.7]{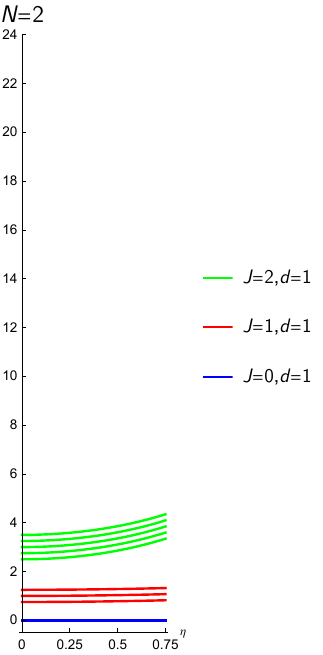}
  \end{minipage}
  \hspace{1mm}
  \begin{minipage}{0.2\textwidth}
    \includegraphics[scale=0.7]{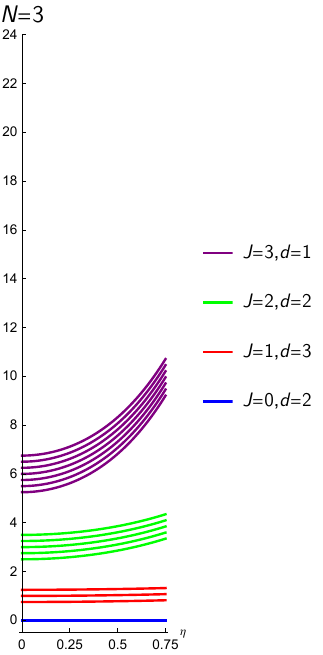}
  \end{minipage}
   \hspace{1mm}
  \begin{minipage}{0.2\textwidth}
    \includegraphics[scale=0.7]{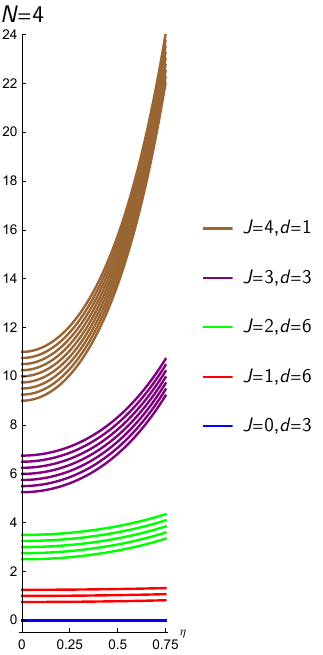}
  \end{minipage}
  \hspace{1mm}
  \begin{minipage}{0.2\textwidth}
    \includegraphics[scale=0.7]{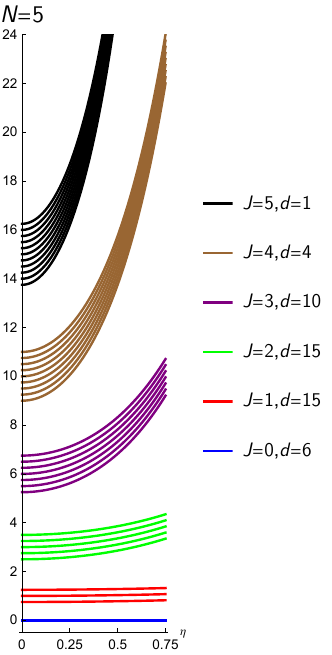}
  \end{minipage}
  \caption{Energy levels (in the units $|I|$) for $N=2,3,4,5$ spins $j=1$ respectively coupled antiferromagnetically as a function of the deformation parameter $\eta$ displaced from the ground level with an external magnetic field $(h=0.25)$.}
  \label{qkskj1eigenD}
\end{figure}

\begin{figure}[H]
\centering
  \begin{minipage}{0.2\textwidth}
    \includegraphics[scale=0.7]{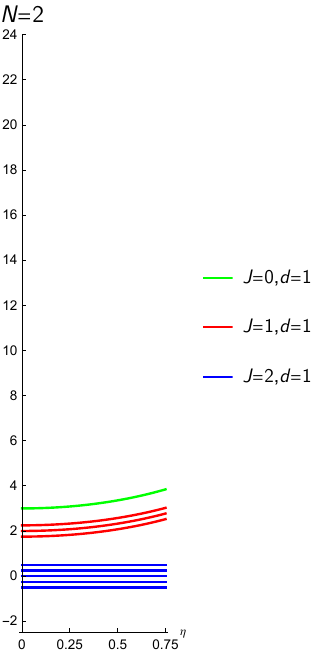}
  \end{minipage}
  \hspace{1mm}
  \begin{minipage}{0.2\textwidth}
    \includegraphics[scale=0.7]{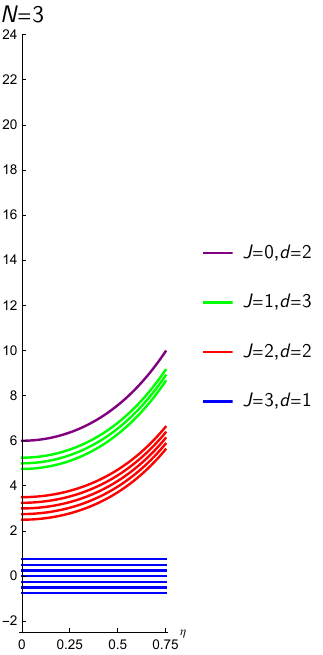}
  \end{minipage}
   \hspace{1mm}
  \begin{minipage}{0.2\textwidth}
    \includegraphics[scale=0.7]{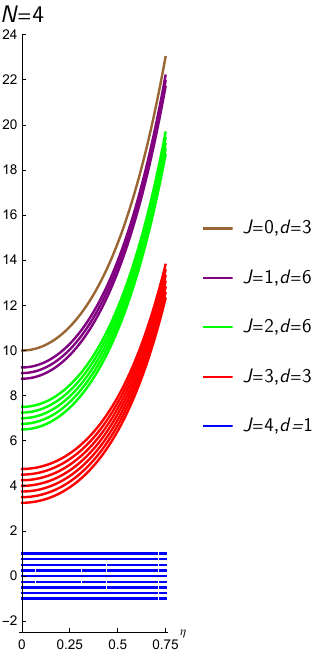}
  \end{minipage}
  \hspace{1mm}
  \begin{minipage}{0.2\textwidth}
    \includegraphics[scale=0.7]{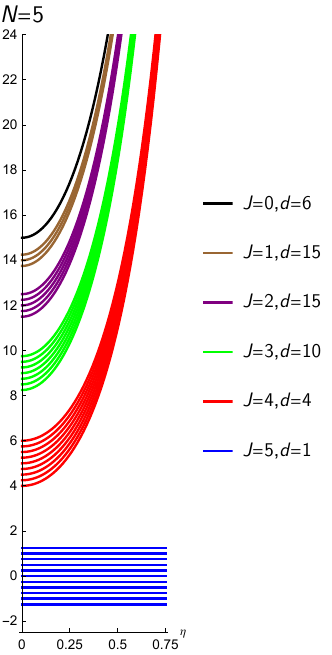}
  \end{minipage}
  \caption{Energy levels (in the units $|I|$) for $N=2,3,4,5$ spins $j=1$ respectively coupled ferromagnetically as a function of the deformation parameter $\eta$ displaced from the ground level with an external magnetic field $(h=0.25)$.}
  \label{qkskj1eigenDFerro}
\end{figure}

The eigenvalues of the $j=1$ $q$-deformed Hamiltonian (\ref{eq:qenergy}) for $N=2,3,4,5$ in the antiferromagnetic and ferromagnetic cases are presented in Figs.~\ref{qkskj1eigenD}, \ref{qkskj1eigenDFerro} respectively. 

Notably, although the global structure of the energy spectrum is preserved for a given $\eta$, the higher-lying energy levels exhibit a rapid growth with respect to $\eta$. As we will see in future sections, this behaviour plays a critical role in determining the thermodynamic properties of the model.

\subsection{Energy densities of the \texorpdfstring{$j=1$}{j=1} KS model and its \texorpdfstring{$q$}{q}-deformation}
\label{sec:qksk}
\begin{figure}[H]
  \begin{minipage}{0.4\textwidth}
    (a)\includegraphics[scale=0.9]{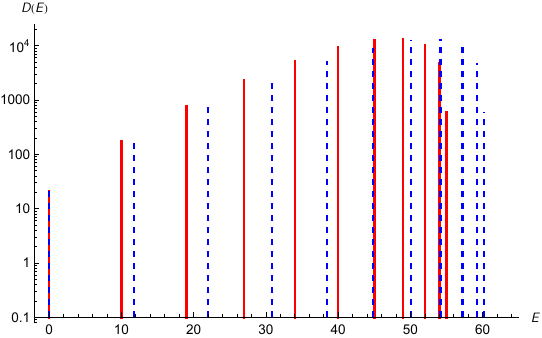}
  \end{minipage}
  \hspace{15mm}
  \begin{minipage}{0.4\textwidth}
    (b)\includegraphics[scale=0.9]{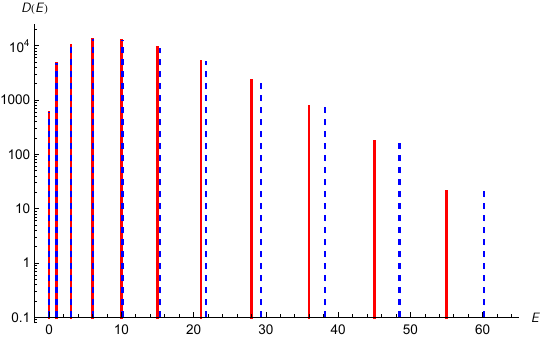}
  \end{minipage}
  \caption{Density of energy levels for $N=10$ as a function of the parameter $\eta=0$ (red) and $\eta=0.1$ (blue) (a) for the ferromagnetic case and (b) for the antiferromagnetic case.}
  \label{kskj1DensityD}
\end{figure}

Regarding the energy distribution, we next compare the deformed density with the undeformed one to assess how the deformation increases the distance between energy states. This comparison is detailed in Fig.~\ref{kskj1DensityD} for the specific case of $N=10$ particles and $\eta=0.1$.

\begin{figure}[H]
  \begin{minipage}{0.4\textwidth}
    (a)\includegraphics[scale=0.9]{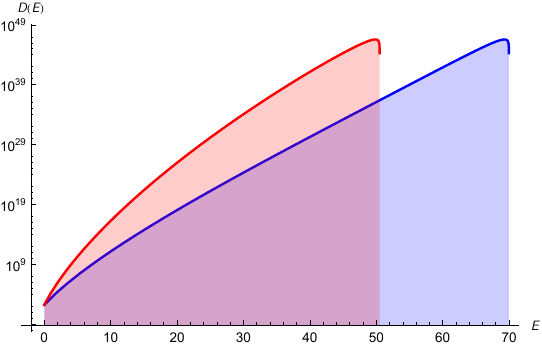}
  \end{minipage}
  \hspace{15mm}
  \begin{minipage}{0.4\textwidth}
    (b)\includegraphics[scale=0.9]{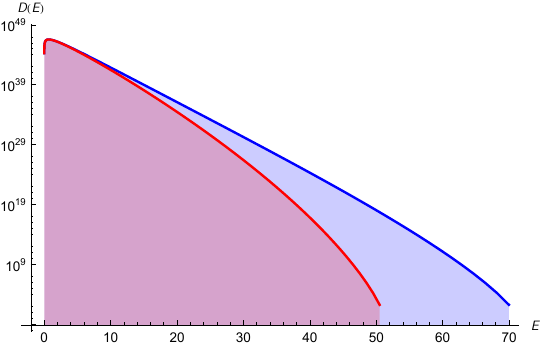}
  \end{minipage}
  \caption{Density of energy levels for $N=100$ as a function of the parameter $\eta=0$ (red) and $\eta=0.02$ (blue) (a) for the ferromagnetic case and (b) for the antiferromagnetic case.}
\end{figure}

\begin{figure}[H]
  \begin{minipage}{0.4\textwidth}
    (a)\includegraphics[scale=0.9]{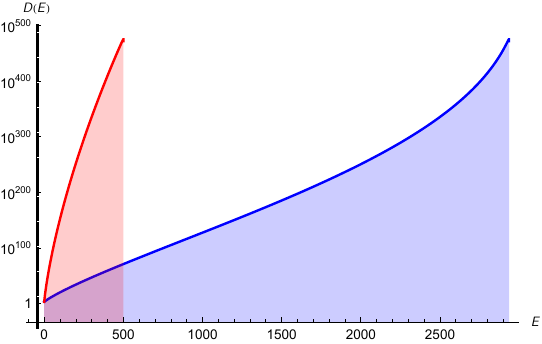}
  \end{minipage}
  \hspace{15mm}
  \begin{minipage}{0.4\textwidth}
    (b)\includegraphics[scale=0.9]{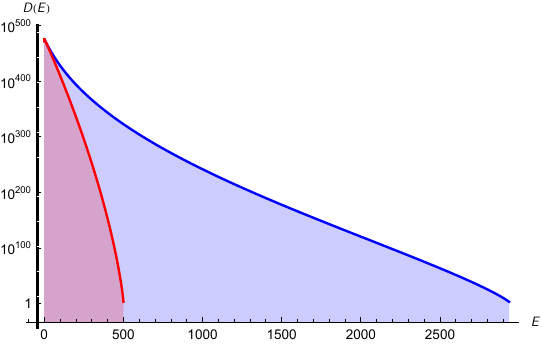}
  \end{minipage}
  \caption{Density of energy levels for $N=1000$ as a function of the parameter $\eta=0$ (red) and $\eta=0.005$ (blue) (a) for the ferromagnetic case and (b) for the antiferromagnetic case.}
\end{figure}

Similar to the behaviour found in the undeformed model, the energy profiles of the ferromagnetic and antiferromagnetic configurations exhibit clear mutual antisymmetry. The main consequence of the deformation is that every energy level experiences a shift while degeneracy remains unperturbed; however, this displacement is not uniform and increases significantly as we transition toward higher energy states associated with larger values of $J$. This progressive scaling implies that the effects of the deformation are far more dominant in high-$J$ sectors.

\section{The ferromagnetic case}
\label{fer}
We begin with the ferromagnetic case, discussing both the undeformed and deformed KS models for a small number of spins as well as in the thermodynamic limit.

\subsection{Small number of spins}
\begin{figure}[H]
    \centering
    \includegraphics[scale=0.9]{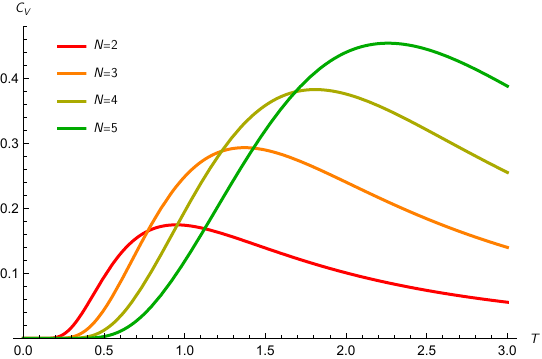}
    \caption{Specific heat as a function of temperature for values of the number of particles $N=2$ red), $3$ (orange), $4$ (yellow) and $5$ (green) for the ferromagnetic case.}
    \label{kskj1cvf}
\end{figure}

We first consider the ferromagnetic case, which corresponds to a positive value of the coupling constant ($I=1$). In Fig.~\ref{kskj1cvf}, we represent the specific heat. For the $j=1/2$ case~\cite{mariscal2025thermodynamics}, we observe that the maximum is larger and shifted to higher temperatures when $N$ increases.

\begin{figure}[H]
    \centering
    \includegraphics[scale=0.9]{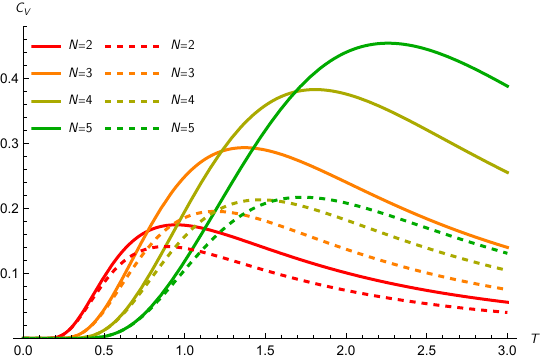}
    \caption{Comparison of the exact specific heat (continuous line) with the most probable levels approximation (dashed line) as a function of temperature for values of number of particles $N=2$ (red), $3$ (orange), $4$ (yellow) and $5$ (green) for the ferromagnetic case.}
    \label{kskj1cvcomp}
\end{figure}

The peaks observed in the previous figure correspond mostly to the energy required to transition from the ground state to the first excited state. This can be easily verified by considering the two most probable levels and comparing the results with the exact expression. This comparison is shown in Fig.~\ref{kskj1cvcomp}.
We see that it is a good estimate for low temperatures, which is partly due to the small number of particles considered. However, any bump, even for $N=2$, is not exactly reproduced, and this difference between the complete expression and the approximation is emphasised when $N$ increases. 

The ferromagnetic case presents two distinct regimes: for a small number of particles, where the coupling is set to $I=1$, the thermodynamic behaviour shifts far outside the physically reasonable temperature interval for any $N>5$. Conversely, the thermodynamic limit can be successfully analysed because the coupling scales as $I=1/N$, which compresses the energy distribution, thereby restoring the thermodynamic properties to a reasonable temperature range. Regarding the approximation of the most probable energy levels, this approach cannot be utilised in either scenario. For small systems, its applicability is strictly constrained by the size limitation, whereas in the thermodynamic limit, the most probable energy levels correspond to those with the lowest degeneracy. Consequently, given the massive total number of energy levels, their individual weights in the partition function become negligible. Thus, this approximation will be only carried out in detail for the antiferromagnetic case.
 
The temperature at which each maximum specific heat appears can be qualitatively determined. Owing to the discussion above, considering the difference between the two lowest energy levels, we find $E_{1}-E_{2}=IN$. The temperature of the specific heat maxima is proportional to this difference (the main contribution to the bumps is due to these two levels, as shown in Fig.~\ref{kskj1cvcomp}), and they are approximately proportional to $N$: $T\propto \Delta E\propto N$. Therefore, the maxima are shifted to higher temperatures as the number of particles increases. Therefore, the number of spins considered here is restricted to  $N=5$, since the peaks  for larger $N$ appear beyond the reasonable range of temperatures. Thus,  the thermodynamic limit with coupling $I=1$ in the ferromagnetic case is not realistic.

The shift of the maxima at higher temperatures as the spin number increases can also be explained from a different perspective. As in the case $j=1/2$, the density of the energy levels decreases as the energy increases in the antiferromagnetic case, as shown in Fig. \ref{kskj1DensityD} (b), while this density increases in the ferromagnetic case, seen in~Fig. \ref{kskj1DensityD} (a). At low energies, the distances between the energy levels are negligible in the antiferromagnetic case. Therefore, it can be treated as a continuum, and the maximum specific heat occurs at similar temperatures. The distances between energy levels are not negligible in the ferromagnetic case because of the spectrum. This discretisation causes the specific heat maxima to occur at larger temperatures for larger $N$.

\begin{figure}[H]
    \centering
    \includegraphics[scale=0.9]{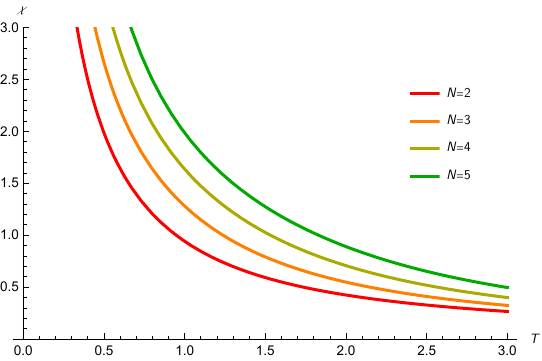}
    \caption{Magnetic susceptibility as a function of temperature for values of the number of particles $N=2$ (red), $3$ (orange), $4$ (yellow) and $5$ (green) for the ferromagnetic case.}
    \label{kskj1xf}
\end{figure}

The magnetic susceptibility is shown in Fig.~\ref{kskj1xf}. At zero temperature, the magnetic susceptibility is infinite. As the most probable ferromagnetic levels have the largest angular momenta $J$, all considered spins are aligned in the up position $(m=1)$ at a zero temperature. This indicates that a phase transition occurs at zero temperature. When the temperature is increased, the remaining levels with smaller total angular momenta become relevant. Thus, a decay $1/(NT)$ is observed at higher temperatures, regardless of the parity of the number of particles. 

\begin{figure}[H]
    \centering
    \includegraphics[scale=0.9]{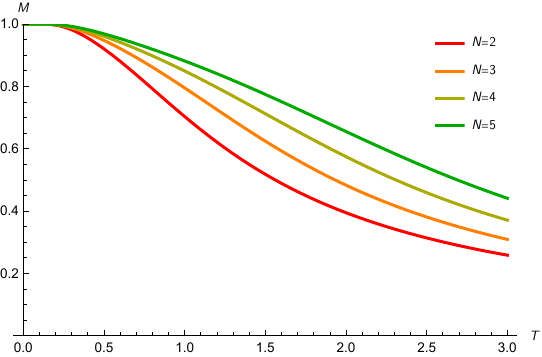}
    \caption{magnetisation as a function of temperature for values of the number of particles $N=2$ (red), $3$ (orange), $4$ (yellow) and $5$ (green) for the ferromagnetic case.}
    \label{kskj1mf}
\end{figure}

The magnetisation in the presence of an external magnetic field (in units $h=\gamma=1$) is shown in Fig.~\ref{kskj1mf}. For any number of particles $N$, the magnetisation at zero temperature is always 1. Because the ground energy state in the ferromagnetic case consists of all spins up at the $J=m=N$ level, the magnetisation is always $M=m/N=Nj/N=1$ at low temperatures. When the temperature is increased, the remaining states start to contribute, and their average makes the magnetisation tend toward zero. 

\begin{figure}[H]
    \centering
    \includegraphics[scale=0.9]{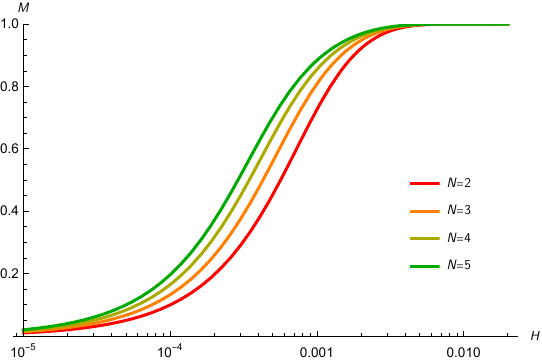}
    \caption{magnetisation as a function of magnetic field for values of the number of particles  $N=2$ (red), $3$ (orange), $4$ (yellow), and $5$ (green) for temperature $T=0.001$, for the ferromagnetic case.}
    \label{ferrotransition}
\end{figure}

Finally, we observed whether the action of an external magnetic field on magnetisation had an observable effect, as shown in Fig.~\ref{ferrotransition} when the temperature is zero. 
A phase transition is observed when the external magnetic field goes from $h=0$ to $h>0$ at $T=0.001$. The closer this value is to zero,  the {more similar to a step function in the $T \to 0$ limit become}. This effect can be easily understood because, for $h=0$ the ground state is shared by every possible state in the energy level $J=N$, the magnetisation is then zero, and when the external magnetic field is slightly increased, the lowest energy state becomes that of $J=m=N$. The magnetisation of this state is $M=1$, which is the highest possible value. Therefore, this value does not change when $h$ increases. In fact, this state becomes even less energetic (whose magnetisation remains $M=1$).

\subsection{Thermodynamic limit}
To ensure the extensivity of the model the thermodynamic limit for $j=1$ requires scaling the coupling constant as $I \rightarrow I/N$ \cite{kac1968statistical}. As stated in \cite{czachor2008energy}, coupling modification causes the energy distribution to compress.  Consequently, thermodynamic characteristics become observable at reduced temperatures, specifically within the range $T\in (0,3)$.

\subsubsection{Thermodynamic properties}
In the absence of an external magnetic field, the ground states correspond to $J=N$ and $J=N-1$. These states are crucial for the analyses of the specific heat and magnetic susceptibility.

\begin{figure}[H]
    \centering
    \includegraphics[scale=0.9]{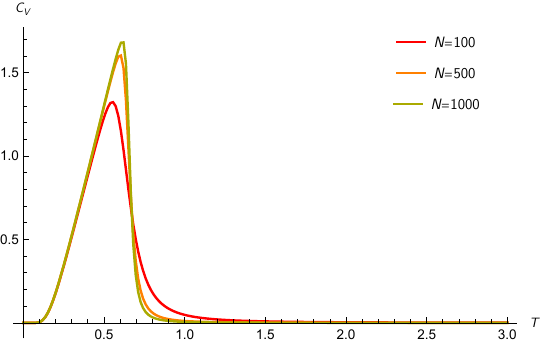}
    \caption{Specific heat as a function of temperature for values of the number of particles $N=100$ (red), $500$ (orange), and $1000$ (yellow), for the interaction constant $I=1/N$ ferromagnetic case.}
    \label{cvlimitferro}
\end{figure}

In Fig. \ref{cvlimitferro}, we observe that the specific heat peaks occur at nearly the same temperature and become slightly larger as the number of particles increases.

\begin{figure}[H]
    \centering
    \includegraphics[scale=0.9]{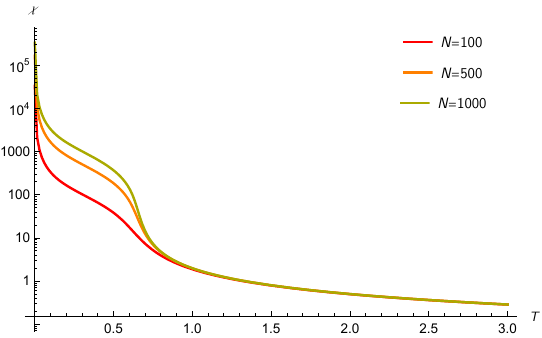}
    \caption{Magnetic susceptibility as a function of temperature for values of the number of particles $N=100$ (red), $500$ (orange), and $1000$ (yellow), for the interaction constant $I=1/N$ ferromagnetic case.}    
    \label{chilimitferro}
\end{figure}

In Fig. \ref{chilimitferro}, magnetic susceptibility exhibits convergence at low temperatures, where it diverges to infinity, and at high temperatures, where it approaches zero. Within the intermediate range, the curve exhibits a Curie transition, which is addressed in a subsequent section. These curves differ exclusively during this transition phase; while it takes place at a nearly identical temperature across all samples, higher susceptibility values are attained in relation to an increased number of particles.

\begin{figure}[H]
    \centering
    \includegraphics[scale=0.9]{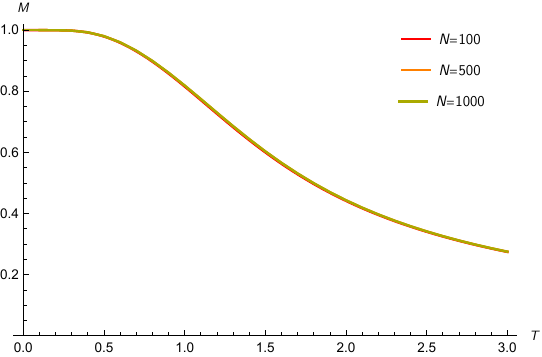}
    \caption{ Magnetisation as a function of temperature for values of the number of particles $N=100$ (red), $500$ (orange), and $1000$ (yellow), for the interaction constant $I=1/N$ in the ferromagnetic case. They all are coincident.}
    \label{mlimitferro}
\end{figure}

In the presence of an external magnetic field, the two ground states are $\ket{J,m}=\ket{N,N}$ and $\ket{J,m}=\ket{N,N-1}$. In Fig. \ref{mlimitferro}, we present the magnetisation. We can see that all the $N$ curves coincide in the thermodynamic limit.

Considering the ground states, it is observed that at zero temperature, the magnetisation reaches its maximum value $M=1$. Regarding its decrease, the excited states depend exclusively on $N$, and because the chosen particle numbers are very large, the differences in magnetisation are minimal, making the different curves appear to coincide.

\subsubsection{Curie temperature}
As in the $j=1/2$ case, we followed the strategy for obtaining the Curie temperature used in \cite{kittel1965development}. This consists of getting this temperature through the association $T \leftrightarrow J_{\text{max}}$. The energy spectrum is given by the following equation:
\begin{equation}
    E_{J}=-\frac{I}{2}(J(J+1)-2N).
\end{equation}
The multiplicity \eqref{eq:degeneracy} involves hypergeometric functions, leading to complex expressions that cannot be dealt with. Therefore, we have used \cite{curtright2017spin} to take the approximation of the the multiplicity of $j=1$ energetic states when $N\rightarrow\infty$: 
\begin{equation}
    D_{N,J}=\frac{(2J+1)\cdot3^{N+\frac{3}{2}}\cdot e^{-\frac{3J(J+1)}{4N}}}{8N^{\frac{3}{2}}\sqrt{\pi}}\cdot\left(1-\frac{21}{16N}\right).
    \label{eq:degJ1aprox}
\end{equation}
The partition function is given by:
\begin{equation}
    Z=\sum_{J=0}^{N}\sum_{m=-J}^{J}D_{N,J}e^{-\beta E_{J}}=\sum_{J=0}^{N}(2J+1)D_{N,J}e^{-\beta E_{J}}
\end{equation}
Neglecting terms of order $1/N$ and $J$, the following function is obtained:
\begin{equation}
    Z_{KS}\cong\sum_{J=0}^{N}\text{exp}\left[\left(\frac{3}{2}+N\right)\ln{(3)}-\ln{\left(8 N^{3/2} \sqrt{\pi}\right)}-\frac{3J(J+1)}{4N}+\frac{\beta I}{2}\left(J(J+1)-2N\right)\right].
\end{equation}
Maximising the exponent with respect to the energetic level $J$, we obtain the following relation:
\begin{equation}
    k_{B}T_{C}=\frac{2NI}{3}\xrightarrow[]{I\to\frac{I}{N}}k_{B}T_{C}=\frac{2I}{3}.
\end{equation}
This is the same result obtained in \cite{bjornberg2020quantum} from different arguments. 

\begin{figure}[H]
  \begin{minipage}{0.4\textwidth}
    (c)\includegraphics[scale=0.7]{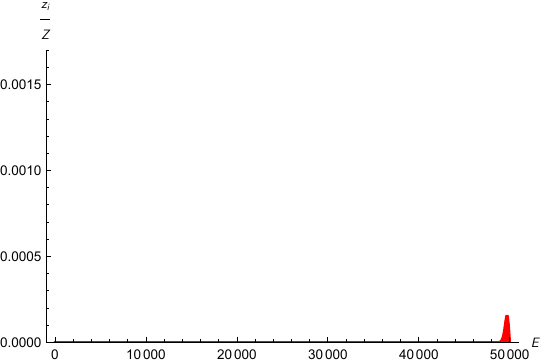}
  \end{minipage}
  \hspace{15mm}
  \begin{minipage}{0.4\textwidth}
    (d)\includegraphics[scale=0.7]{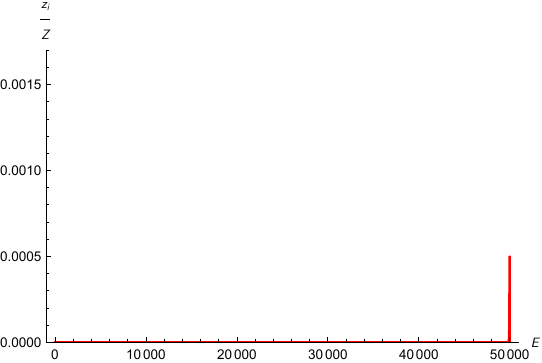}
  \end{minipage}
  \caption{Normalized weight of each energy level in the partition function for number of particles $N=10^5$  and  temperature (a) $T=0.500$, (b) $T=0.600$, (c) $T=T_{C}=0.666$ and (d) $T=0.700$, for the interaction constant $I=1/N$ ferromagnetic case.}
  \label{ngrande}
\end{figure}

To visualise what happens to the partition function when the Curie transition occurs, we show its evolution in Figure \ref{ngrande}. 
\begin{figure}[H]
  \begin{minipage}{0.4\textwidth}
    (a)\includegraphics[scale=0.7]{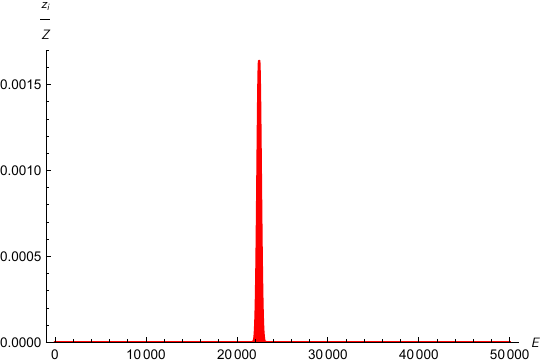}
  \end{minipage}
  \hspace{15mm}
  \begin{minipage}{0.4\textwidth}
    (b)\includegraphics[scale=0.7]{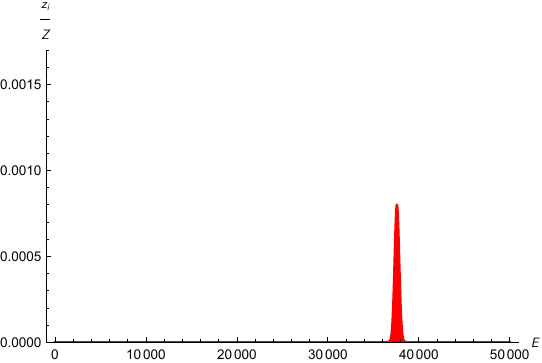}
  \end{minipage}
\end{figure}
Before the Curie temperature, there is a peak at the energy levels of high $J$, which corresponds to ferromagnets. This peak shifts to the energy levels of smaller $J$. When the Curie temperature is exceeded, the peak appears at the lowest $J$ energy levels, which is associated with paramagnets. Thus, the Curie transition can be considered as the point at which ferromagnets start to behave as paramagnets, corresponding to the smallest peak in Figure \ref{ngrande}.

\subsection{\texorpdfstring{$q$}{q}-deformation of small number of spins}
\label{qfer}
For the undeformed scenario, the derivation of analytical expressions using the most probable level approximation is not useful because of the limitations of the ferromagnetic density profile. The lowest energetic levels in the ferromagnetic case are those with the highest total angular momentum $J$. Owing to the shape of the spectrum shown in Fig.~\ref{kskj1DensityD} (a), the action of the deformation is more noticeable than in the antiferromagnetic case because the $q$ deformation of a number is larger as the number increases, as long as the number is greater than 1. Consequently, the number of particles considered does not exceed $N=5$ and the deformation parameters we handle are $\eta=0,0.3$. These limits were introduced to observe all the thermodynamic quantities in the temperature range considered, $T\in (0,3)$. Therefore, we focus on numerically studying the thermodynamic properties for different values of $q$. Negative values of $\eta$ were not considered because the model properties were symmetric with respect to $\eta=0$. 

\begin{figure}[H]
  \begin{minipage}{0.4\textwidth}
    (a)\includegraphics[scale=0.7]{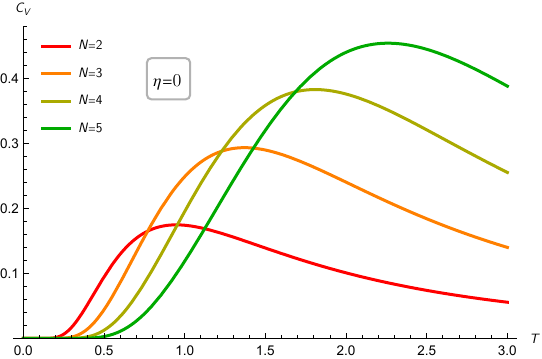}
  \end{minipage}
  \hspace{15mm}
  \begin{minipage}{0.4\textwidth}
    (b)\includegraphics[scale=0.7]{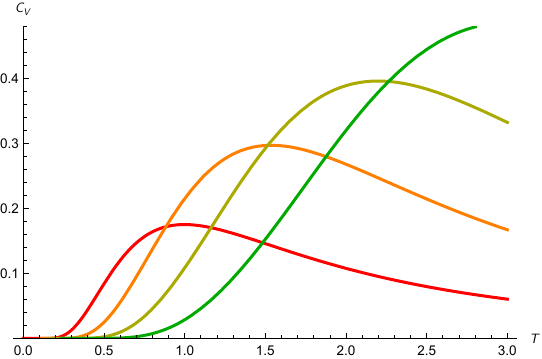}
  \end{minipage}
  \caption{Specific heat as a function of temperature for values of the number of particles $N=2$ (red), $3$ (orange), $4$ (yellow), and $5$ (green) and of the parameter (a) $\eta=0 $ and (b) $\eta=0.3$, for the ferromagnetic case.}
  \label{qkskj1cvf}
\end{figure}

The specific heat behaviour is presented in Fig.~\ref{qkskj1cvf}, where the maxima shift toward higher temperatures as the deformation parameter increases. Concurrently, the peak values exhibit a slight enhancement with increasing $q$. Beyond these shifts, pronounced variations emerge at significantly elevated temperatures.
As expected, Fig.~\ref{qkskj1cvf} (a), corresponding to $\eta=0$, is identical to the undeformed case. As the most probable levels have the largest angular momenta $J$ (so these energies depend explicitly on $N$), we observe that the temperatures at which specific heat maxima occur are equally separated at higher temperatures for contiguous $N$. When deformation is introduced, this gap widens, but in the same way for every pair of curves. Therefore, the number of particles larger than $N=5$ is outside the reasonable temperature range.

\begin{figure}[H]
  \begin{minipage}{0.4\textwidth}
    (a)\includegraphics[scale=0.7]{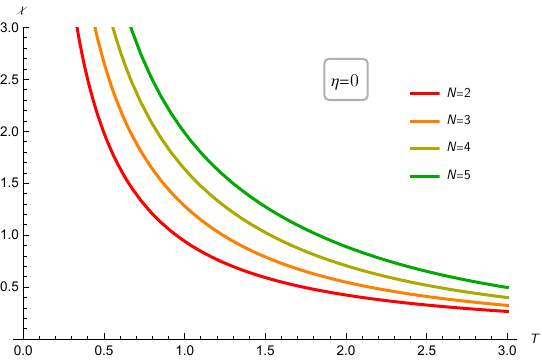}
  \end{minipage}
  \hspace{15mm}
  \begin{minipage}{0.4\textwidth}
    (b)\includegraphics[scale=0.7]{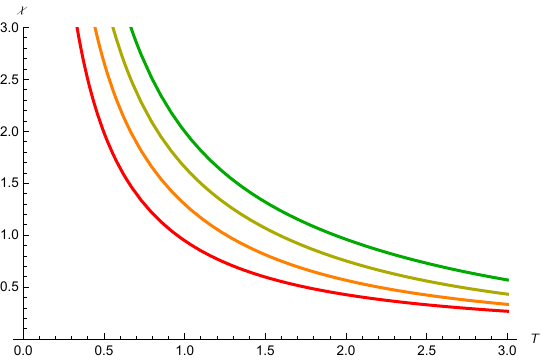}
  \end{minipage}
  \caption{Magnetic susceptibility as a function of temperature for values of the number of particles $N=2$ (red), $3$ (orange), $4$ (yellow), and $5$ (green) and of the parameter (a) $\eta=0$ and (b) $\eta=0.3$, for the ferromagnetic case.}
  \label{qkskj1xf}
\end{figure}

Regarding the magnetic susceptibility, Fig.~\ref{qkskj1xf} was obtained for different values of the deformation parameter $\eta$. As in the undeformed scenario, every spin can sense the magnetic field, even when the temperature is null. This sensitivity remains more present, or in other words, takes longer to vanish with temperature the higher the parameter $\eta$. The decay $1/NT$ is also observed for the deformed case. Hence, no noticeable changes are observed in the magnetic susceptibility.

\begin{figure}[H]
  \begin{minipage}{0.4\textwidth}
    (a)\includegraphics[scale=0.7]{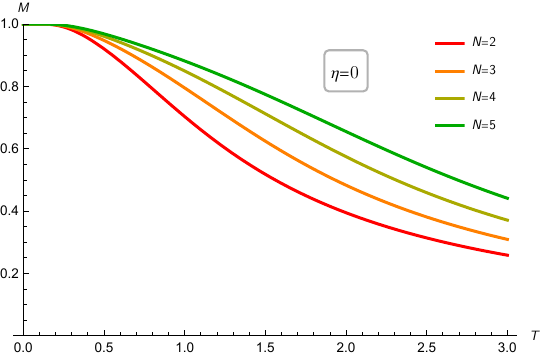}
  \end{minipage}
  \hspace{15mm}
  \begin{minipage}{0.4\textwidth}
    (b)\includegraphics[scale=0.7]{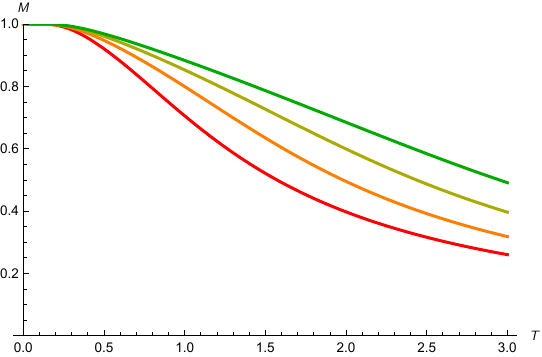}
  \end{minipage}
  \caption{magnetisation as a function of temperature for values of the number of particles $N=2$ (red), $3$ (orange), $4$ (yellow), $5$ (green) and of the parameter (a) $\eta=0 $ and (b) $\eta=0.3$, for the $h=\gamma=1$ ferromagnetic case.}
  \label{qkskj1mf}
\end{figure}

Finally, we computed the magnetisation for a certain external magnetic field, as shown in Fig.~\ref{qkskj1mf}. Again, no significant changes were observed with increasing deformation. As the deformation parameter increases, the magnetisation convexity becomes slightly more pronounced. This behaviour arises because the introduction of deformation breaks the uniformity of the coupling, meaning that each individual spin no longer experiences an identical interaction from all other spins.

\subsection{\texorpdfstring{$q$}{q}-deformation of the thermodynamic limit}
\label{subsec:therlimit}
We next consider the thermodynamic limit in the deformed scenario. Since the deformed energy eigenvalues are expressed in terms of $q$-numbers, they exhibit exponential growth. Consequently, to guarantee the extensivity of the model for an arbitrary $N$, the most straightforward approach involves scaling the coupling constant as $I \rightarrow I/N$ (paralleling the undeformed KS model) alongside the deformation parameter as $\eta \rightarrow \eta/N$ \cite{ballesteros2025quantum}.

In the undeformed case, a connection between the properties analysed and the Curie temperature was established. In the deformed case, new strategies must be developed to derive analytical expressions for this critical temperature.

\subsubsection{Thermodynamic properties}
As in the undeformed scenario, we analyze the thermodynamic limit of the deformed case. While the thermodynamic properties in the undeformed case show little to no dependence on the number of particles, the deformed case reveals a noticeable separation between the curves corresponding to different particle numbers. Furthermore, this effect becomes more significant as the deformation increases. 

The influence of deformation is significantly stronger than in the antiferromagnetic case, making it essential to carefully choose the deformation parameters. Here, we used $4/N$, which is sufficient to show the differences between the deformed and undeformed models. However, this inverse proportionality does not uniformly shift all curves; instead, it helps mitigate some of the differences between the deformed curves.
\begin{figure}[H]
  \begin{minipage}{0.4\textwidth}
    (a)\includegraphics[scale=0.7]{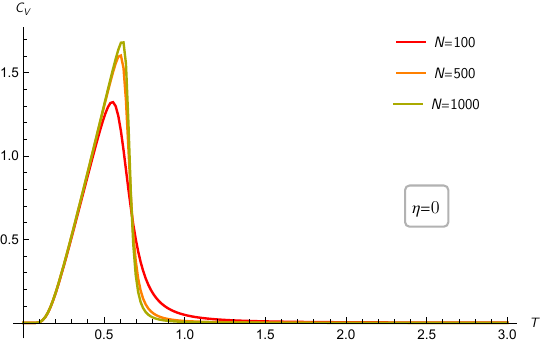}
  \end{minipage}
  \hspace{15mm}
  \begin{minipage}{0.4\textwidth}
    (b)\includegraphics[scale=0.7]{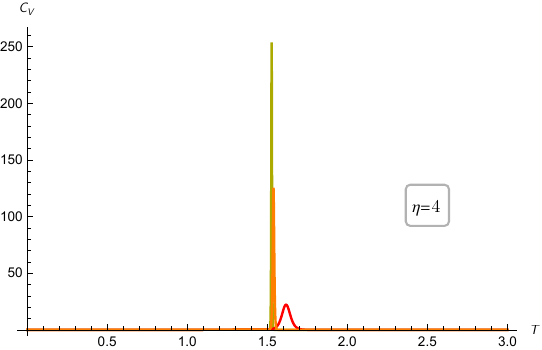}
  \end{minipage}
  \caption{Specific heat as a function of temperature for values of the number of particles $N=100$ (red), $500$ (orange), and $1000$ (yellow) for the interaction constant $I=1/N$ ferromagnetic case. The values of the parameter $q=e^{\eta/N}$ are: (a) $\eta=0$ and (b) {$\eta=4$}.}
\end{figure}
Regarding specific heat, the peaks reach significantly higher values and occur at higher temperatures, although they also become narrower. This behavior results from deformation broadening the energy distribution, which reduces the influence of highly excited states while enhancing the role of ground states within the considered temperature range.
\begin{figure}[H]
  \begin{minipage}{0.4\textwidth}
    (a)\includegraphics[scale=0.7]{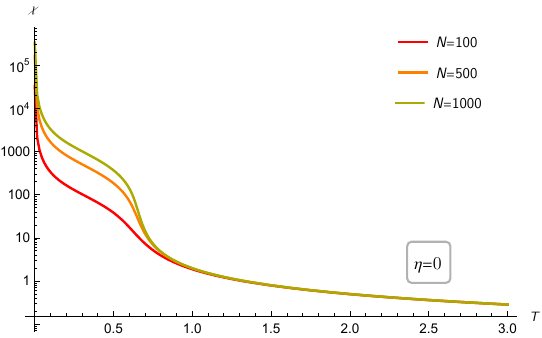}
  \end{minipage}
  \hspace{15mm}
  \begin{minipage}{0.4\textwidth}
    (b)\includegraphics[scale=0.7]{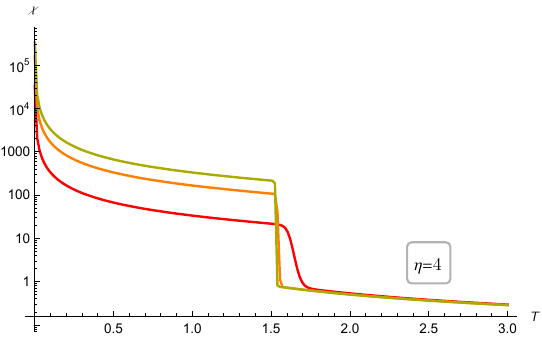}
  \end{minipage}
  \caption{Magnetic susceptibility as a function of temperature for values of the number of particles $N=100$ (red), $500$ (orange), and $1000$ (yellow) for the interaction constant $I=1/N$ ferromagnetic case. The values of the parameter $q=e^{\eta/N}$ are: (a) $\eta=0$ and (b) $\eta=4$.}
  \label{qchilimit} 
\end{figure}

For magnetic susceptibility, the introduction of deformation alters the convergence of the different curves observed in the undeformed case. This convergence, which occurs both before and after the Curie temperature in the undeformed case, is shifted to smaller temperatures before and progressively higher temperatures after the transition due to deformation. This effect becomes more pronounced as the deformation increases. Regarding the Curie transition, it has already been noted that it takes place at higher temperatures as the deformation grows. In terms of its shape, the transition becomes progressively less smooth, with the curve appearing increasingly more vertical as the deformation increases. As far as the different curves, it is observed that those corresponding to smaller values of $N$ experience a more significant shift of the Curie transition to higher temperatures. Consequently, the relation inversely proportional to $N$ of deformation exerts a stronger influence than the deformation itself.
\begin{figure}[H]
  \begin{minipage}{0.4\textwidth}
    (a)\includegraphics[scale=0.7]{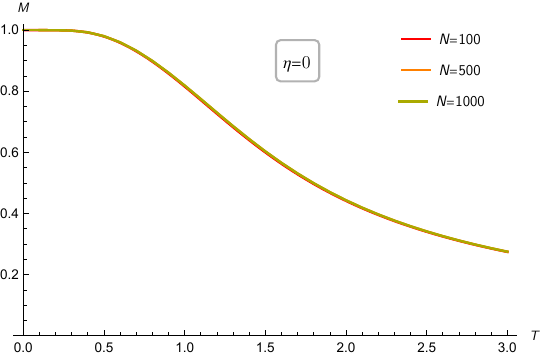}
  \end{minipage}
  \hspace{15mm}
  \begin{minipage}{0.4\textwidth}
    (b)\includegraphics[scale=0.7]{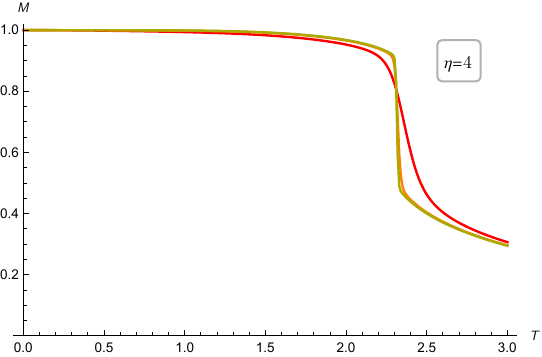}
  \end{minipage}
  \caption{magnetisation as a function of temperature for values of the number of particles $N=100$ (red), $500$ (orange), and $1000$ (yellow) for the interaction constant $I=1/N$ and ${h\gamma=1}$ ferromagnetic case. The values of the parameter $q=e^{\eta/N}$ are: (a) $\eta=0$ and (b) $\eta=4$.}
\end{figure}
As is well known, magnetisation, like susceptibility, provides insight into the Curie transition. Therefore, the transition observed in magnetisation will shift in the same manner as in susceptibility when deformation is introduced. In the undeformed case, all the curves nearly coincide, but in the deformed case, they do not. While the deformations are not identical for each curve, if they were, the differences would be even more pronounced. Unlike the antiferromagnetic case, the ground states remain unchanged in the deformed case, which results in a similar behavior at low temperatures for all deformation values.

\subsubsection{Curie temperature}
In the thermodynamic study of the $q$-deformed Kittel-Shore model for spin-$1/2$ particles presented in \cite{ballesteros2025quantum}, the determination of the Curie temperature initially required a detailed preliminary analysis of the magnetic susceptibility behavior as a function of temperature across various deformation parameters. Conversely, for the current case involving spin $j=1$, this behaviour has already been clearly illustrated in Figure \ref{qchilimit}, which displays the susceptibility response in the thermodynamic limit under increasing deformation. From the results presented there, it is straightforwardly observed that the system retains the same fundamental qualitative feature found in the lower-spin counterpart: the Curie temperature shifts toward higher values as the deformation parameter increases.

This correspondence reinforces the notion that quantum deformation consistently enhances the energy gaps between the ground and first excited states, regardless of the spin quantum number of the constituent particles. Similar to the $j=1/2$ case in \cite{ballesteros2025quantum}, the model for spin $j=1$ demonstrates in Figure \ref{qchilimit} that the ferromagnetic transition is driven to higher temperatures owing to the non-linearities induced by the $U_q(\mathfrak{su}(2))$ quantum group symmetry. With this shared qualitative conclusion established, we now proceed to formalise the quantitative analysis of the Curie temperature for the $j=1$ system by examining the partition function and the relevant asymptotic approximations in the macroscopic limit.

\begin{figure}[H]
    \centering
    \includegraphics[scale=1.0]{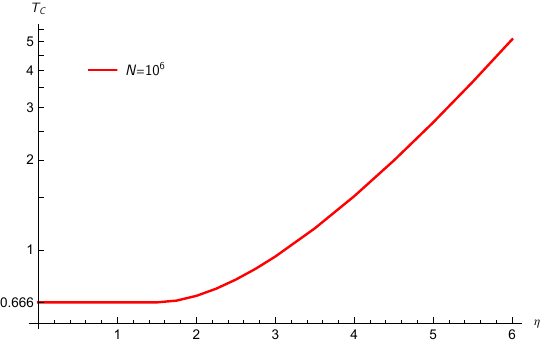}
    \caption{Curie temperature of the $q$-KS model as a function of the deformation parameter $\eta=\log q$ for $N=10^6$.}
    \label{Cureta}
\end{figure}

The behaviour of the Curie temperature, determined numerically as the transition temperature of the magnetic susceptibility, is shown in Figure \ref{Cureta} for $N=10^6$ as a function of deformation.

In order to  obtain an analytical estimate of the function $T_C(\eta)$ represented in Figure \ref{Cureta}, we consider the partition function of the $q$-KS model in the form
\begin{equation}
    Z_N^\eta=\sum_{i}^{} z^\eta(E^\eta_i) =\sum_{p=0}^{p_m} z_p^\eta \, ,
\end{equation}
and perform an exhaustive numerical analysis of the distribution of the normalized contributions $z^\eta(E_i)/Z_N^\eta$ for different temperatures and deformation parameters.

\begin{figure}[H]
  \begin{minipage}{0.4\textwidth}
    (c)\includegraphics[scale=0.7]{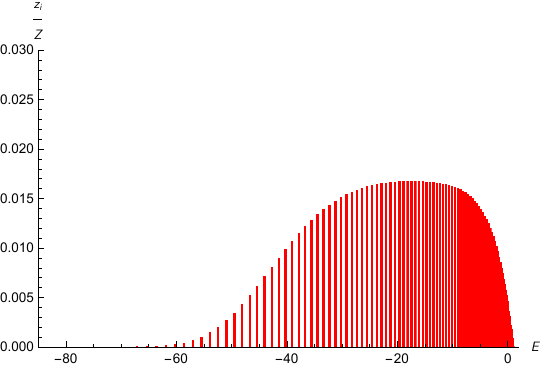}
  \end{minipage}
  \hspace{15mm}
  \begin{minipage}{0.4\textwidth}
    (d)\includegraphics[scale=0.7]{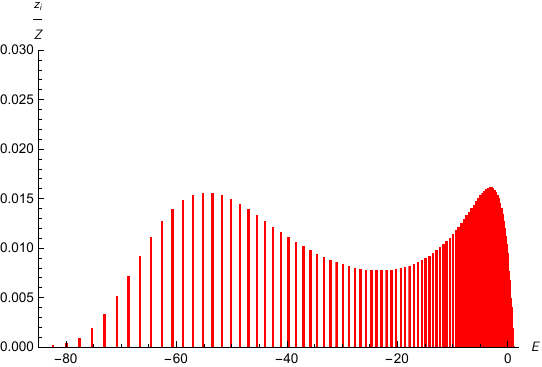}
  \end{minipage}
  \caption{Normalized contribution of each energy level $E_{i}$ in the partition function for number of particles $N=100$, deformation parameter $q=e^{\eta/N}$ for (a) $\eta=0$, (b) $\eta=1.5$, (c) $\eta=2$, (d) $\eta=2.5$ and Curie temperature $T_{C}(\eta)$ given by (a) $T=T_{C}(0)=0.669$, (b) $T=T_{C}(1.5)=0.750$, (c) $T=T_{C}(2)=0.702$, (d) $T=T_{C}(2.5)=0.853$ (ferromagnetic case).}
  \label{picos}
\end{figure}

Following the procedure of \cite{ballesteros2025quantum}, Figure \ref{picos} displays the normalized contributions $z^\eta(E^\eta_i)/Z_N^\eta$ for $N=100$ under different $\eta$ values, all at the Curie temperature $T_C(\eta)$, which is determined numerically via the discontinuity in the magnetic susceptibility. It reproduces the same pattern observed for $j=1/2$: a larger $\eta$ broadens the profile until it splits into two separate peaks, passing through an intermediate plateau.
\begin{figure}[H]
  \begin{minipage}{0.4\textwidth}
    (a)\includegraphics[scale=0.7]{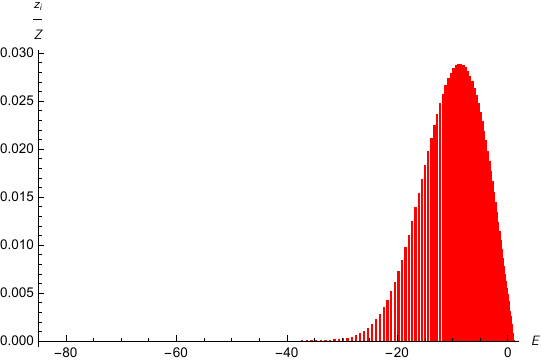}
  \end{minipage}
  \hspace{15mm}
  \begin{minipage}{0.4\textwidth}
    (b)\includegraphics[scale=0.7]{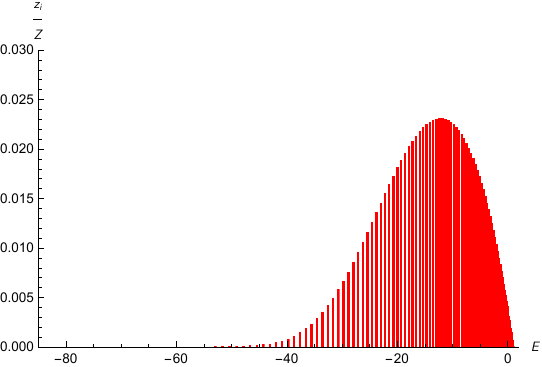}
  \end{minipage}
\end{figure}

\begin{figure}[H]
  \begin{minipage}{0.4\textwidth}
    (c)\includegraphics[scale=0.7]{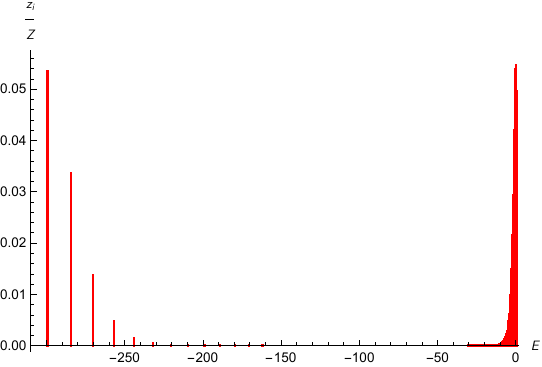}
  \end{minipage}
  \hspace{15mm}
  \begin{minipage}{0.4\textwidth}
    (d)\includegraphics[scale=0.7]{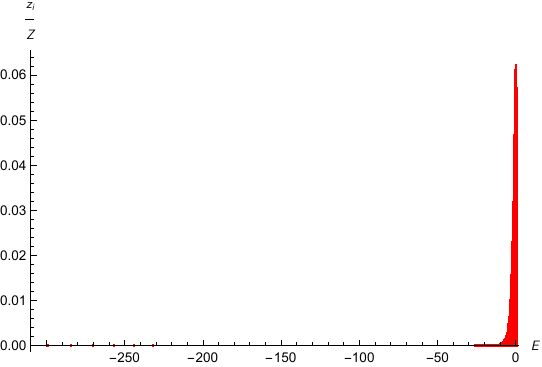}
  \end{minipage}
  \caption{Normalized contribution of each energy level $E_{i}$ in the partition function for number of particles $N=100$, deformation parameter $q=e^{\eta/N}$ of $\eta=5$ and temperatures (a) $T=0.660$, (b) $T=2.900$, (c) $T=T_{C}=2.939$, and (d) $T=3.200$, for the ferromagnetic case.}
  \label{picos2}
\end{figure}

\begin{figure}[H]
  \begin{minipage}{0.4\textwidth}
    (c)\includegraphics[scale=0.7]{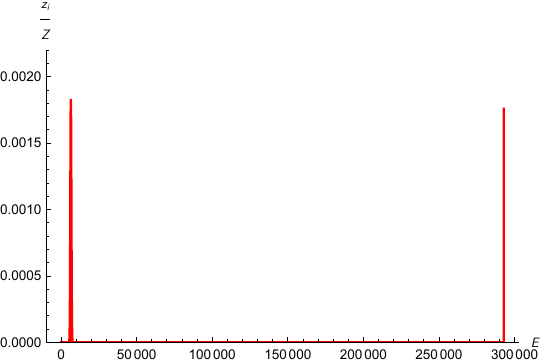}
  \end{minipage}
  \hspace{15mm}
  \begin{minipage}{0.4\textwidth}
    (d)\includegraphics[scale=0.7]{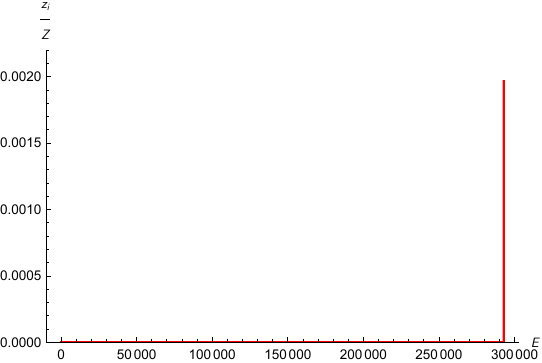}
  \end{minipage}
  \caption{Normalized contribution of each energy level $E_i$ in the partition function for $N=10^5$, deformation parameter $\eta=5$ and temperatures (a) $T=2.60$, (b) $T=2.66$, (c) $T=T_{C}(5)=2.66635$, and (d) $T=2.70$, for the ferromagnetic case.}
  \label{fpdef3}
\end{figure}

This temperature dependence is illustrated in Figure~\ref{picos2} for $N=100$ and a fixed large enough deformation $\eta=5$. In agreement with the spin-$1/2$ case, two maxima develop as $T$ increases, where the dominant one gradually moves toward higher energies. Notably, the Curie temperature matches the exact point where both peaks of the distribution exhibit identical values. Lastly, the distributions computed for $N=10^5$ and presented in Figure \ref{fpdef3} ensure that this striking behavior persists in the thermodynamic limit.
\begin{figure}[H]
  \begin{minipage}{0.4\textwidth}
    (a)\includegraphics[scale=0.7]{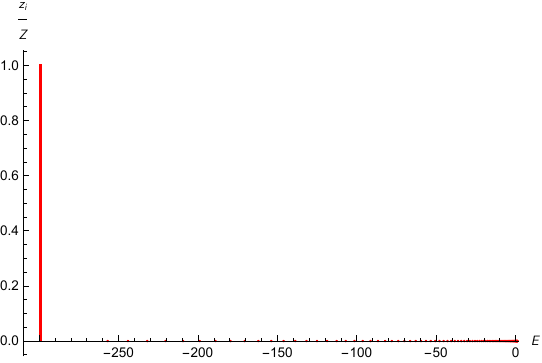}
  \end{minipage}
  \hspace{15mm}
  \begin{minipage}{0.4\textwidth}
    (b)\includegraphics[scale=0.7]{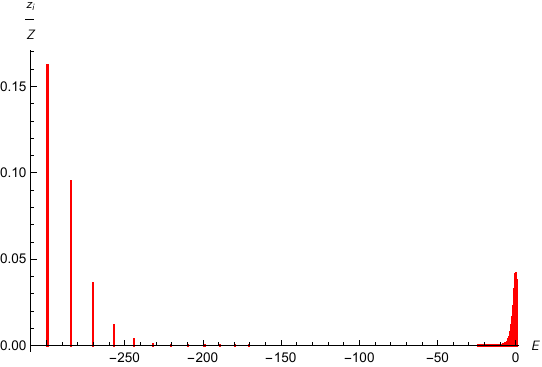}
  \end{minipage}
\end{figure}

\begin{figure}[H]
  \begin{minipage}{0.4\textwidth}
    (a)\includegraphics[scale=0.7]{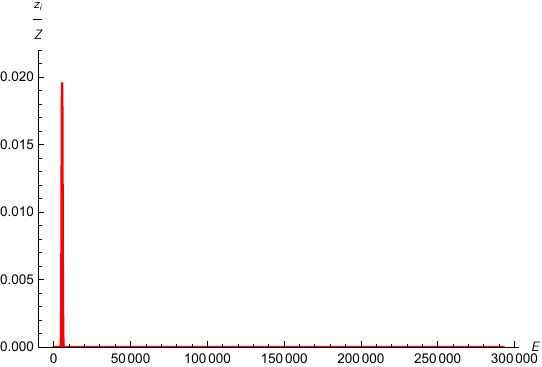}
  \end{minipage}
  \hspace{15mm}
  \begin{minipage}{0.4\textwidth}
    (b)\includegraphics[scale=0.7]{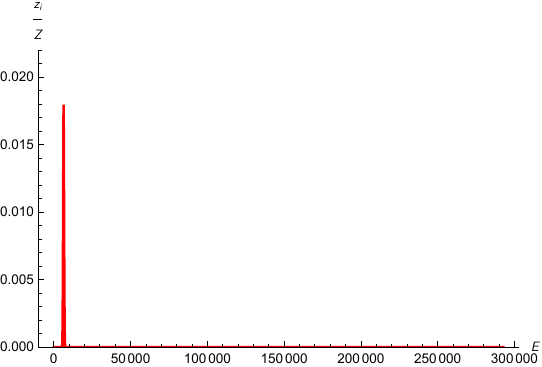}
  \end{minipage}
\end{figure}

Given that the behaviour of the partition function components at the Curie temperature for $j=1$ is identical to that of the $j=1/2$ case, the same analytical strategies can be deployed here. Consequently, this equivalence allows us to derive closed analytical expressions for $T_C(\eta)$ by reproducing the algebraic procedures established in our prior work \cite{ballesteros2025quantum}.

Taking the Hamiltonian \eqref{qcas} without an external magnetic field, we have the following energy spectrum:
\begin{equation}
    E_{J}^{q}=-\frac{I}{2}\left(\left[J\right]_{q}\left[J+1\right]_{q}-N\left[2\right]_{q}\right),
\end{equation}
where $p=N-J$. So, the deformed partition function is:
\begin{equation}
    Z_{KS}^{q}=\sum_{J=0}^{N}\sum_{m=-J}^{J} A\cdot \text{exp}\left[\frac{\beta I}{2}\left(\left[J\right]_{q}\left[J+1\right]_{q}-N\left[2\right]_{q}\right)\right]=\sum_{J=0}^{N}(2J+1)\cdot A\cdot \text{exp}\left[\frac{\beta I}{2}\left(\left[J\right]_{q}\left[J+1\right]_{q}-N\left[2\right]_{q}\right)\right],
    \label{eq:FPJ1}
\end{equation}
where
\begin{equation*}
    A=\frac{\Gamma (N+1) \left((j+N+1) \, _2\tilde{F}_1\left(\frac{1}{2} (-j-N),\frac{1}{2} (-j-N+1);1-j;4\right)-\, _2\tilde{F}_1\left(\frac{1}{2} (-j-N-1),\frac{1}{2} (-j-N);-j;4\right)\right)}{\Gamma (j+N+2)},
\end{equation*}
and ${}_2\tilde{F}_1$ denotes the regularized hypergeometric function (given by the quotient of the standard function ${}_{2}F_{1}$ and the Gamma function).

To derive an analytical estimation for the Curie temperature in terms of $\eta$ for the spin-1 model, we must account for the qualitatively different regimes exhibited by $Z_N^\eta$, which remain consistent with the framework established for $j=1/2$. As Figure~\ref{picos} illustrates, the partition function contributions clearly distinguish between a ``unimodal'' regime for small $\eta$ and a ``bimodal'' one for larger deformations. Looking at Figure \ref{Cureta}, this structural transition is well-defined in the thermodynamic limit, where the unimodal range corresponds approximately to $\eta\in(0,1.5)$ and the bimodal behavior takes over for $\eta>1.5$.

For small and intermediate deformation parameters ($\eta \in (0,5)$), a piecewise fit to the data in Figure~\ref{Cureta} yields the following functions:
\begin{align}
    T_{C}(\eta)=&0.6666 \hspace{11.65cm} \eta\in(0,1.5),   \label{tc1}\\
    T_{C}(\eta)=&1.008800-0.512063\eta+0.226084\eta^2-0.025093\eta^3-0.000339\eta^4+0.000617\eta^5 \hspace{1cm} \eta\in(1.5,5),
    \label{tc2}
\end{align}
where the correlation coefficient is $R^2=0.9999999$ (and the root mean squared error is $0.00019$).

When the deformation becomes large enough to result in the appearance of two peaks in the partition function, a different strategy is employed. The first maxima will occur closer to $p=0$ and the second to $p=N$ the larger the deformation and the number of particles is. We can now make an approximation for obtaining an expression of the Curie temperature. Following the discussion above, we can impose the first maximum of \eqref{eq:FPJ1} for $p=0$ and the second for $p=N$. This approximation is better the higher the deformation and the greater the number of particles. Taking this approach analytically, we obtain the following equation:
\begin{equation}
    T_{C}=\frac{I[N]_{q}[N+1]_{q}}{2k_{B} N \left(\log \left(\frac{N! \, _2F_1\left(\frac{1-N}{2},\frac{2-N}{2};2;4\right)}{(N-1)!}-\, _2F_1\left(\frac{1-N}{2},-\frac{N}{2};1;4\right)\right)-\log (2 N+1) \right)}.
    \label{eq:TCdef1}
\end{equation}
where, as usual, $q=e^{\eta/N}$. This framework yields an excellent approximation for the Curie temperature in systems characterized by a large number of particles and a pronounced deformation parameter. Moreover, by taking the thermodynamic limit of (\ref{eq:TCdef1}), we deduce that
\begin{equation}
    \lim_{N\rightarrow \infty} T_{C}(\eta) = \frac{I}{2k_B \log 3} \left( \frac{\sinh(\eta/2)}{\eta/2} \right)^2\thicksim\frac{e^\eta}{\eta^2\log{9}}
\end{equation}

\subsection{Finite-Size Scaling Analysis}
\label{subsec:ffsferro}
In order to determine the critical behavior of the system in the thermodynamic limit ($N \to \infty$) for the individual spin $j=1$ case, a systematic finite-size scaling analysis was performed for different values of the deformation parameter $\eta \in [1, 10]$. The numerical procedure relies on the exact evaluation of the canonical partition function for intermediate system sizes, specifically $N \in \{10000, 30000, 60000, 100000\}$. For each size $N$ and parameter $\eta$, the specific heat and the magnetic susceptibility are computed to locate and analyse their corresponding peaks.

The extraction of the scaling exponents is carried out through a multi-step statistical procedure. First, to optimize the localization of the maxima, the expressions of $T_C(\eta)$ given in Eqs.~\ref{tc2} and \ref{eq:TCdef1} are used as an analytical seed. Within a narrow temperature window around this seed, a fine mesh and parabolic interpolation are implemented to accurately extract the precise pseudo-critical peak temperature $T_{\max}(N)$ and the corresponding maximum peak heights. Subsequently, the Curie temperature in the thermodynamic limit, $T_C(\infty)$, is extrapolated via a linear regression of $T_{\max}(N)$ against $1/N$. Finally, the scaling exponents $p(C_V)$ and $p(\chi)$ are extracted via a linear regression in a double-logarithmic scale of the maximum observable values against the system size $N$:
\begin{equation}
    \ln(\text{Observable}_{\max}) = p \ln(N) + B,
\end{equation}
where $p$ is the scaling exponent (the slope of the fit) and $B$ is a constant, implying that $\text{Observable}_{\max} \propto N^p$. The reliability and accuracy of these regressions are quantified through the coefficient of determination $R^2$.

The global results obtained from this systematic scaling analysis for $j=1$ are summarised in Table~\ref{tab:fss_results}.
\begin{table}[H]
\centering
\small
\begin{tabular}{ccccccc}
\hline
$\eta$ & $T_C(\infty)$ & $R^2(T_C)$ & $p(C_v)$ & $R^2(C_v)$ & $p(\chi)$ & $R^2(\chi)$ \\
\hline
1 & 0.6614 & 0.9768 & 0.0386 & 0.9873 & 0.9074 & 0.9991 \\
2 & 0.7011 & 0.9989 & 0.9984 & 0.9996 & 0.9997 & $\approx$1 \\
3 & 0.9497 & 0.9995 & 0.9636 & 0.9989 & 0.9999 & $\approx$1 \\
4 & 1.5140 & 0.9996 & 1.0010 & $\approx$1 & 0.9999 & $\approx$1 \\
5 & 2.6757 & 0.9997 & 1.0000 & $\approx$1 & 0.9999 & $\approx$1 \\
6 & 5.0815 & 0.9997 & 1.0005 & $\approx$1 & 0.9999 & $\approx$1 \\
7 & 10.1717 & 0.9997 & 1.0007 & $\approx$1 & 0.9999 & $\approx$1 \\
8 & 21.1877 & 0.9998 & 1.0007 & $\approx$1 & 0.9999 & $\approx$1 \\
9 & 45.5216 & 0.9998 & 1.0007 & $\approx$1 & 0.9999 & $\approx$1 \\
10 & 100.2420 & 0.9998 & 1.0007 & $\approx$1 & 0.9999 & $\approx$1 \\
\hline
\end{tabular}
\caption{Global results of the finite-size scaling analysis for different values of the deformation parameter $\eta$ and for the individual spin $j=1$ case.}
\label{tab:fss_results}
\end{table}

The quantitative analysis of the data reveals a clear evolution in the scaling behaviour as the deformation parameter $\eta$ increases. For weak deformation ($\eta = 1$), the specific heat exponent is near-zero ($p(C_v) \approx 0.0384$), indicating a negligible or logarithmic growth of the peak with system size, while $p(\chi) \approx 0.6311$ exhibits a distinct sub-linear regime. A sharp crossover occurs at $\eta \ge 2$, where $p(C_v)$ abruptly jumps and firmly stabilises around $p(C_v) \approx 1.0$ with excellent quality of fit ($R^2 \approx 1.0$). Similarly, the magnetic susceptibility exponent fluctuates around $p(\chi) \approx 1.0$ for intermediate values and asymptotically converges to unity ($p(\chi) \to 1.00$ with $R^2 \to 1.0$) in the strongly deformed limit ($\eta \ge 8$). Consequently, for $\eta \ge 2$, both thermodynamic response functions scale linearly with $N$ in the finite-size scaling regime. The crossover observed around $\eta \approx 2$, which triggers this extensive divergence, is directly related to a structural change in the energy distribution. While for small deformation the Boltzmann weights exhibit a single peak, corresponding to a standard thermodynamic behaviour, increasing $\eta$ leads to a flattening of the distribution followed by the emergence of a bimodal structure. 

Compared to the $j=1/2$ case, the Curie temperature is more accurately fitted by the previously specified $1/N$ scaling law instead of a $1/\sqrt{N}$ behaviour. 
Additionally, the scaling exponents exhibit the same regime structure, although the convergence to 1 for both thermodynamic magnitudes is reached with smaller deformations than in the $j=1/2$ case. This is because its energy scaling is exponentially larger, and the number of states for $j=1$ is greater than that for $j=1/2$.

\section{The antiferromagnetic case}
\label{anti}
In this section, we discuss the antiferromagnetic case ($I=-1$), where an analogous analysis to the ferromagnetic case is performed.

\subsection{Small number of spins}
\begin{figure}[H]
    \centering
    \includegraphics[scale=0.9]{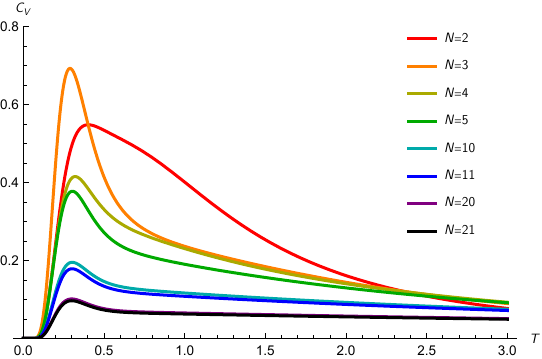}
    \caption{Specific heat as a function of temperature for values of the number of particles $N=2$ (red), $3$ (orange), $4$ (yellow), $5$ (green), $10$ (cyan), $11$ (blue), $20$ (purple), and $21$ (black) for the antiferromagnetic case.}
    \label{kskj1cvp}
\end{figure}

In the $j=1$ case, there is no difference between the even and odd $N$ cases. This is in contrast to the case $j=1/2$~\cite{al1998exact}. The reason is that the global angular momenta $J$ go from 0 to $N$, regardless of whether $N$ is odd or even. Therefore, the two lowest energy bands (that can be read from the energy spectrum of~\eqref{eq:energy}) are the same for both cases ($J=0$ and $J=1$), while this is not the case for the $j=1/2$ model ($J=0$ and $J=1$ for $N$ even and $J=1/2$ and $J=3/2$ for $N$ odd). Taking this fact into account, we now study the thermodynamic properties for the undeformed case. First, the specific heat is represented in Fig.~\ref{kskj1cvp}.
The specific heat exhibits a homogeneous behavior. The corresponding maximum occurs at approximately the same temperature (around $T\approx 0.3$), since the difference between the two smallest energy levels is always $|I|$, and decreases as the number of spins increases. For high temperatures, a decay $1/T^2$ is observed, especially marked when $N$ is small in the temperature interval considered. As in the case $j=1/2$, the energy levels amplitude is proportional to $N^{2}$, so the same asymptotic behavior is observed~\cite{al1998exact}. This behavior is seen for every value of $N$ except for $N=2$. Moreover, this maximum is smaller than the $N=3$ one.

\begin{figure}[H]
    \centering
    \includegraphics[scale=0.9]{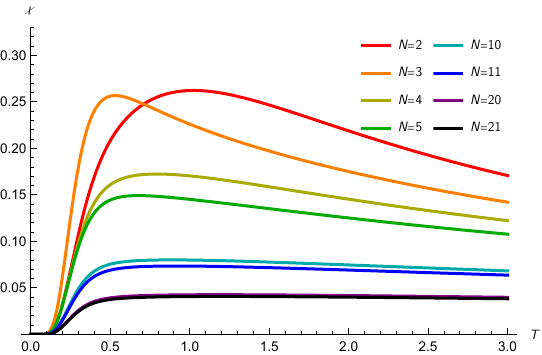}
    \caption{Magnetic susceptibility as a function of temperature for values of the number of particles $N=2$ (red), $3$ (orange), $4$ (yellow), $5$ (green), $10$ (cyan), $11$ (blue), $20$ (purple), and $21$ (black) for the antiferromagnetic case.}
    \label{kskj1xp}
\end{figure}

The magnetic susceptibility is shown in Fig.~\ref{kskj1xp}. The lowest energy level for any number of particles is always the one corresponding to $J=0$, which does not split into two in the presence of an external magnetic field. This means that the magnetic susceptibility is zero at zero temperature because the spins cannot sense the external field. At high temperatures, a decay $1/T$ is observed, especially marked in this temperature interval for those curves with the smallest number of particles.

The maxima are shifted to lower temperatures as the number of particles increases. Here, there is a slightly different behaviour of $N=3$ with respect to the rest of the spins. The maximum of this curve is placed at a slightly lower temperature than all $N$ considered. The ratio between the multiplicities of the ground and first excited states presents a local minimum for $N=3$ (the global minimum is obtained for the thermodynamic limit with an infinite number of particles). Thus, reaching the first excited level in this case ($N=3$) requires less energy than in any other case.

\begin{figure}[H]
    \centering
    \includegraphics[scale=0.9]{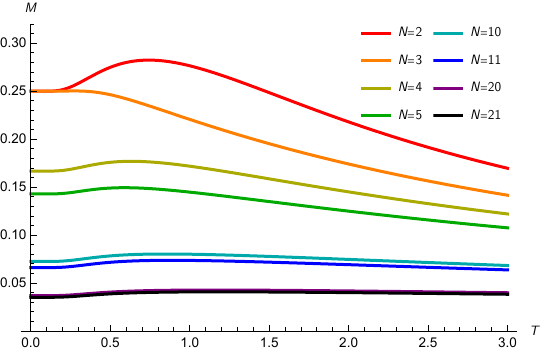}
    \caption{magnetisation as a function of temperature for values of the number of particles $N=2$ (red), $3$ (orange), $4$ (yellow), $5$ (green), $10$ (cyan), $11$ (blue), $20$ (purple), and $21$ (black), with $h=\gamma=1$ for antiferromagnetic case.}
    \label{kskj1mp}
\end{figure}

The magnetisation for an external magnetic field $h=1$ (in units $\gamma=1$) is shown in Fig.~\ref{kskj1mp}. These curves approach zero as the number of particles and temperature increase. At low temperatures, a bump is observed for any number of particles. In this temperature range, only the ground and first excited states dominate. The ground state is shared by the $\ket{0,0}$ and $\ket{1,1}$ states, and the first excited state is shared by the two states, $\ket{1,0}$ and $\ket{2,2}$  for any number of particles. Thus, the initial value of the magnetisation is defined by both the ground states. When the temperature is increased, the first excited energy states become relevant, implying an increase in the magnetisation at higher temperatures. This is also true even for $N=3$, for which the magnetisation at low temperatures follows a slightly different behaviour than for any other $N$, as shown in Fig.~\ref{kskj1mp}. Unlike the other cases, the magnetisation for \(N=3\) does not exhibit a pronounced take-off region at low temperatures; instead, this take-off is barely visible before the curve begins to decrease. This occurs because the exact combinatorial degeneracy of the ground state produces a baseline magnetisation that is almost identical to the average magnetisation of the first excited states, resulting in a negligible net increase when thermal excitations begin. Finally, the magnetisation becomes zero when the other states become relevant as the temperature increases.

The characteristic properties of these magnitudes are observed at low temperatures. At high temperatures, all the aforementioned properties tend to zero. In the range of small temperatures, the most probable levels become relevant, which are the least energetic. These levels correspond to the smallest angular momenta $J$ in the antiferromagnetic case. As in~\cite{mariscal2025thermodynamics}, we restrict ourselves to the two most probable levels $J=0,1$. Considering these two terms, the partition function takes the following expression:
\begin{equation}
    Z\approx\frac{1}{2} e^{\frac{-h+N-1}{T}} \left(-\left((N-1) N \left(e^{h/T}+e^{\frac{2 h}{T}}+1\right) \, f_{3}\right)+2 e^{\frac{h+1}{T}} \, f_{1}+2 N \left(e^{h/T}+e^{\frac{2 h}{T}}-e^{\frac{h+1}{T}}+1\right) \, f_{2}\right)
\end{equation}

Owing to the use of this approximation, we are able to obtain simple and manageable analytical expressions. In particular, we obtain for the specific heat
\begin{equation}
    C_{V}\approx\frac{6 e^{1/T} \left(N \, f_{2}-\, f_{1}\right) \left((N-1) \, f_{3}-2 \, f_{2}\right)}{T^2 \left(-2 e^{1/T} \, f_{1}+2 N \left(e^{1/T}-3\right) \, f_{2}+3 (N-1) N \, f_{3}\right){}^2}.
    \label{eq:cv1p}
\end{equation}
The magnetic susceptibility reads
\begin{equation}
    \chi\approx\frac{2 (N-1) \, f_{3}-4 \, f_{2}}{T \left(-2 e^{1/T} \, f_{1}+2 N \left(e^{1/T}-3\right) \, f_{2}+3 (N-1) N \, f_{3}\right)},
    \label{eq:chi1p}
\end{equation}
and finally, the magnetisation is
\begin{equation}
    M\approx-\frac{\left(e^{2/T}-1\right) \left(2 \, f_{2}-(N-1) \, f_{3}\right)}{(N-1) N \left(e^{1/T}+e^{2/T}+1\right) \, f_{3}-2 e^{2/T} \, f_{1}-2 N \left(e^{1/T}+1\right) \, f_{2}}.
     \label{eq:m1p}
\end{equation}
We used the notation
\begin{align}
 f_1={}_2F_1\left(\frac{1-N}{2},-\frac{N}{2};1;4\right) \qquad 
 f_2={}_2F_1\left(\frac{1-N}{2},1-\frac{N}{2};2;4\right) \qquad 
 f_3={}_2F_1\left(1-\frac{N}{2},\frac{3}{2}-\frac{N}{2};3;4\right). 
\end{align}

\begin{figure}[H]
    \centering
    \includegraphics[scale=0.9]{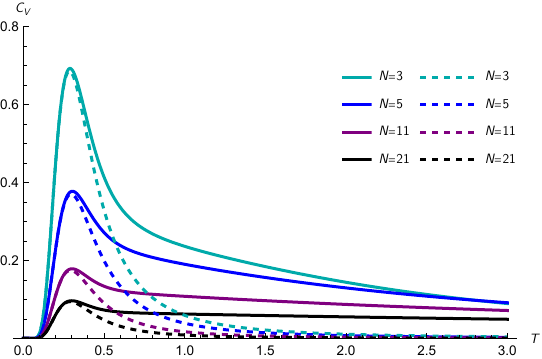}
  \caption{Comparison of the exact specific heat (continuous line) with the most probable levels approximation (dashed line) as a function of temperature for values of the odd number of particles $N=3$ (cyan), $5$ (blue), $11$ (purple), and $21$ (black) for the antiferromagnetic case.}
  \label{qkskj1comparisoncv}
\end{figure}

\begin{figure}[H]
    \centering
    \includegraphics[scale=0.9]{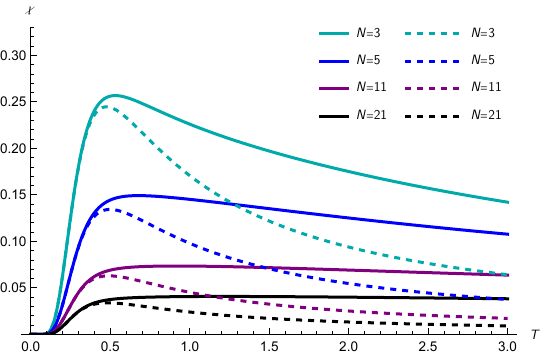}
  \caption{Comparison of the exact magnetic susceptibility (continuous line) with the most probable levels approximation (dashed line) as a function of temperature for values of the odd number of particles $N=3$ (cyan), $5$ (blue), $11$ (purple), and $21$ (black) for the antiferromagnetic case.}
  \label{qkskj1comparisonchi}
\end{figure}

\begin{figure}[H]
    \centering
    \includegraphics[scale=0.9]{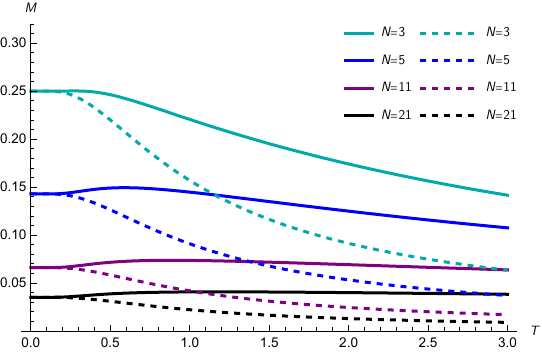}
  \caption{Comparison of the exact magnetisation (continuous line) with the most probable levels approximation (dashed line) as a function of temperature for values of the odd number of particles $N=3$ (cyan), $5$ (blue), $11$ (purple), and $21$ (black) for the antiferromagnetic case.}
  \label{qkskj1comparisonm}
\end{figure}

In order to compare the accuracy of this approximation, a comparison between the two most probable levels and the whole partition function is made. The three thermodynamic quantities are evaluated exactly and using the aforementioned approximation in Figs.~\ref{qkskj1comparisoncv},~\ref{qkskj1comparisonchi}, and \ref{qkskj1comparisonm}. It is worth emphasizing that no distinction in thermodynamic properties exists between even and odd values of $N$, contrary what happens for $j=1/2$~\cite{mariscal2025thermodynamics}. For this reason, we just leave the odd case.
We can see that the approximation is particularly good for small temperatures. This approach is very reasonable for large numbers of particles and is significantly better for the specific heat and magnetic susceptibility than for magnetisation. This is due to the fact that the bump is due to the state $\ket{2,2}$, which is not considered in this approximation.

We analyse the thermodynamic properties using the approximated analytical expressions. From Eq.~\eqref{eq:cv1p}, it is clear that the specific heat is dominated by the term $1/(T^{2} e^{1/T})$. As in the case $j=1/2$, it can be seen that the specific heat shows a decay of the form $1/T^{2}$~\cite{mariscal2025thermodynamics}. However, this decreasing form is observed only at high temperatures, because the exponential term dominates at low temperatures. Studying the form of Eq.~\eqref{eq:chi1p}, magnetic susceptibility goes zero when $T\rightarrow 0$, since the dominant term is $1/(T e^{1/T})$. Here, we cannot deduce any phase transition yet. On the other hand, we can model the magnetisation behaviour at low temperatures, especially when $T\rightarrow 0$. This limit is given by
\begin{equation}    
\lim_{T\rightarrow 0}M=-\frac{\left(N^2+3 N+2\right) \left((N-1) \, f_{3}-2 \, f_{2}\right)}{2 \left(N^2+3 N+2\right) \, f_{1}-2 (N-1) N \left((4 N+1) \, f_{4}-3 \, f_{5}\right)},
\end{equation}
where we used the additional notation
\begin{align} 
_2F_1\left(\frac{3-N}{2},1-\frac{N}{2};2;4\right)=f_4, \qquad _2F_1\left(\frac{3-N}{2},1-\frac{N}{2};1;4\right)=f_5.
\label{eq:notation}
\end{align}
Contrary to the case $j=1/2$, for $j=1$, there is no distinction between the odd and even $N$ cases. Therefore, the ground state in the absence of an external magnetic field will always be $\ket{0,0}$ for any number of particles $N$. This means that the ground state remains at $\ket{0,0}$ for small magnetic field values and also for any $N$. Then, the magnetisation in this range is zero. This is true until $h=1$, where the lowest energy state is now shared between $\ket{0,0}$ and $\ket{1,1}$ states, so a phase transition happens at this value of the external magnetic field, because there is a sudden change in  the magnetisation:
\begin{equation}
  M_{0}=0 \qquad \to \qquad  M_{*}=\frac{m}{N}=\frac{i}{N\cdot i+N\cdot k},
\end{equation}
where $k$ and $i$ are the degeneracies of the $J=0$ and $J=1$ levels, respectively. This last expression of the magnetisation at zero temperature for external magnetic field $h=1$ can be expressed in terms of hypergeometric functions making the limit of \ref{eq:m1p}
\begin{equation}
    M_{*}=\frac{(N+1) (N+2) \left(2 \, f_{2}-(N-1) \, f_{3}\right)}{2 (N+1) (N+2) \, f_{1}+2 (N-1) N \left(3 \, f_{5}-(4 N+1) \, f_{4}\right)},
\end{equation}
where we used the previous notation \ref{eq:notation}. This magnetisation remains constant when the external magnetic field increases up to $h=2$, for which the ground state is now shared between the ${\ket{1,1}}$ and ${\ket{2,2}}$ states. Again, a new phase transition takes place, and the magnetisation is increased. In this way, magnetisation grows as the external magnetic field is increased. This increasing ends when the magnetisation reaches its maximum value, $M=1$, for $h\geq N$.

\begin{figure}[H]
    \centering
    \includegraphics[scale=0.9]{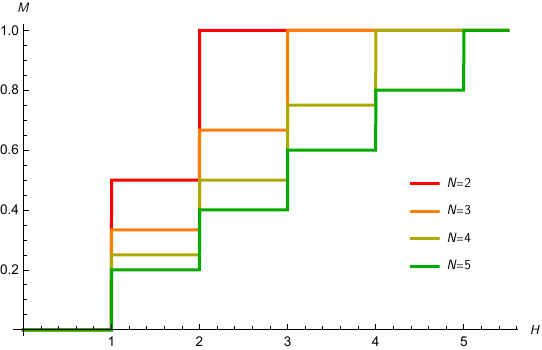}
    \caption{magnetisation as a function of magnetic field for values of the number of particles $N=2$ (red), $3$ (orange), $4$ (yellow), and $5$ (green) for the temperature $T=0^{+}$ antiferromagnetic case.}
    \label{kskj1magh}
\end{figure}

One can then observe $N$ phase transitions, one for each number of particles considered, happening at every unit of magnetic field. The external magnetic field at which these phase transitions take place is therefore given by
\begin{align}
     h^{*}_{o}\left(n\right)=n, \qquad  &\text{with} \qquad n=1,2,\dots ,  N.
     \label{qkskj1ms}
\end{align}
Fig.~\ref{kskj1magh} shows this behavior for $N=2,3,4,5$.

\subsection{Thermodynamic limit}
Now, we analyze the specific heat, magnetic susceptibility and magnetisation in the thermodynamic limit. We have chosen a number of particles $N=100,500,1000$, large enough to study the thermodynamic limit. Here, the coupling $I$ is substituted by $I/N$ in order to make the limit $\lim_{N\rightarrow\infty}E/V$ finite~\cite{le2004equilibrium}.

\begin{figure}[H]
    \centering
    \includegraphics[scale=0.9]{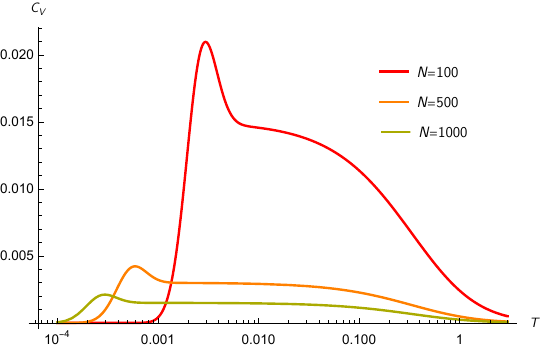}
    \caption{Specific heat as a function of temperature for values of the number of particles $N=100$ (red), $500$ (orange), and $1000$ (yellow), for the antiferromagnetic case.}
    \label{kskcvlt}
\end{figure}

In Fig. \ref{kskcvlt}, we observe that specific heat behavior is quite similar to the $j=1/2$ case for even $N$ \cite{mariscal2025thermodynamics}. The specific heat vanishes at zero temperature, reaches a maximum, and then a saturation effect appears. The curves are rescaled as a function of the number of particles, with the maxima shifted to lower temperatures and decreasing in magnitude.

One can observe in Fig.~\ref{kskcvlt} that the specific heat tends to zero in the thermodynamic limit, as expected from statistical physics. For these large amounts of particles, the decay $1/T^{2}$ is not preserved in the temperature range considered. The energy distribution of the energy levels is compressed and proportional to $N$, as said in \cite{al1998exact} for $j=1/2$, so the temperature range considered ($T\in(0,3)$) is sufficient to obtain the majority of the highest energy levels. Thus, the system can hardly absorb more energy outside of the temperature range considered.

The most probable levels of specific heat in the thermodynamic limit are still those with $J=0,1$ for any $N$. Taking the approach of considering just the two most probable levels, the analytic expressions of the approximation of the specific heat in the thermodynamic limit is
\begin{equation}
    C_{V}^{TL}\approx\frac{6 \left(N f_{2}-f_{1}\right) \left((N-1) f_{3}-2 f_{2}\right) e^{\frac{1}{N T}}}{N^2 T^2 \left(-2 f_{1} e^{\frac{1}{N T}}+2 N f_{2} \left(e^{\frac{1}{N T}}-3\right)+3 (N-1) N f_{3}\right){}^2}.
\end{equation}

\begin{figure}[H]
    \centering
    \includegraphics[scale=0.9]{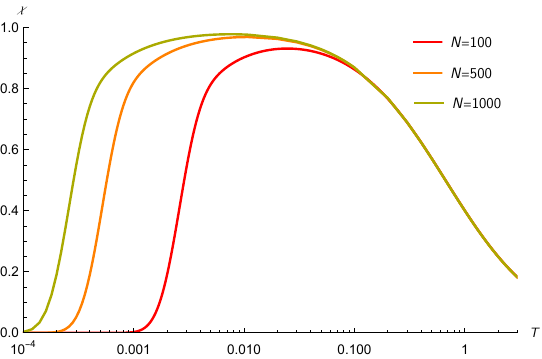}
    \caption{Magnetic susceptibility as a function of temperature for values of the number of particles $N=100$ (red), $500$ (orange), and $1000$ (yellow), for the antiferromagnetic case.}
    \label{kskchilt}
\end{figure}

In Fig. \ref{kskchilt}, the magnetic susceptibility is represented in the thermodynamic limit. As in the case $j=1/2$, we observe that every curve converges to each other in the thermodynamic limit when the temperature is sufficiently high.

\begin{figure}[H]
    \centering
    \includegraphics[scale=0.9]{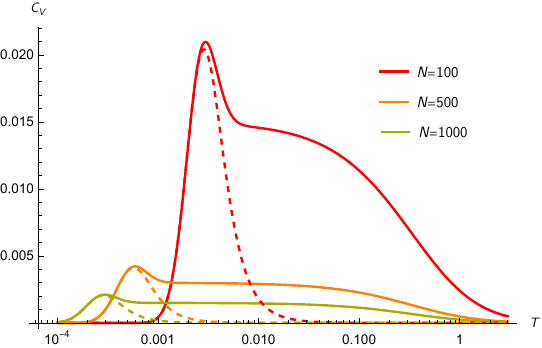}
    \caption{Comparison of the exact specific heat (continuous line) with the most probable levels approximation (dashed line) as a function of temperature for values of the number of particles $N=100$ (red), $500$ (orange), and $1000$ (yellow), for the antiferromagnetic case.}
    \label{cvltapprox}
\end{figure}

\begin{figure}[H]
    \centering
    \includegraphics[scale=0.9]{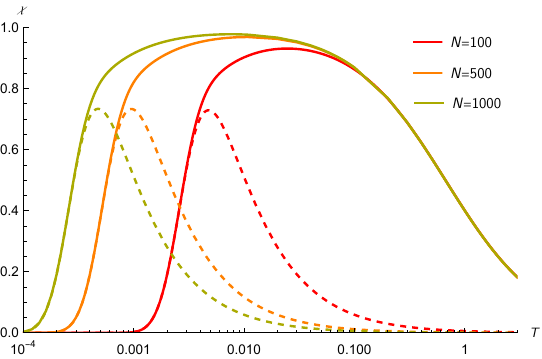}
    \caption{Comparison of the exact magnetic susceptibility (continuous line) with the most probable levels approximation (dashed line) as a function of temperature for values of the number of particles $N=100$ (red), $500$ (orange), and $1000$ (yellow), for the antiferromagnetic case.}
    \label{chiltapprox}
\end{figure}

The most probable levels of magnetic susceptibility in the thermodynamic limit do not change once an external magnetic field is applied because this thermodynamic property is evaluated for a null field. Taking just these two levels ($J=0,1$), the analytic expression of the approximation of the magnetic susceptibility is given by
\begin{equation}
    \chi^{LT}\approx-\frac{2 \left(2 f_{2}-N f_{3}+f_{3}\right)}{T \left(3 N^2 f_{3}+2 N f_{2} e^{\frac{1}{N T}}-2 f_{1} e^{\frac{1}{N T}}-6 N f_{2}-3 N f_{3}\right)}.
\end{equation}
These approximations are particularly accurate in the regime governed by the first energy level transition. In Fig \ref{cvltapprox}, we observe that this transition marks the first peak of the specific heat.
In Fig. \ref{chiltapprox}, we see that the first energy level transition coincides with the initial onset of the increase in magnetic susceptibility.

\begin{figure}[H]
    \centering
    \includegraphics[scale=0.9]{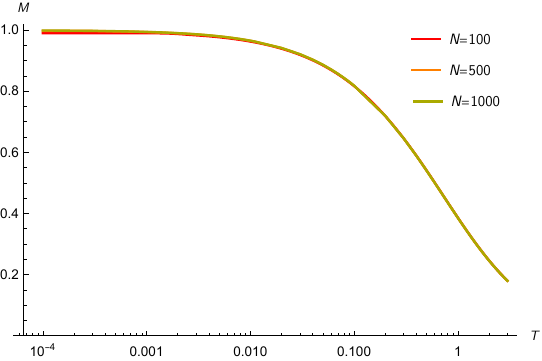}
    \caption{magnetisation as a function of temperature for values of the number of particles $N=100$ (red), $500$ (orange), and $1000$ (yellow), for the antiferromagnetic case.}
    \label{kskmlt}
\end{figure}

In Fig. \ref{kskmlt}, we show the magnetisation in the thermodynamic limit. All the curves seem to be coincident, as in the $j=1/2$ case \cite{mariscal2025thermodynamics}. However, the curves are not exactly coincident. Contrary to previous quantities, the most probable states change in the magnetisation in the thermodynamic limit, as an external magnetic field is considered. The lowest energy is shared by ${\ket{N,N}}$ and ${\ket{N-1,N-1}}$ states. Thus, the magnetisation at zero temperature is given by an average that considers each state's weights. The ${\ket{N,N}}$ state gives a magnetisation $M^{(N)}=1$ for any $N$. However, the state ${\ket{N-1,N-1}}$ gives a different magnetisation, $M^{(N-1)}=1-1/N$, which is a function of $N$. Therefore, every curve is different because it has a different value of $N$. The state ${\ket{N-1,N-1}}$ generates a magnetisation closer to unity as larger would be $N$, but never gives $M^{*}=1$ for $N$ finite. The reason why we see all the curves together is that $N$ is so large that we cannot appreciate neither the difference between each curve nor that the magnetisation is not one at $T=0$.

\begin{figure}[H]
    \centering
    \includegraphics[scale=0.9]{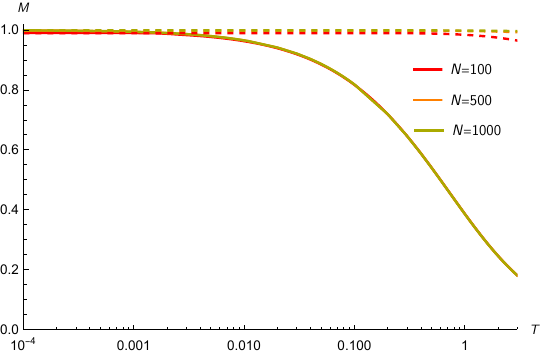}
    \caption{Comparison of the exact magnetisation (solid line) with the most probable levels approximation (dashed line) as a function of temperature for values of the number of particles $N=100$ (red), $500$ (orange), and $1000$ (yellow), for the antiferromagnetic case.}
    \label{kskmltaprox}
\end{figure}

Attending to the two most probable levels ($J=N-1,N$), we obtain the analytic expression of the approximation of the magnetisation in the thermodynamic limit
\begin{equation}
    M^{TL}\approx\frac{F+G}{H},
\end{equation}
where
\begin{equation*}
    F=(N-1) \left[e^{\frac{3 N}{T}} \left((N-1) e^{1/T}-N\right)+e^{\frac{N+1}{T}} \left(N \left(e^{1/T}-1\right)+1\right)\right]-N,
\end{equation*}
\begin{equation*}
    G=(N+1) e^{1/T}+e^{\frac{2 N+1}{T}} \left(N \left(e^{1/T}-1\right)-1\right),
\end{equation*}
\begin{equation*}
    H=N \left(e^{1/T}-1\right) \left((N-1) e^{\frac{3 N}{T}}-(N-1) e^{\frac{N+1}{T}}+e^{\frac{2 N+1}{T}}-1\right).
\end{equation*}
In Fig. \ref{kskmltaprox}, one can observe that the first energy level transition directly corresponds to the initial plateau of the magnetisation curve.

In all previous graphs of the thermodynamic limit, it is important to note that the maxima are shifted to the left for increasing values of $N$. For large $N$, all thermodynamic properties~\eqref{eq:cv1p}-\eqref{eq:m1p} converge to the same value, independently of the value of $N$, up to a factor $1/N$. However, the interaction coupling $I$ is replaced by $I/N$ in the thermodynamic limit, which causes the displacement of the maxima. Therefore, the behaviour of these quantities in the thermodynamic limit follows a tendency when $N$ increases. Moreover, the approximation considered by taking the two most probable levels is accurate for temperatures below the transition to the second excited levels (corresponding to the peak in the specific heat).  Specifically, this interval corresponds to the fundamental transition between energy levels, which manifests as the first maximum in the specific heat, governs the initial rising branch of the magnetic susceptibility, and dictates the initial stabilisation zone of the magnetisation.

\subsection{\texorpdfstring{$q$}{q}-deformation of the antiferromagnetic case}
\label{qanti}
Following the same strategy as in the undeformed case, the two most probable levels ($J=0,1$) approximation is taken into account in order to obtain analytical expressions of the thermodynamic quantities, which yield precise results in the regime of low temperatures (in a better way than in the undeformed case). Then, to observe this behaviour, we will show in the plots the exact results. While in the case $j=1/2$ the term proportional to $N$ in Hamiltonian Eq.~\eqref{eq:qenergy} goes to 1 for large values of $q$~\cite{mariscal2025thermodynamics}, for $j=1$ this term goes to infinity ($[1]_{q}[2]_{q}\to \infty$ when $\eta (q)\to \infty$). Consequently, the changes due to the deformation parameter are more noticeable in this case. Taking just these two lowest energy levels, the partition function takes the following expression:
\begin{align}
    Z_{q}\approx & \frac{1}{2} e^{\left(\frac{(q+1) \left(-2 h \sqrt{q}+N (q-1)-q+1\right)}{2 (q-1) \sqrt{q} T}\right)} \left(-\left((N-1) N \, f_{3} \left(e^{\frac{2 h}{(q-1) T}}+e^{\frac{2 h q}{(q-1) T}}+e^{\frac{h (q+1)}{(q-1) T}}\right)\right)+2 \, f_{1} e^{\frac{(q+1) \left(2 h \sqrt{q}+q-1\right)}{2 (q-1) \sqrt{q} T}}+\right.\notag \\
    &\left.+2 N \, f_{2} \left(e^{\frac{2 h}{(q-1) T}}+e^{\frac{2 h q}{(q-1) T}}+e^{\frac{h (q+1)}{(q-1) T}}-e^{\frac{(q+1) \left(2 h \sqrt{q}+q-1\right)}{2 (q-1) \sqrt{q} T}}\right)\right).
\end{align}

Using the previous equation, we are able to obtain for the specific heat 
\begin{equation}
    C_{V,q}\approx\frac{3 (q+1)^2 \left(N \, f_{2}-\, f_{1}\right) \left((N-1) \, f_{3}-2 \, f_{2}\right) e^{\frac{q+1}{2 \sqrt{q} T}}}{2 q T^2 \left(-2 \, f_{1} e^{\frac{q+1}{2 \sqrt{q} T}}+2 N \, f_{2} \left(e^{\frac{q+1}{2 \sqrt{q} T}}-3\right)+3 (N-1) N \, f_{3}\right){}^2},
    \label{eq:cv_qparaj1}
\end{equation}
the magnetic susceptibility
\begin{equation}
    \chi_{q}\approx\frac{2 \left((N-1) \, f_{3}-2 \, f_{2}\right)}{2 T \left(N \, f_{2}-\, f_{1}\right) e^{\frac{q+1}{2 \sqrt{q} T}}+3 N T \left((N-1) \, f_{3}-2 \, f_{2}\right)},
    \label{eq:qsuscep}
\end{equation}
and the magnetisation
\begin{equation}
    M_{q}\approx\frac{(N+2)\left(2 \, f_{2}-(N-1) \, f_{3}\right) \left(e^{\frac{2}{(q-1) T}}-e^{\frac{2 q}{(q-1) T}}\right)}{(N+2)\left(2 \, f_{1}-2 N \, f_{2}\right) e^{-\frac{q^{3/2}+2 q-\frac{1}{\sqrt{q}}+2}{2 T(1- q)}}+N \left((N+5) \, f_{2}-3 (N-1) \, f_{4}\right) e^{\frac{q+1}{(q-1) T}} \left(2 \cosh \left(\frac{1}{T}\right)+1\right)}. 
    \label{eq:qmagentization}
\end{equation}

The larger the deformation parameter, the more energy and temperature required to jump to the next energetic level. This can be seen in Fig.~\ref{kskj1DensityD} (a), and also in~\eqref{eq:qenergy} owing to the properties of the $q$-numbers. Because $[a+1]_{q}+[b+1]_{q} > [a]_{q}+[b]_{q}+2$ for $q>1$, the difference between the energy levels becomes greater for larger values of the deformation parameter. Thus, these previous expressions reproduce the exact thermodynamic properties more faithfully in the considered temperature range, because the behaviour of these two lowest energy levels is dominant for larger temperature ranges.

\begin{figure}[H]
  \begin{minipage}{0.4\textwidth}
    (a)\includegraphics[scale=0.7]{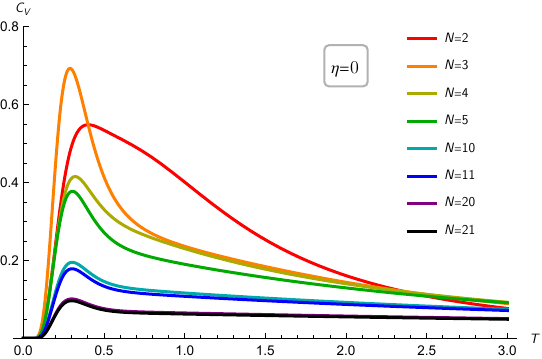}
  \end{minipage}
  \hspace{15mm}
  \begin{minipage}{0.4\textwidth}
    (b)\includegraphics[scale=0.7]{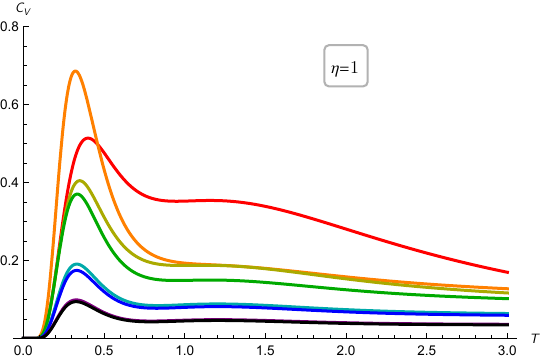}
  \end{minipage}
  \caption{Specific heat as a function of temperature for values of the number of particles $N=2$ (red), $3$ (orange), $4$ (yellow), $5$ (green), $10$ (cyan), $11$ (blue), $20$ (purple), and $21$ (black) and of the parameter (a) $\eta=0$ and (b) $\eta=1$, for the antiferromagnetic case.}
  \label{qkskj1cvpd}
\end{figure}

We start by reproducing the specific heat, represented in Fig.~\ref{qkskj1cvpd}, including the undeformed case, for some value of $\eta$. We have chosen $\eta=1$, to see the behavior changes with care. 
Obviously, Fig.~\ref{qkskj1cvpd} (a), corresponding to $\eta=0$, is the same as the undeformed case. When the deformation parameter is greater, the first peak of specific heat slightly moves to higher temperatures (barely noticeable in the plots). Moreover, certain bumps are observed at higher temperatures associated with the transition from the first excited state to the second one. Noticeably, these bumps appear for higher temperatures as $\eta$ increases, being more pronounced and smaller with increasing values of the deformation parameter. All of these phenomena can be explained from the deformed energy spectrum represented in Fig.~\ref{kskj1DensityD}. The lowest energy levels ($J=0,1$ in the figure) have less energy when the deformation parameter is increased (fact that cannot be appreciated in the figure). This explains why we see the first peak of the specific heat always at approximately the same temperature. Regarding the bumps, the first excited level $(J=1)$ has a lower energy for any value for the deformation parameter, while the second level $(J=2)$ depends on the values of $N$ and $\eta$. That is why we see some curves that are sharper than others. In both cases, an increase in deformation results in a greater difference between the levels, and this explains why the bumps appear for larger temperatures. Moreover, the lowest energy states become more likely (less energetic), so the deformed specific heat has a greater contribution of them than in the undeformed case (at the considered temperature range). This implies that the specific heat shows smaller bumps for greater values of $\eta$, being the greatest in the undeformed case. 

\begin{figure}[H]
  \begin{minipage}{0.4\textwidth}
    (a)\includegraphics[scale=0.7]{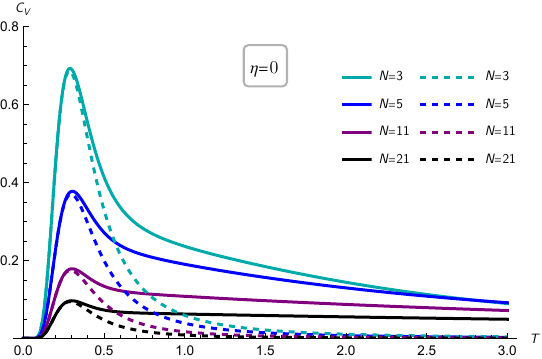}
  \end{minipage}
  \hspace{15mm}
  \begin{minipage}{0.4\textwidth}
    (b)\includegraphics[scale=0.7]{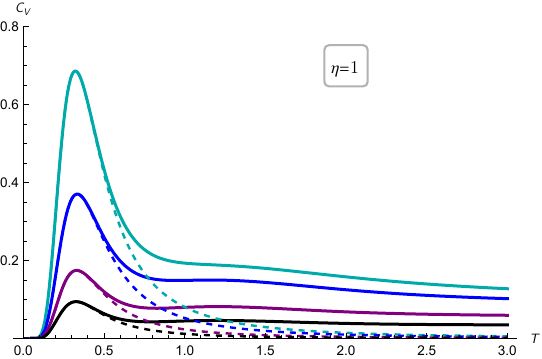}
  \end{minipage}
  \caption{Comparison of the exact specific heat (continuous line) with the most probable levels approximation (dashed line) as a function of temperature for odd values of the number of particles $N=3$ (cyan), $5$ (blue), $11$ (purple), and $21$ (black) for the antiferromagnetic case. The values of the parameter $q=e^{\eta/N}$ are: (a) $\eta=0$ and (b) $\eta=1$.}
  \label{comparcv}
\end{figure}

The specific heat calculated using the most probable levels approximation for the deformed case is shown in Fig. \ref{comparcv}. As can be seen here and in the following, the approximation is better for all the thermodynamic properties in the deformed case, given that it follows the exact curve over a longer range of temperature. The physical origin of this enhanced agreement across all deformed thermodynamic properties is the deformation-induced widening of the energy level spacing. This increased separation shifts thermodynamic transitions; most notably, the higher-energy excitations, toward progressively higher temperatures. Consequently, in the undeformed scenario, there is a greater weight on transitions that are not considered in the approximation. Conversely, in the deformed case, these transitions shift to higher temperatures, enabling the approximation to remain more accurate over a wider interval.

\begin{figure}[H]
  \begin{minipage}{0.4\textwidth}
    (a)\includegraphics[scale=0.7]{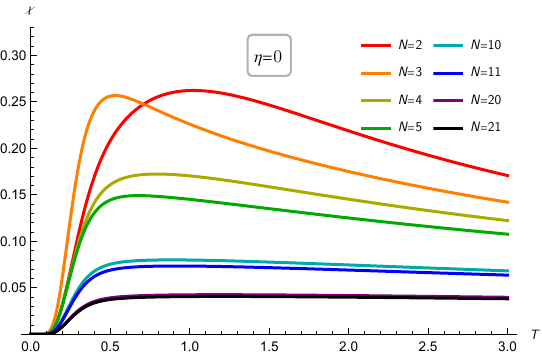}
  \end{minipage}
  \hspace{15mm}
  \begin{minipage}{0.4\textwidth}
    (b)\includegraphics[scale=0.7]{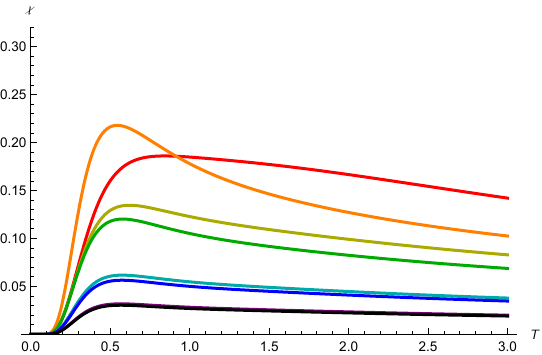}
  \end{minipage}  \caption{Magnetic susceptibility as a function of temperature for values of the number of particles $N=2$ (red), $3$ (orange), $4$ (yellow), $5$ (green), $10$ (cyan), $11$ (blue), $20$ (purple), and $21$ (black) and of the parameter (a) $\eta=0$ and (b) $\eta=1$, for the antiferromagnetic case.}
    \label{qkskj1xpd}
\end{figure}

Magnetic susceptibility for different values of $\eta$ is shown in Fig.~\ref{qkskj1xpd}. The maxima become smaller as the deformation parameter becomes larger. The explanation is the same as that for the specific heat. The energy gaps between the excited states and the ground state increase exponentially when the deformation parameter grows, making thermal excitations less probable at a given temperature. Consequently, the observed peaks, corresponding to the transition from the ground state to the first excited state, have a greater contribution of the lowest energy states, and, especially, from the ground state ${\ket{0,0}}$, which cannot sense the external magnetic field. This globally implies a lower sensitivity to the magnetic field, so the magnetic susceptibility is smaller for greater values of $\eta$.
In the $\eta=1$ representation, it makes really clear how unmarked the $N=3$ case is. This peak also decreases with deformation, but does so differently from the rest $N$ considered. Indeed, the peak for $N=2$ decreases with deformation at a faster rate than that of $N=3$. At approximately $\eta \approx 0.5$, the $N=2$ peak begins to lie below the one for $N=3$. Therefore, the sensitivity to the magnetic field for $N=3$ is the highest for $\eta\gtrsim 0.5$. This is due to the fact that $N=3$ is a special case due to its multiplicity. We showed above that the ratio of the multiplicity of the ground state and the multiplicity of the first excited state possesses a local minimum at $N=3$. Thus, the magnetic susceptibility for $N=3$ is not too influenced by the value of the deformation parameter. The first excited state of $N=3$ has the highest probability weight and that is why for any other $N$ the magnetic susceptibility decreases faster.

\begin{figure}[H]
  \begin{minipage}{0.4\textwidth}
    (a)\includegraphics[scale=0.7]{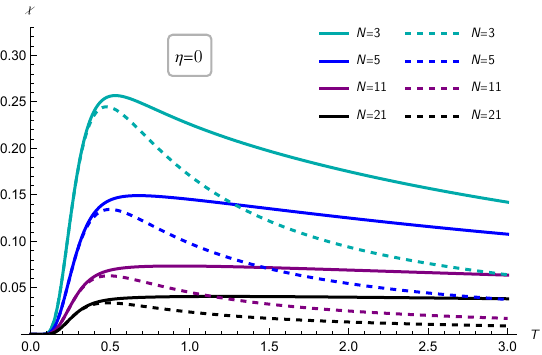}
  \end{minipage}
  \hspace{15mm}
  \begin{minipage}{0.4\textwidth}
    (b)\includegraphics[scale=0.7]{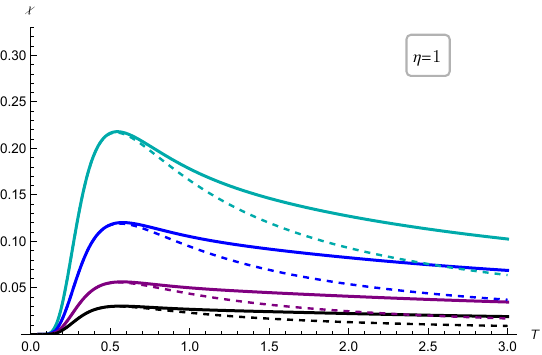}
  \end{minipage}
  \caption{Comparison of the exact magnetic susceptibility (continuous line) with the most probable levels approximation (dashed line) as a function of temperature for odd values of the number of particles $N=3$ (cyan), $5$ (blue), $11$ (purple), and $21$ (black) for the antiferromagnetic case. The values of the parameter $q=e^{\eta/N}$ are: (a) $\eta=0$ and (b) $\eta=1$.}
  \label{comparchi}
\end{figure}

The magnetic susceptibility obtained through the most probable states approximation for the deformed system is illustrated in Fig. \ref{comparchi}.  Again, the approximation is also better for the deformed scenario; this improvement is even more pronounced, as it can reproduce the structure up to and including the peak, which was not achieved in the undeformed case.

\begin{figure}[H]
  \begin{minipage}{0.4\textwidth}
    (a)\includegraphics[scale=0.7]{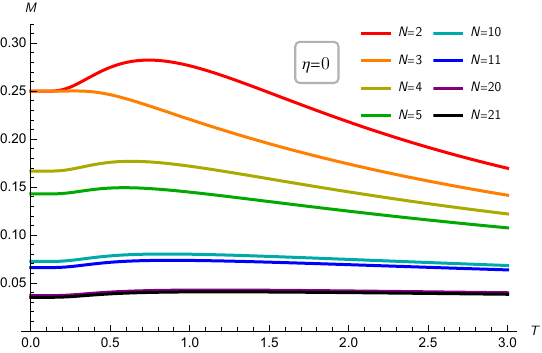}
  \end{minipage}
  \hspace{15mm}
  \begin{minipage}{0.4\textwidth}
    (b)\includegraphics[scale=0.7]{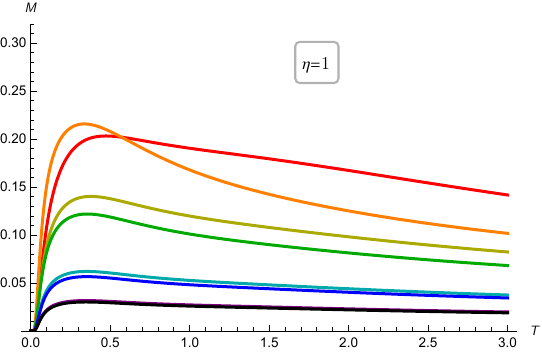}
  \end{minipage}
  \caption{magnetisation as a function of temperature for values of the number of particles $N=2$ (red), $3$ (orange), $4$ (yellow), $5$ (green), $10$ (cyan), $11$ (blue), $20$ (purple), and $21$ (black) and of the parameter (a) $\eta=0$ and (b) $\eta=1$ for the $h=\gamma=1$ antiferromagnetic case. }
  \label{qkskj1mpd}
\end{figure}

Finally, the magnetisation is represented in Fig.~\ref{qkskj1mpd} for an external magnetic field. We observe that maxima decrease and are shifted to larger temperatures as the deformation parameter increases. Again, this behavior can be understood from the deformed energy spectrum. The maxima become smaller, because the deformed lowest-energy states are more likely than the undeformed ones. The displacement of the maxima is due to the fact that the deformation makes the energetic difference between states larger, so the required temperature at which the transition appears becomes greater.
An important remark about these representations is that magnetisation at zero temperature has a nonzero value for the undeformed case and a zero value for for any $\eta \neq 0$. The reason lies in the chosen magnetic field $h=\gamma=1$. This value allows the ground energy state to be shared between the ${\ket{0,0}}$ and ${\ket{1,1}}$  states, whose average magnetisation is not null for the undeformed case (independently of the parity of $N$). However, the ground state becomes just the ${\ket{0,0}}$ when a deformation becomes relevant, no matter how small it is for any $\eta \neq 0$. Consequently, the magnetisation is null at zero temperature for $\eta>0$.

The evolution of magnetisation with respect to the deformation parameter is quite homogeneous. The case $N=2$ exemplifies this trend. In the undeformed case, there is a non-vanishing magnetisation at zero temperature. For higher temperatures, we observe a peak corresponding to the transition from the ground state, shared by ${\ket{0,0}}$ and ${\ket{1,1}}$, to the first excited state, shared by ${\ket{1,0}}$ and ${\ket{2,2}}$. Then, the magnetisation gets a higher value (its maximum value for $N=2$). When the deformation parameter is slightly increased, a rearrangement of the most probable states occurs. Consequently, the magnetisation is zero at zero temperature, since the ground state is no longer shared. For different values of the deformation parameter, the curves clearly display several bumps. This can be understood from the energy of different levels as a function of the parameter $\eta$: the deformed ground state is ${\ket{0,0}}$, the first excited state is ${\ket{1,1}}$, the second state ${\ket{1,0}}$, and the third depends on the particular value of $\eta$. From Eq.~\eqref{eq:qenergy}, one can easily check that the ${\ket{2,2}}$ state is less energetic than the ${\ket{1,-1}}$ state for $\eta\in(0,0.842)$; but this is not true for $\eta>0.842$, so the magnetisation does not increase: when the temperature grows enough to obtain the ${\ket{2,2}}$ state, whose magnetisation is larger, there are also states contributing to the increase in the magnetisation, so there are no new bumps. This reasoning can be followed for any number of particles.

\begin{figure}[H]
  \begin{minipage}{0.4\textwidth}
    (a)\includegraphics[scale=0.7]{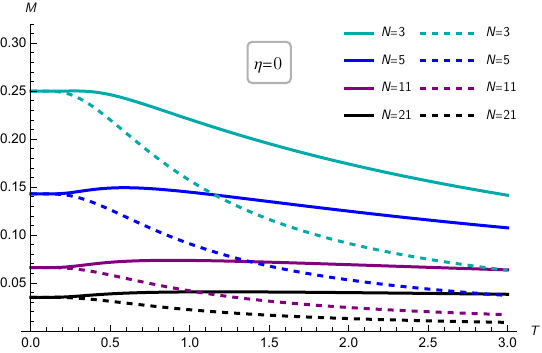}
  \end{minipage}
  \hspace{15mm}
  \begin{minipage}{0.4\textwidth}
    (b)\includegraphics[scale=0.7]{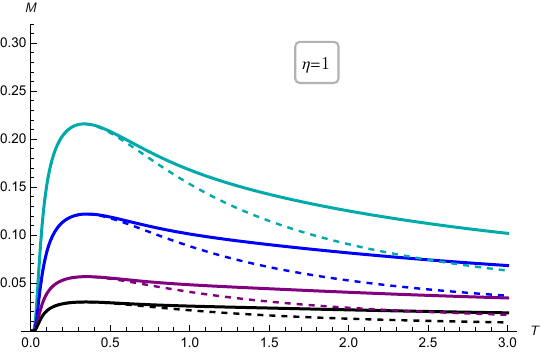}
  \end{minipage}
  \caption{Comparison of the exact magnetisation (continuous line) with the most probable levels approximation (dashed line) as a function of temperature for values of the odd number of particles $N=3$ (cyan), $5$ (blue), $11$ (purple), and $21$ (black) for the $h=\gamma=1$ antiferromagnetic case. The values of the parameter $q=e^{\eta/N}$ are: (a) $\eta=0$ and (b) $\eta=1$.}
  \label{comparmag}
\end{figure}

The magnetisation evaluated using the most probable states approximation for the deformed system is shown in Fig. \ref{comparmag}. Finally, regarding the magnetisation, the approximation again demonstrates a substantial improvement in the deformed case, successfully capturing both the peak structure and a portion of its subsequent decay. 

\begin{figure}[H]
  \begin{minipage}{0.4\textwidth}
    (a)\includegraphics[scale=0.7]{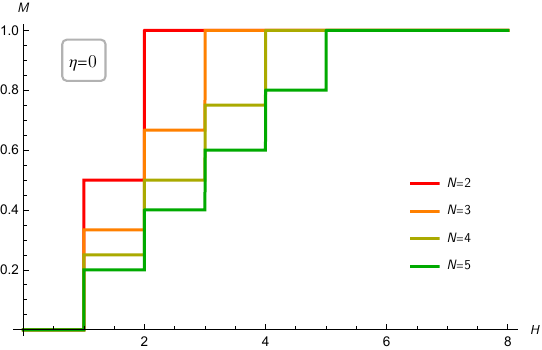}
  \end{minipage}
  \hspace{15mm}
  \begin{minipage}{0.4\textwidth}
    (b)\includegraphics[scale=0.7]{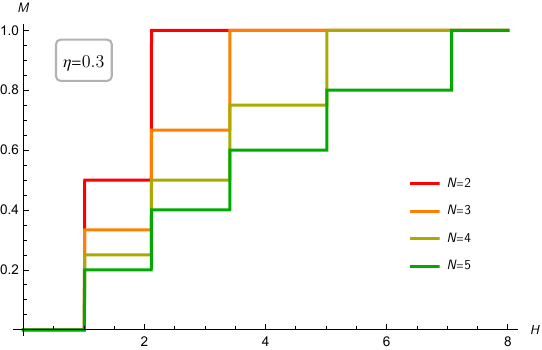}
  \end{minipage}
  \caption{magnetisation as a function of magnetic field for values of the number of particles $N=2$ (red), $3$ (orange), $4$ (yellow), and $5$ (green) and of the parameter (a) $\eta=0$ and (b) $\eta=0.3$ for the temperature $T=0.0001$ antiferromagnetic case.}
  \label{qkskj1transition}
\end{figure}

To better understand why magnetisation takes those values at zero temperature, we have analysed this property as a function of the external magnetic field in Fig.~\ref{qkskj1transition} for several deformation values. We chose $\eta = 0.3$ to better appreciate the changes in the field interval considered, $H\in(0,8)$. 

Several phase transitions are observed for each $N$. The number of these phenomena is the same as in the undeformed case: there are $N$ phase transitions, regardless of whether $N$ is odd or even. The difference from the undeformed case is due to the spacing between the transitions, because the energies for different angular momenta are not given by~\eqref{eq:energy} but by~\eqref{eq:qenergy}. As before, by computing the difference between contiguous energy levels from~\eqref{eq:qenergy} it is easy to obtain the values of the external magnetic field at which every phase transition takes place 
\begin{align}
    H^{*}\left(n\right)=\frac{[2\cdot n]_{q}}{2}, \qquad  &\text{with} \qquad n=1,2,\dots ,  N.
    \label{qkskj1msde}
\end{align}
Taking the limit $q\to1$ one find the same expression~\eqref{qkskj1ms} obtained previously for the undeformed case.

In Fig.~\ref{qkskj1transition}, we can observe that the external magnetic field needed for a transition to take place is greater for greater values of the deformation parameter. The changes in these critical values are based on the differences in the energy levels, as previously discussed. 

\subsection{\texorpdfstring{$q$}{q}-deformation of the thermodynamic limit}
\label{subsec:therlimitanti}
From the undeformed scenario, we can analyze the thermodynamic limit of the deformed case.
\begin{figure}[H]
  \begin{minipage}{0.4\textwidth}
    (a)\includegraphics[scale=0.7]{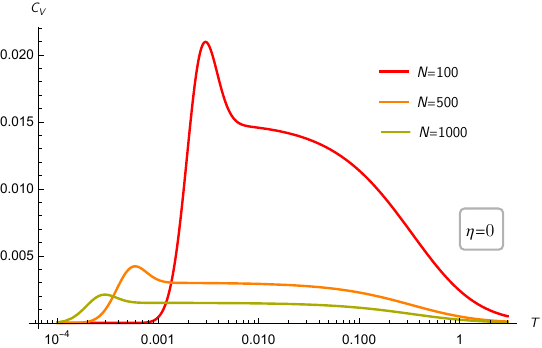}
  \end{minipage}
  \hspace{15mm}
  \begin{minipage}{0.4\textwidth}
    (b)\includegraphics[scale=0.7]{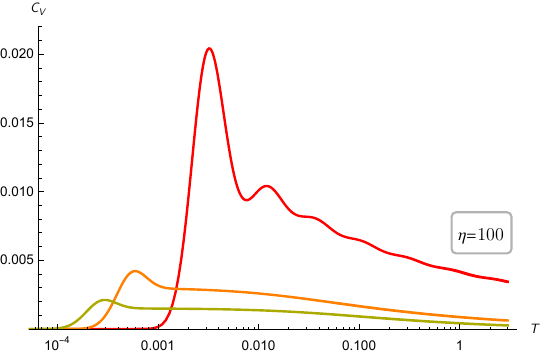}
  \end{minipage}
  \caption{Specific heat as a function of temperature for values of the number of particles $N=100$ (red), $N=500$ (orange), $N=1000$ (yellow) for the interaction constant $I=1/N$ antiferromagnetic case. The values of the parameter $q=e^{\eta/N}$ are: (a) $\eta=0$ and (b) $\eta=100$.}
  \label{qkskj1transitioncv}
\end{figure}   

It can be seen that, as in the undeformed case, the maxima are shifted to the left for increasing values of the number of particles (for the same $q$), following a similar behaviour as for $j=1/2$ \cite{mariscal2025thermodynamics}. Thus, we observe that the specific heat does not show many changes, except for the bumps that appear with the deformation. These bumps become particularly sharp at intermediate temperatures for $\eta=1$ (Fig. \ref{qkskj1transitioncv} (b)). The reason for the appearance of bumps is the transitions between consecutive $J$ energetic levels and the properties of $q$-numbers. The temperature at which the bumps of the specific heat appear is proportional to the difference between consecutive energetic levels. In the deformed case, the energetic levels are given by $q$ integer numbers. Consequently, the temperature demanded to jump to the next energy level becomes higher as the deformation parameter increases (this is shown in Fig. \ref{kskj1DensityD}). Thus, all the energy states, which overlap in the undeformed case, are considerably separated in Fig. \ref{qkskj1transitioncv} (b), allowing us to better see every transition. The bumps become smaller in magnitude as the deformation parameter grows, since the energy differences between the ground and excited states are greater for larger values of the deformation parameter. Therefore, in the considered range of temperature, there is a greater weight of the ground states, showing smaller bumps for greater values of $\eta$.

In the deformed case, the most probable levels for any number of particles are the same as those in the undeformed scenario ($J=0,1$ for any $N$). Consequently, we can also obtain the analytic expressions using this approximation of the specific heat in the deformed case. 
\begin{equation}
    C_{V,q}^{TL}\approx\frac{3 (q+1)^2 \left(N f_{2}-f_{1}\right) \left((N-1) f_{3}-2 f_{2}\right) e^{\frac{q+1}{2 N \sqrt{q} T}}}{2 N^2 q T^2 \left(-2 f_{1} e^{\frac{q+1}{2 N \sqrt{q} T}}+2 N f_{2} \left(e^{\frac{q+1}{2 N \sqrt{q} T}}-3\right)+3 (N-1) N f_{3}\right){}^2}.
\end{equation}

\begin{figure}[H]
  \begin{minipage}{0.4\textwidth}
    (a)\includegraphics[scale=0.7]{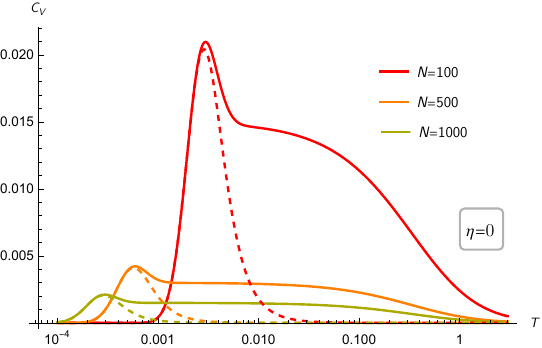}
  \end{minipage}
  \hspace{15mm}
  \begin{minipage}{0.4\textwidth}
    (b)\includegraphics[scale=0.7]{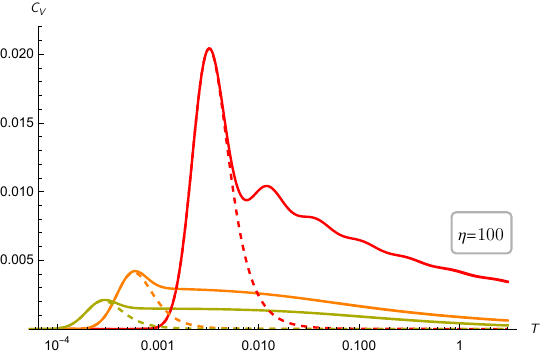}
  \end{minipage}
  \caption{Comparison of the exact specific heat (continuous line) with the most probable levels approximation (dashed line) as a function of temperature for values of the number of particles $N=100$ (red), $N=500$ (orange), $N=1000$ (yellow) for the interaction constant $I=1/N$ antiferromagnetic case. The values of the parameter $q=e^{\eta/N}$ are: (a) $\eta=0$ and (b) $\eta=100$.}
  \label{cvltdefapprox}
\end{figure}

In Fig. \ref{cvltdefapprox}, we compare this approximation of the most probable levels in the deformed case against the exact specific heat. As previously discussed, this approximation performs significantly better in the deformed case, becoming increasingly accurate for larger values of the deformation parameter $\eta$.

\begin{figure}[H]
  \begin{minipage}{0.4\textwidth}
    (a)\includegraphics[scale=0.7]{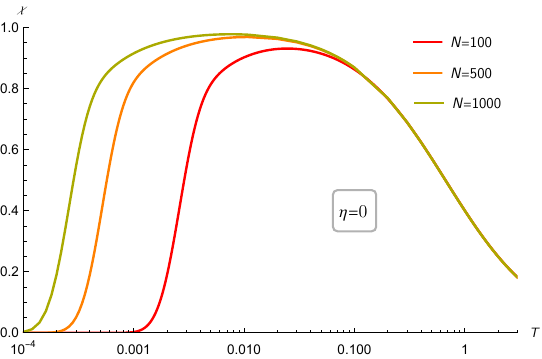}
  \end{minipage}
  \hspace{15mm}
  \begin{minipage}{0.4\textwidth}
    (b)\includegraphics[scale=0.7]{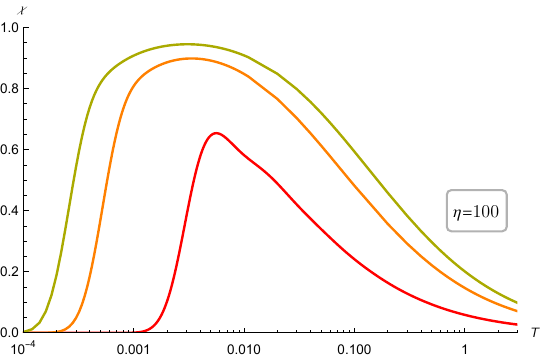}
  \end{minipage}
  \caption{Magnetic susceptibility as a function of temperature for values of the number of particles $N=100$ (red), $N=500$ (orange), $N=1000$ (yellow) for the interaction constant $I=1/N$ antiferromagnetic case. The values of the parameter $q=e^{\eta/N}$ are: (a) $\eta=0$ and (b) $\eta=100$.}
  \label{qkskj1transitionchi}
\end{figure}

In Fig. \ref{qkskj1transitionchi}, we show the deformed magnetic susceptibility. The convergence of this quantity in the undeformed case is clearly attenuated here. As deformation is introduced, the curves require more temperature to converge and practically do so for a zero value of the susceptibility. Moreover, as in the undeformed case, the curves are shifted to the left for increasing values of $N$ (for the same $q$). An appreciable difference is the decrease in the magnitude of the susceptibility as a function of the deformation parameter. This is the same as for the specific heat: the energy difference between the ground and excited states increases as the values of the deformation parameter increase. 

Similar to the undeformed case, the most probable levels of magnetic susceptibility are the same as those of the specific heat ($J=0,1$). This fact enables the derivation of an analytical expression for the magnetic susceptibility in the deformed case.
\begin{equation}
    \chi^{TL}_{q}\approx-\frac{2 \left(2 f_{2}-N f_{3}+f_{3}\right)}{T \left(3 N^2 f_{3}+2 N f_{2} e^{\frac{\sqrt{q}}{2 N T}+\frac{1}{2 N \sqrt{q} T}}-2 f_{1} e^{\frac{\sqrt{q}}{2 N T}+\frac{1}{2 N \sqrt{q} T}}-6 N f_{2}-3 N f_{3}\right)}.
\end{equation}

\begin{figure}[H]
  \begin{minipage}{0.4\textwidth}
    (a)\includegraphics[scale=0.7]{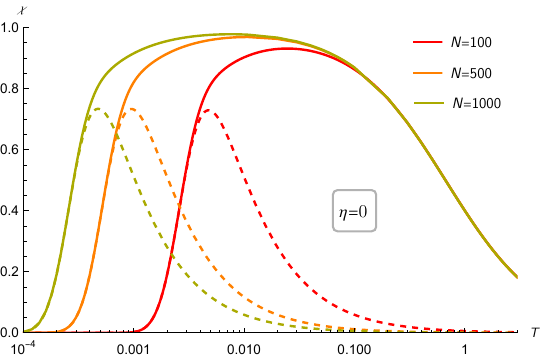}
  \end{minipage}
  \hspace{15mm}
  \begin{minipage}{0.4\textwidth}
    (b)\includegraphics[scale=0.7]{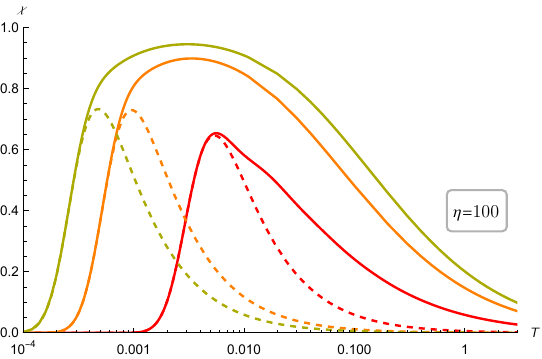}
  \end{minipage}
  \caption{Comparison of the exact magnetic susceptibility (continuous line) with the most probable levels approximation (dashed line) as a function of temperature for values of the number of particles $N=100$ (red), $N=500$ (orange), $N=1000$ (yellow) for the interaction constant $I=1/N$ antiferromagnetic case. The values of the parameter $q=e^{\eta/N}$ are: (a) $\eta=0$ and (b) $\eta=100$.}
  \label{chiltdefapprox}
\end{figure}

In order to see the accuracy of this approximation, we represent in Fig. \ref{chiltdefapprox} its comparison with exact magnetic susceptibility. Again, the approximation exhibits a slightly better performance in the deformed case, for the reasons explained above.

\begin{figure}[H]
  \begin{minipage}{0.4\textwidth}
    (a)\includegraphics[scale=0.7]{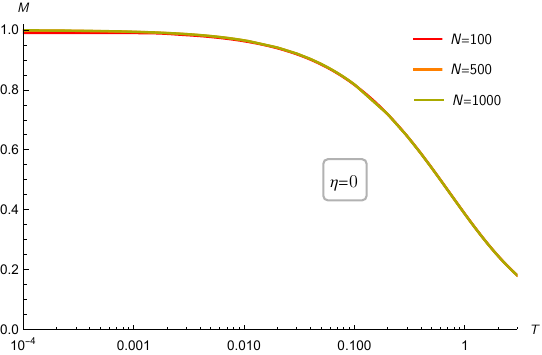}
  \end{minipage}
  \hspace{15mm}
  \begin{minipage}{0.4\textwidth}
    (b)\includegraphics[scale=0.7]{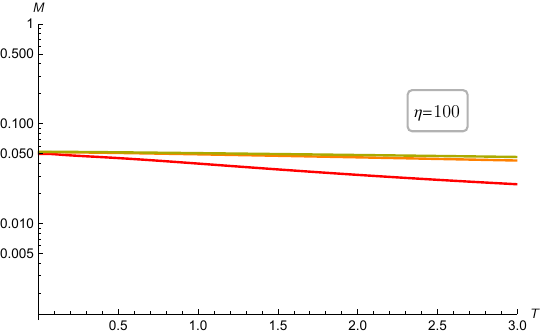}
  \end{minipage}
  \caption{Magnetisation as a function of temperature for values of the number of particles $N=100$ (red), $N=500$ (orange), $N=1000$ (yellow) for the interaction constant $I=1/N$ and $h\gamma=1$ antiferromagnetic case. The values of the parameter $q=e^{\eta/N}$ are: (a) $\eta=0$ and (b) $\eta=100$. }
  \label{qkskj1transitionm}
\end{figure}

In Fig. \ref{qkskj1transitionm}, we represent the deformed magnetisation.

The most probable states of the deformed magnetisation are not the same as those of the specific heat or undeformed magnetisation states. In fact, they are not even the same states for each value of the deformation parameter. There exists a competition between $J$, $\eta$, and $N$ to reduce the energy of each of the states. To describe them, we derive an equation that allows us to obtain the most probable states from the deformation parameter and number of particles. This is obtained by minimizing the energy expression with an external magnetic field \eqref{eq:qenergy} (for $\gamma=h=1$ and $m=J$) as a function of $J$
\begin{equation}
J_{MP}=m_{MP}=\frac{N\cdot\mathop{\mathrm{arcsinh}}\left(\frac{2 N^2 (\cosh (\eta/N)-1)}{\eta }\right)}{\eta }-\frac{1}{2}.
    \label{eq:MostProbable}
\end{equation}
Note that this expression is the same one obtained for the $j=1/2$ case studied in \cite{mariscal2025thermodynamics}.

The two most probable energy levels are those whose angular momentum is that of the two nearest integers for any $N$. From them, we can also write the analytical approximation used for the undeformed case. However, because the most probable levels depend on $N$, we cannot write a generic expression as previously.

Owing to the dependency of the most probable states on the number of particles and the value of the deformation parameter, the deformed magnetisation has different values at $T=0$ for different deformation parameters and different $N$, as shown in Fig. \ref{qkskj1transitionm} (b). These magnetisation values at zero temperature can be calculated using \eqref{eq:MostProbable}: $M(T=0)=m_{MP}/N$.  We show the computation of the most probable states (magnetisation at zero temperature) for three deformation values in Table~\ref{tablamp}.
\begin{table}[H]
    \centering
    \begin{tabular}{c c c c}
    \hline
    $N$   & |$J_{MP},m_{MP}\rangle_{q=e^{10/N}}$ & |$J_{MP},m_{MP}\rangle_{q=e^{50/N}}$  & |$J_{MP},m_{MP}\rangle_{q=e^{100/N}}$\\ \hline
    100 & |$29,29\rangle$   & |$9,9\rangle$   &  |$5,5\rangle$  \\ \hline
    500 & |$149,149\rangle$ & |$46,46\rangle$ &  |$26,26\rangle$  \\ \hline
    1000 &|$299,299\rangle$ & |$92,92\rangle$ &  |$52,52\rangle$  \\ \hline
    \end{tabular}
    \caption{Data for different number of particles of the most probable state in presence of an external magnetic field {$h\gamma=1$}.}
    \label{tablamp}
\end{table}

In addition, it is observed that the magnetisation becomes smaller and that the decrease in the magnitude is considerably attenuated as the deformation parameter increases. Both phenomena are due to the fact that the system requires more energy to reach the excited states. Consequently, the magnetisation behaviour is dominated by the two most probable levels. 

\subsection{Finite-size scaling}
\label{subsec:fssanti}
An elegant feature of this model in the antiferromagnetic regime ($I < 0$) is that its low-temperature scaling behavior is identical to that of the even standard spin-$1/2$ ($j=1/2$) case \cite{mariscal2025thermodynamics}. Physically, this equivalence stems from a robust collective universality. At low temperatures, the thermal behaviour is entirely dictated by low-lying collective excitations near the $J=0$ ground state. 

While the total dimension of the Hilbert space depends on the constituent spin magnitude ($3^N$ for the trinomial $j=1$ case versus $2^N$ for $j=1/2$), this difference only manifests as a constant shift in the configurational entropy, $\Delta S \sim N \ln(2S+1)$. Because the specific heat $C_V$ is proportional to the second temperature derivative of the free energy, this constant contribution disappears. Furthermore, for $J \ll N$, the degeneracy factors $g(J)$ share the same form up to an overall scaling constant, which cancels out when computing thermal averages such as $C_V$ and the magnetic susceptibility $\chi$. 

Consequently, the individual spin details are completely washed out in the thermodynamic limit, yielding identical universal scaling curves. This collective behaviour is also remarkably resilient to $q$-deformation: as long as we remain in the numerically stable regime where $\eta/N \ll 1$, the algebraic distortion of the low-energy states is negligible, and the system asymptotically flows to the same standard $su(2)$ scaling limits.

In the thermodynamic limit, the scaling variable is $s=NT$. The specific-heat maximum and its height scale as
\[
T_{\max}(N)\sim \frac{0.296}{N}, \qquad C_V^{\max}(N)\sim \frac{2.115}{N},
\]
yielding a finite limit for $NC_V^{\max}$. The magnetic susceptibility at $T_{\max}$ remains finite, $\chi(T_{\max})\to 0.53365\ldots$.

Table~\ref{tab:even-eta} confirms this for different values of $\eta$. For $\eta/N\ll 1$, $NT_{\max}$ and $NC_V^{\max}$ stabilise rapidly. A log-log analysis yields scaling exponents $p_{C_V}=-1$ and $p_{\chi}=0$.

\begin{table}[H]
\centering
\begin{tabular}{c c c c c}
\hline
$\eta$ & $N$ & $NT_{\max}$ & $NC_V^{\max}$ & $\chi(T_{\max})$ \\
\hline
\multirow{8}{*}{$0$} 
        & $10^3$          & $0.2966$ & $2.1105$  & $0.533757$ \\
        & $3\cdot 10^3$   & $0.2965$ & $2.11383$ & $0.534573$ \\
        & $6\cdot 10^3$   & $0.2965$ & $2.11466$ & $0.534852$ \\
        & $10^4$          & $0.2965$ & $2.115$   & $0.534963$ \\
        & $10^5$          & $0.296$  & $2.11544$ & $0.533635$ \\
        & $3\cdot 10^5$   & $0.296$  & $2.11547$ & $0.533646$ \\
        & $6\cdot 10^5$   & $0.296$  & $2.11548$ & $0.533649$ \\
        & $10^6$          & $0.296$  & $2.11548$ & $0.53365$  \\
\hline
\multirow{8}{*}{$50$} 
        & $10^3$          & $0.2967$ & $2.11007$ & $0.533572$ \\
        & $3\cdot 10^3$   & $0.2965$ & $2.11378$ & $0.53452$  \\
        & $6\cdot 10^3$   & $0.2965$ & $2.11465$ & $0.534838$ \\
        & $10^4$          & $0.2965$ & $2.11499$ & $0.534958$ \\
        & $10^5$          & $0.296$  & $2.11544$ & $0.533635$ \\
        & $3\cdot 10^5$   & $0.296$  & $2.11547$ & $0.533646$ \\
        & $6\cdot 10^5$   & $0.296$  & $2.11548$ & $0.533649$ \\
        & $10^6$          & $0.296$  & $2.11548$ & $0.53365$  \\
\hline
\multirow{8}{*}{$100$} 
        & $10^3$          & $0.2967$ & $2.10881$ & $0.532138$ \\
        & $3\cdot 10^3$   & $0.2965$ & $2.11364$ & $0.53436$  \\
        & $6\cdot 10^3$   & $0.2965$ & $2.11461$ & $0.534798$ \\
        & $10^4$          & $0.2965$ & $2.11498$ & $0.534944$ \\
        & $10^5$          & $0.296$  & $2.11544$ & $0.533635$ \\
        & $3\cdot 10^5$   & $0.296$  & $2.11547$ & $0.533646$ \\
        & $6\cdot 10^5$   & $0.296$  & $2.11548$ & $0.533649$ \\
        & $10^6$          & $0.296$  & $2.11548$ & $0.53365$  \\
\hline
\end{tabular}
{\caption{Asymptotic results for even $N$ and several deformation parameters $\eta$, restricted to the numerically stable regime $\eta/N \ll 1$.}
\label{tab:even-eta}}
\end{table}

\section{Conclusions}
\label{sec:conclusions}
In this work, we have constructed and analyzed the quantum group deformation of the Kittel–Shore (KS) model for spin-1 particles by exploiting the underlying $U_q(\mathfrak{su}(2))$ coalgebra symmetry, whose complete integrability is fully preserved under $q$-deformation via the $N$-th coproduct map. Building upon the exact energy spectrum and multi-spin eigenvectors, the second (and central part) of this paper investigates the main thermodynamic properties of the spin-1 KS model alongside the physical implications of its $q$-deformation. By rigorously analysing the specific heat, magnetic susceptibility, magnetisation, and phase transitions for deformed Hamiltonian, complemented by a detailed evaluation of the Curie temperature in the ferromagnetic regime, this study elucidates how quantum deformation reshapes the collective behaviour of the system.

In this context, explicit analytical expressions for the $j=1$ Kittel–Shore Hamiltonian are derived by extending the foundational results previously established for the $j=1/2$ case \cite{ballesteros2025quantum}, where the general extension to an arbitrary spin $j$ was formally demonstrated. A detailed examination of the resulting energy level distributions reveals the same behaviour as the case $j=1/2$: while the state degeneracies remain entirely unaffected by the deformation, the energy spectrum systematically stretches and broadens as the deformation parameter departs from the undeformed limit ($q \to 1$). This nonlinear expansion of the spectrum alters the density of states without breaking the multiplet degeneracy structure, providing a microscopic basis for the novel thermodynamic features observed in the system.

In the ferromagnetic regime, the model displays the physical limitations previously reported for the $j=1/2$ case, although they manifest at even smaller system sizes. While for $j=1/2$ the model exhibited an unphysical shift of the specific heat peak toward high temperatures ($T \ge 3$) for $N > 9$ (and $I=1$), in the spin-1 case, this threshold drops to $N > 5$, a feature further aggravated by $q$-deformation. As in the $j=1/2$ model, the large energy gap between the lowest-lying states (which grows with $N$) combined with their small degeneracy relative to the excited states prevents low-level analytical approximations from capturing moderate-temperature behaviour, causing the specific heat response to fall outside a realistic thermal range for larger $N$.

When taking the thermodynamic limit, the previous constraints for a large number of particles disappear because of the coupling change ($I\rightarrow I/N$) and the deformation rescaling ($\eta\rightarrow \eta/N$). Then, the thermodynamic limit has been extensively analysed. In the undeformed case, the behaviour is almost entirely independent of the number of particles: the peaks in the specific heat are similar in magnitude and occur at the same temperature; the Curie transition in magnetic susceptibility occurs at nearly the same temperature; and the magnetisation curves are virtually identical. In stark contrast, the deformed case shows a dramatic change in behaviour. This highlights the need for careful selection of deformation parameters. In this study, we chose exceptionally small deformations, much smaller than those used in the antiferromagnetic case to observe gradual changes. In all three thermodynamic functions, the introduction of deformation leads to a significant shift toward higher temperatures.

In addition to the other thermodynamic properties, the Curie temperature was also analysed in the ferromagnetic case. In the undeformed scenario, we presented various representations of the magnetic susceptibility to visualise the transitions. Reproducing the Kittel's method, we found the analytical expression of Curie temperature for individual spin $j=1$. In the deformed scenario, we observed that the Curie temperature increases as the deformation grows. To derive an approximate analytical expression, we utilised the relationship between the partition function and the critical temperature, as derived in the $j=1/2$ case, developing several approaches to obtain analytical expressions for both small and large values of the deformation parameter.

In the antiferromagnetic case, by considering the most probable levels, we were able to find a reasonable and useful approximation (valid at low temperatures); thus, analytical expressions for both undeformed and deformed scenarios were obtained, from which we could understand the behaviour of the thermodynamic quantities and phase transitions. This approximation based on considering the two most probable energy levels is more accurate for the deformed case because the separation between the energy levels increases with higher values of the deformation parameter, implying transitions to shift to higher temperatures. Moreover, we were able to find (exact) analytical expressions for the magnetic field at which phase transitions occur for both deformed and undeformed cases. 

We observe that the analytical expressions of the thermodynamic quantities considered here for a small number of particles converge to the same value, up to a factor $1/N$, for both the undeformed and deformed cases. However, upon taking the thermodynamic limit, both the deformation parameter and the interaction coupling $I$ and the deformation parameter $\eta$ are rescaled with the system size ($\eta \to \eta/N$ and $I \to I/N$). It is precisely the rescaling of these constants that modifies the energy scale of the system, causing this small-$N$ convergence to disappear and revealing distinct thermodynamic behaviours for the deformed and undeformed models in the infinite-size limit.

As highlighted in the preliminary analysis of the energy level distributions, the $q$-deformation reshapes the density of states by non-linearly expanding the energy spectrum and pushing excited states toward higher energies. This spectral stretching has a profound effect on the thermodynamic response. In the undeformed case, the specific heat displays smooth behaviour, and the magnetic susceptibility curves for different system sizes collapse onto a single universal curve; however, this pattern breaks down under deformation. In particular, for sufficiently large values of the deformation parameter, the specific heat develops a series of distinct bumps or shoulders at intermediate temperatures. These structures—absent in both the undeformed thermodynamic limit and small deformed systems—are Schottky-like anomalies. They originate from the sequential thermal activation across the widened energy gaps: as the temperature increases, low-lying multiplets with large initial Boltzmann weights give way to higher energy levels whose activation requires crossing the $q$-deformed energy thresholds.

The most radical changes were observed in the magnetisation. In the undeformed case, the magnetisation curves for different system sizes almost completely overlap and start near full saturation at zero temperature. This behaviour arises because the thermodynamics is dominated by the states with the highest total quantum numbers $J$ and $m$, rendering the thermal decay essentially independent of $N$. However, under $q$-deformation, this universal behaviour breaks down: the curves for different $N$ separate significantly and even originate at distinct values at zero temperature. This is because the most probable states—and, in particular, the ground state—shift depending on both the system size $N$ and the deformation parameter. In close analogy with the $j=1/2$ case, an explicit analytical formula for these ground-state levels has been derived, providing precise tracking of how the deformation modifies the ground-state structure and its corresponding zero-temperature magnetisation.

Furthermore, a systematic finite-size scaling analysis was performed to characterise the critical behaviour and elucidate how the spin-1 system approaches the thermodynamic limit. In the ferromagnetic regime, the scaling of both the specific heat and the magnetic susceptibility with $N$ at the critical temperature follows the same behaviour observed for $j=1/2$: in both thermodynamic properties, a discrepancy with the fit in the low-deformation range is exhibited, whereas beyond a certain threshold, the fit becomes accurate and remains stable thereafter. In the antiferromagnetic case, the low-temperature finite-size response is completely analogous to that of the $j=1/2$ model with an even number of particles. Notably, our analysis demonstrates that the $j=1$ model reproduces the exact same universal scaling exponents as the even $N$ of the $j=1/2$ case: it asymptotically converges to a universal scaling law independently of $q$, where the specific heat peak decays as $N^{-1}$ and the magnetic susceptibility approaches a size-independent limit. This proves that the overall structure of the scaling laws is remarkably robust not only against algebraic $q$-deformations, but also under extensions to higher spin representations.

Finally, several promising avenues remain open for future investigation. First, a thorough exploration of the complex deformation parameter regime is worth studying: specifically, when $q$ lies on the unit circle ($\vert{}q\vert{}=1$) but not a root of unity, because the theory of representations changes dramatically \cite{biedenharn1995quantum}. From a phenomenological perspective, the site-dependent interactions naturally generated by the $q$-coproduct offer a flexible framework to fit anisotropic exchange couplings in real physical systems, such as families of molecular magnets, magnetic clusters, and quantum dot arrays. Furthermore, the deformation parameter $\eta$ could be reinterpreted as an effective quantitative measure of spatial inhomogeneities, environmental noise, decoherence, or experimental errors in small-scale quantum spin simulators and processors. In particular, while tetrahedral particles (the family of $\text{Cu}_2\text{Te}_2\text{O}_5X_2$ compounds) have been modelled successfully \cite{mariscal2025thermodynamics}, extending this approach to the $j=1$ case could provide an excellent candidate for studying triangular molecules among many other configurations. This specific geometric configuration is better suited to the $j=1$ case because, unlike the $j=1/2$ case, it introduces a richer structure of energy levels, which prevents the coupling constant from absorbing the role of the deformation parameter. Exploring these open problems will bridge the gap between algebraic quantum group symmetries and real-world quantum materials and devices.

\section*{Acknowledgments}
The authors appreciate useful discussions with Nicolás Cordero. 
This work has been partially supported by Agencia Estatal de Investigaci\'on (Spain) under grant PID2019-106802GB-I00/AEI/10.13039/501100011033, by the Regional Government of Castilla y Le\'on (Junta de Castilla y Le\'on, Spain), and by the Spanish Ministry of Science and Innovation MICIN and the European Union NextGenerationEU (PRTR C17.I1).

\appendix
\section{Coefficients of the multiplicity of levels}
\label{sec:appendix}
In this appendix, we show how the result of Eq.~\eqref{eq:degeneracy} is obtained. This can be derived by the multinomial theorem: 
\begin{equation}
(x_{1}+x_{2}+\cdot\cdot\cdot+x_{m})^{N}=\sum_{k_{1}+k_{2}+\cdot\cdot\cdot+k_{m}=N;k_{1},k_{2},\cdot\cdot\cdot,k_{m}\geq 0}\binom{N}{k_{1},k_{2},...,k_{m}}\prod_{t=1}^{m}x_{t}^{k_{t}},
\end{equation}
where
\begin{equation}
    \binom{N}{k_{1},k_{2},...,k_{m}}=\frac{N!}{k_{1}!k_{2}!\cdot\cdot\cdot k_{m}!}
\end{equation}
is a multinomial coefficient. In the case considered in this paper, which is $j=1$, we have from Eq.~\eqref{eq:degeneracy_Morse} the next multinomial expression 
\begin{equation}
    \left(x+1+\frac{1}{x}\right)^{N},
\end{equation}
so the coefficients we look for can be expressed as:
\begin{equation}
    \Omega(N,t,k)=\frac{N!}{t!k! (N-t-k)!},
\end{equation}
where $t$ is the power of $x$ and $k$ the corresponding one of $1/x$. As these powers are mixed in the multinomial expression it is preferable to change these indexes, so the global angular momenta $J$ appears. The coefficient with power $J$, that is, the one of $x$, is defined by the corresponding term with power $t-k$. Following this argument, the generic coefficient of $x$ can be expressed as:
\begin{equation}
    \Omega(N,J)=\sum_{k=0}^{\frac{N-J}{2}}\frac{N!}{(J+k)!k!(N-J-2k)!},
\end{equation}
or, in terms of hypergeometric functions,
\begin{equation}
    \Omega(N,J)=\frac{N! \, _2F_1\left(\frac{J-N}{2},\frac{1}{2} (J-N+1);J+1;4\right)}{J! (N-J)!}=  {}_2F_1\left(\frac{J-N}{2},\frac{1}{2} (J-N+1);J+1;4\right)\binom{N}{J}.
\end{equation}

\bibliographystyle{abbrvurlmendeley}
\bibliography{references}

@article{kac1968statistical,
  title={Statistical physics, phase transitions and superfluidity},
  author={Kac, M},
  journal={Brandeis University Summer Institute in Theoretical Physics},
  volume={1},
  pages={241--305},
  year={1968}
}

@article{mariscal2025thermodynamics,
  author  = {Mariscal, V. and Relancio, J. J.},
  title   = {Thermodynamics of the $q$-deformed {K}ittel--{S}hore model},
  journal = {Int. J. Mod. Phys. B},
  year    = {2026},
  month   = {09},
  note    = {Accepted, to be published}
}

@book{majid2000foundations,
   author = {Shahn Majid},
   city = {Cambridge},
   doi = {10.1017/CBO9780511613104},
   isbn = {9780521460323},
   month = {12},
   publisher = {Cambridge University Press},
   title = {Foundations of $\text{Q}$uantum $\text{G}$roup $\text{T}$heory},
   url = {https://www.cambridge.org/core/product/identifier/9780511613104/type/book},
   year = {1995},
}

@book{biedenharn1995quantum,
   author = {L C Biedenharn and M A Lohe},
   doi = {10.1142/2815},
   isbn = {978-981-02-2331-1},
   month = {8},
   publisher = {World Scientific},
   title = {Quantum $\text{G}$roup $\text{S}$ymmetry and $\text{Q}$-$\text{T}$ensor $\text{A}$lgebras},
   url = {https://www.worldscientific.com/worldscibooks/10.1142/2815},
   year = {1995},
}

@article{curtright1991quantum,
   author = {T. L. Curtright and G. I. Ghandour and C. K. Zachos},
   doi = {10.1063/1.529410},
   issn = {0022-2488},
   issue = {3},
   journal = {Journal of Mathematical Physics},
   month = {3},
   pages = {676-688},
   title = {Quantum algebra deforming maps, $\text{C}$lebsch–$\text{G}$ordan coefficients, coproducts, $\text{R}$ and $\text{U}$ matrices},
   volume = {32},
   url = {https://pubs.aip.org/jmp/article/32/3/676/229295/Quantum-algebra-deforming-maps-Clebsch-Gordan},
   year = {1991},
}

@article{czachor2002verification,
   author = {Andrzej Czachor},
   doi = {10.1016/S0921-4526(01)01056-0},
   issn = {09214526},
   issue = {1-2},
   journal = {Physica B: Condensed Matter},
   month = {1},
   pages = {56-60},
   title = {A verification of the random-phase-approximation using exact thermodynamic functions for the $\text{K}$ittel–$\text{S}$hore–$\text{K}$ac model magnet},
   volume = {311},
   url = {https://linkinghub.elsevier.com/retrieve/pii/S0921452601010560},
   year = {2002},
}

@article{van1993heisenberg,
   author = {A. J. van der Sijs},
   doi = {10.1103/PhysRevB.48.7125},
   issn = {0163-1829},
   issue = {10},
   journal = {Physical Review B},
   month = {9},
   pages = {7125-7133},
   title = {Heisenberg models and a particular isotropic model},
   volume = {48},
   url = {https://link.aps.org/doi/10.1103/PhysRevB.48.7125},
   year = {1993},
}

@article{czachor2008energy,
   author = {A. Czachor},
   doi = {10.12693/APhysPolA.113.1161},
   issn = {0587-4246},
   issue = {4},
   journal = {Acta Physica Polonica A},
   month = {4},
   pages = {1161-1169},
   title = {Energy $\text{S}$pectrum for the $\text{S}$ystem of $\text{N}$ $\text{I}$sing $\text{S}$pins with $\text{I}$dentical $\text{S}$pin-$\text{S}$pin $\text{C}$oupling $\text{K}$/$\text{N}$ - $\text{A}$natomy of $\text{P}$hase $\text{T}$ransition},
   volume = {113},
   url = {http://przyrbwn.icm.edu.pl/APP/PDF/113/a113z406.pdf},
   year = {2008},
}

@article{morse1932theory,
   author = {Philip M. Morse},
   doi = {10.1126/science.76.1971.326},
   issn = {0036-8075},
   issue = {1971},
   journal = {Science},
   month = {10},
   pages = {326-328},
   title = {The $\text{T}$heory of $\text{E}$lectric and $\text{M}$agnetic $\text{S}$usceptibilities. $\text{B}$y $\text{J}$. $\text{H}$. van $\text{V}$leck. $\text{O}$xford $\text{U}$niversity $\text{P}$ress, 384 pages, 1932.},
   volume = {76},
   year = {1932},
}

@article{kittel1965development,
   author = {C. Kittel and H. Shore},
   doi = {10.1103/PhysRev.138.A1165},
   issn = {0031-899X},
   issue = {4A},
   journal = {Physical Review},
   month = {5},
   pages = {A1165-A1169},
   title = {Development of a $\text{P}$hase $\text{T}$ransition for a $\text{R}$igorously $\text{S}$olvable $\text{M}$any-$\text{B}$ody $\text{S}$ystem},
   volume = {138},
   url = {https://link.aps.org/doi/10.1103/PhysRev.138.A1165},
   year = {1965},
}

@article{al1998exact,
   author = {Housni Al-Wahsh and Miklós Urbán and Andrzej Czachor},
   doi = {10.1016/S0304-8853(97)01147-5},
   isbn = {03048853/98},
   issn = {03048853},
   issue = {2},
   journal = {Journal of Magnetism and Magnetic Materials},
   month = {6},
   pages = {L144-158},
   title = {Exact solutions for a model antiferromagnet with identical coupling between spins},
   volume = {185},
   url = {https://linkinghub.elsevier.com/retrieve/pii/S0304885397011475},
   year = {1998},
}

@book{le2004equilibrium,
   author = {Michel Le Bellac and Fabrice Mortessagne and G. George Batrouni},
   doi = {10.1017/CBO9780511606571},
   isbn = {9780521528955},
   month = {4},
   pages = {112-113},
   publisher = {Cambridge University Press},
   title = {Equilibrium and $\text{N}$on-$\text{E}$quilibrium $\text{S}$tatistical $\text{T}$hermodynamics},
   year = {2004},
}

@article{curtright2017spin,
   author = {T.L. Curtright and T.S. Van Kortryk and C.K. Zachos},
   doi = {10.1016/j.physleta.2016.12.006},
   issn = {03759601},
   issue = {5},
   journal = {Physics Letters A},
   month = {2},
   pages = {422-427},
   title = {Spin multiplicities},
   volume = {381},
   year = {2017},
}

@article{ballesteros2025quantum,
   author = {A Ballesteros and I Gutierrez-Sagredo and V Mariscal and J J Relancio},
   doi = {10.1088/1751-8121/ae1272},
   issn = {1751-8113},
   issue = {44},
   journal = {Journal of Physics A: Mathematical and Theoretical},
   month = {11},
   pages = {445202},
   title = {Quantum group deformation of the $\text{K}$ittel–$\text{S}$hore model},
   volume = {58},
   year = {2025}
}

@article{botet1983large,
   author = {R. Botet and R. Jullien},
   doi = {10.1103/PhysRevB.28.3955},
   issn = {0163-1829},
   issue = {7},
   journal = {Physical Review B},
   month = {10},
   pages = {3955-3967},
   title = {Large-size critical behavior of infinitely coordinated systems},
   volume = {28},
   year = {1983}
}

@article{botet1982size,
   author = {R. Botet and R. Jullien and P. Pfeuty},
   doi = {10.1103/PhysRevLett.49.478},
   issn = {0031-9007},
   issue = {7},
   journal = {Physical Review Letters},
   month = {8},
   pages = {478-481},
   title = {Size $\text{S}$caling for $\text{I}$nfinitely $\text{C}$oordinated $\text{S}$ystems},
   volume = {49},
   year = {1982}
}

@inbook{cardy1988current,
   author = {John L. Cardy},
   doi = {10.1016/B978-0-444-87109-1.50006-6},
   pages = {1-7},
   title = {Introduction to Theory of Finite-Size Scaling},
   publisher = {Elsevier},
   volume = {2},
   year = {1988}
}

@article{bjornberg2016free,
   author = {J. E. Björnberg},
   doi = {10.1063/1.4959238},
   issn = {0022-2488},
   issue = {7},
   journal = {Journal of Mathematical Physics},
   month = {7},
   title = {The free energy in a class of quantum spin systems and interchange processes},
   volume = {57},
   year = {2016}
}

@article{bjornberg2020quantum,
   author = {Jakob E. Björnberg and Jürg Fröhlich and Daniel Ueltschi},
   doi = {10.1007/s00220-019-03634-x},
   issn = {0010-3616},
   issue = {3},
   journal = {Communications in Mathematical Physics},
   month = {5},
   pages = {1629-1663},
   title = {Quantum Spins and Random Loops on the Complete Graph},
   volume = {375},
   year = {2020}
}

@article{fannes1980equilibrium,
   author = {M. Fannes and H. Spohn and A. Verbeure},
   doi = {10.1063/1.524422},
   issn = {0022-2488},
   issue = {2},
   journal = {Journal of Mathematical Physics},
   month = {2},
   pages = {355-358},
   title = {Equilibrium states for mean field models},
   volume = {21},
   year = {1980}
}

@article{papanicolaou1986ground,
   author = {N. Papanicolaou},
   doi = {10.1016/0375-9601(86)90246-X},
   issn = {03759601},
   issue = {2},
   journal = {Physics Letters A},
   month = {5},
   pages = {89-93},
   title = {Ground-state properties of spin-1 nematics},
   volume = {116},
   year = {1986}
}

@article{jakab2018bilinear,
   author = {Dávid Jakab and Gergely Szirmai and Zoltán Zimborás},
   doi = {10.1088/1751-8121/aaa92b},
   issn = {1751-8113},
   issue = {10},
   journal = {Journal of Physics A: Mathematical and Theoretical},
   month = {3},
   pages = {105201},
   title = {The bilinear–biquadratic model on the complete graph},
   volume = {51},
   year = {2018}
}

@article{toth1990phase,
   author = {Bálint Tóth},
   doi = {10.1007/BF01027300},
   issn = {0022-4715},
   issue = {3-4},
   journal = {Journal of Statistical Physics},
   month = {11},
   pages = {749-764},
   title = {Phase transition in an interacting Bose system. An application of the theory of $\text{V}$entsel' and $\text{F}$reidlin},
   volume = {61},
   year = {1990}
}

@article{penrose1991bose,
   author = {O. Penrose},
   doi = {10.1007/BF01029210},
   issn = {0022-4715},
   issue = {3-4},
   journal = {Journal of Statistical Physics},
   month = {5},
   pages = {761-781},
   title = {Bose-$\text{E}$instein condensation in an exactly soluble system of interacting particles},
   volume = {63},
   year = {1991}
}

@article{ryan2023class,
   author = {Kieran Ryan},
   doi = {10.1093/imrn/rnac034},
   issn = {1073-7928},
   issue = {7},
   journal = {International Mathematics Research Notices},
   month = {3},
   pages = {6078-6131},
   title = {On a Class of Orthogonal-Invariant Quantum Spin Systems on the Complete Graph},
   volume = {2023},
   year = {2023}
}

@article{alon2021mean,
   author = {Gil Alon and Gady Kozma},
   doi = {10.1214/20-AIHP1067},
   issn = {0246-0203},
   issue = {3},
   journal = {Annales de l'Institut Henri Poincaré, Probabilités et Statistiques},
   month = {7},
   title = {The mean-field quantum $\text{H}$eisenberg ferromagnet via representation theory},
   volume = {57},
   year = {2021}
}

@article{bjornberg2023heisenberg,
   author = {J.E. Björnberg and H. Rosengren and K. Ryan},
   doi = {10.1016/j.aam.2023.102572},
   issn = {01968858},
   journal = {Advances in Applied Mathematics},
   month = {10},
   pages = {102572},
   title = {Heisenberg models and $\text{S}$chur–$\text{W}$eyl duality},
   volume = {151},
   year = {2023}
}

@book{vladimir1986drinfeld,
   author = {V. G. Drinfeld},
   city = {Providence RI},
   pages = {798-820},
   publisher = {American Mathematical Society},
   title = {Quantum Groups},
   year = {1987}
}

@article{jimbo1985q,
   author = {Michio Jimbo},
   doi = {10.1007/BF00704588},
   issn = {0377-9017},
   issue = {1},
   journal = {Letters in Mathematical Physics},
   month = {7},
   pages = {63-69},
   title = {A $q$-$\text{D}$ifference $\text{A}$nalogue of \textit{U}$(g)$ and the $\text{Y}$ang-$\text{B}$axter $\text{E}$quation},
   volume = {10},
   year = {1985}
}

@article{sklyanin1988boundary,
   author = {E K Sklyanin},
   doi = {10.1088/0305-4470/21/10/015},
   issn = {0305-4470},
   issue = {10},
   journal = {Journal of Physics A: Mathematical and General},
   month = {5},
   pages = {2375-2389},
   title = {Boundary conditions for integrable quantum systems},
   volume = {21},
   year = {1988}
}

@article{pasquier1990common,
   author = {V. Pasquier and H. Saleur},
   doi = {10.1016/0550-3213(90)90122-T},
   issn = {05503213},
   issue = {2-3},
   journal = {Nuclear Physics B},
   month = {1},
   pages = {523-556},
   title = {Common structures between finite systems and conformal field theories through quantum groups},
   volume = {330},
   year = {1990}
}

@article{gomez1996quantum,
  title={Quantum Groups in Two-Dimensional Physics},
  author={G{\'o}mez, Cisar and Ruiz-Altaba, Mart{\'\i}n and Sierra, German},
  journal={Quantum Groups in Two-Dimensional Physics},
  pages={475},
  year={1996}
}

@article{kulish1991general,
   author = {P P Kulish and E K Sklyanin},
   doi = {10.1088/0305-4470/24/8/009},
   issn = {0305-4470},
   issue = {8},
   journal = {Journal of Physics A: Mathematical and General},
   month = {4},
   pages = {L435-L439},
   title = {The general $\textit{U}_{q}(sl(2))$ invariant $\text{XXZ}$ integrable quantum spin chain},
   volume = {24},
   year = {1991}
}

@article{martin1994blob,
  title={The blob algebra and the periodic $\text{T}$emperley-$\text{L}$ieb algebra},
  author={Martin, Paul and Saleur, Hubert},
  journal={Letters in mathematical physics},
  volume={30},
  number={3},
  pages={189--206},
  year={1994},
  publisher={Springer}
}

@article{martin1993algebraic,
  title={On an algebraic approach to higher dimensional statistical mechanics},
  author={Martin, Paul and Saleur, Hubert},
  journal={Communications in mathematical physics},
  volume={158},
  number={1},
  pages={155--190},
  year={1993},
  publisher={Springer}
}

@article{lamers2022spin,
   author = {Jules Lamers and Vincent Pasquier and Didina Serban},
   doi = {10.1007/s00220-022-04318-9},
   issn = {0010-3616},
   issue = {1},
   journal = {Communications in Mathematical Physics},
   month = {7},
   pages = {61-150},
   title = {Spin-$\text{R}$uijsenaars, $q$-$\text{D}$eformed $\text{H}$aldane–$\text{S}$hastry and $\text{M}$acdonald $\text{P}$olynomials},
   volume = {393},
   year = {2022}
}

@article{hakobyan1996spin,
   author = {T. Hakobyan and A. Sedrakyan},
   doi = {10.1016/0370-2693(95)01320-2},
   issn = {03702693},
   issue = {4},
   journal = {Physics Letters B},
   month = {6},
   pages = {250-254},
   title = {Spin chain $\text{H}$amiltonians with affine $\textit{U}_{q}g$ symmetry},
   volume = {377},
   year = {1996}
}

@article{matushko2022elliptic,
   author = {M Matushko and A Zotov},
   doi = {10.1088/1361-6544/aca510},
   issn = {0951-7715},
   issue = {1},
   journal = {Nonlinearity},
   month = {1},
   pages = {319-353},
   title = {Elliptic generalisation of integrable $q$-deformed anisotropic $\text{H}$aldane–$\text{S}$hastry long-range spin chain},
   volume = {36},
   year = {2023}
}

@article{klabbers2024deformed,
   author = {Rob Klabbers and Jules Lamers},
   doi = {10.21468/SciPostPhys.17.6.155},
   issn = {2542-4653},
   issue = {6},
   journal = {SciPost Physics},
   month = {12},
   pages = {155},
   title = {The deformed $\text{I}$nozemtsev spin chain},
   volume = {17},
   year = {2024}
}

@article{ballesteros1998systematic,
   author = {Angel Ballesteros and Orlando Ragnisco},
   doi = {10.1088/0305-4470/31/16/009},
   issn = {0305-4470},
   issue = {16},
   journal = {Journal of Physics A: Mathematical and General},
   month = {4},
   pages = {3791-3813},
   title = {A systematic construction of completely integrable $\text{H}$amiltonians from coalgebras},
   volume = {31},
   year = {1998},
}

\end{document}